\documentclass[a4paper,11pt]{article}
\pdfoutput=1 % if your are submitting a pdflatex (i.e. if you have
\usepackage{jinstpub} % for details on the use of the package, please
\usepackage{graphicx}
\usepackage{comment}
\usepackage{longtable}
\usepackage{color}
\usepackage{caption}
\usepackage{subcaption}
\usepackage{siunitx} %lets you use angstroms with \si{\angstrom}5_Te5
\usepackage{datetime2}
\usepackage{url}
\usepackage{psfrag}
\usepackage{graphics}
\usepackage{graphicx}
\usepackage{float}
\usepackage{epsfig,amsmath,euscript,array}
\usepackage{multicol}
\usepackage{fancyhdr}
\usepackage{lineno}
\def\be{\begin{equation}}
\def\ee{\end{equation}}
\def\ben{\begin{enumerate}}
\def\een{\end{enumerate}}
\def\bi{\begin{itemize}}
\def\ei{\end{itemize}}
\def\bs{\begin{slide}}
\def\es{\end{slide}}
\def\bea{\begin{eqnarray} }
\def\eea{\end{eqnarray} }
\def\bc{\begin{center} }
\def\ec{\end{center} }

\title{\boldmath The Straw Tracking Detector for the Fermilab Muon $g-2$ Experiment}

\author[a,1]{B.~T.~King,\note{Deceased.}}
\author[a]{T.~Albahri,} 
\author[b]{S.~Al-Kilani,}
\author[c]{D.~Allspach,}
\author[c]{D.~Beckner,}
\author[d]{A.~Behnke,}
\author[a]{T.~J.~V.~Bowcock,}
\author[d]{D.~Boyden,}
\author[e]{R.~M.~Carey,}
\author[a]{J.~Carroll,}
\author[c]{B.~C.~K.~Casey,}
\author[c]{S.~Charity,}
\author[b]{R.~Chislett,}
\author[d]{M.~Eads,}
\author[d]{A.~Epps,}
\author[e]{S.~B.~Foster,}
\author[e]{D.~Gastler,}
\author[b]{S.~Grant,}
\author[a]{T.~Halewood-Leagas,}
\author[c]{K.~Hardin,}
\author[e]{E.~Hazen,}
\author[b]{G.~Hesketh,}
\author[a]{D.~J.~Hollywood,}
\author[a]{T.~Jones,}
\author[c]{C.~Kenziora,}
\author[f,2]{A.~Keshavarzi,\note{Corresponding author.}}
\author[c]{M.~Kiburg,}
\author[e]{N.~Kinnaird,}
\author[c]{J.~Kintner,}
\author[f]{M.~Lancaster,}
\author[c]{A.~Luc\`{a},}
\author[b]{G.~Lukicov,}
\author[d]{G.~Luo,}
\author[a]{L.~Mapar,}
\author[a]{S.~J.~Maxfield,}
\author[c,e]{J.~Mott,}
\author[b]{E.~Motuk,}
\author[e]{H.~Mourato,}
\author[d]{N.~Pohlman,}
\author[a]{J.~Price,}
\author[e]{B.~L.~Roberts,}
\author[e,3]{D.~Sathyan,\note{also at Department of Physics, University of Maryland, College Park, MD 20742, USA.}}
\author[d]{M.~Shenk,}
\author[a]{D.~Sim,}
\author[b,4]{T.~Stuttard,\note{also at Neils Bohr Institute, University of Copenhagen, 2100 K\o{}benhaven, Denmark.}}
\author[f]{G.~Sweetmore,}
\author[e]{G.~Thayer,}
\author[a]{K.~Thomson,}
\author[a]{W.~Turner,}
\author[b]{D.~Vasilkova,}
\author[e]{J.~Velho,}
\author[c]{E.~Voirin,}
\author[c]{T.~Walton,}
\author[b]{M.~Warren,}
\author[c]{L.~Welty-Rieger,}
\author[a]{M.~Whitley}
\author[a]{and M.~Wormald}

\affiliation[a]{Department of Physics, University of Liverpool, Liverpool, L69 3BX, United Kingdom}
\affiliation[b]{Department of Physics and Astronomy, University College London, London, WC1E 6BT, United Kingdom}
\affiliation[c]{Fermi National Accelerator Laboratory, Batavia, IL 60510, USA}
\affiliation[d]{Northern Illinois University, DeKalb, IL 60115, USA}
\affiliation[e]{Physics Department, Boston University, Boston, MA 02215, USA}
\affiliation[f]{Department of Physics and Astronomy, University of Manchester, Manchester, M13 9PL, United Kingdom}

\emailAdd{alexander.keshavarzi@manchester.ac.uk}

\abstract{The Muon $g-2$ Experiment at Fermilab uses a gaseous straw tracking
detector to make detailed measurements of the stored muon beam profile, which are 
essential for the experiment to achieve its uncertainty goals. Positrons from 
muon decays spiral inward and pass through the tracking detector
before striking an electromagnetic calorimeter. The tracking detector is therefore located 
inside the vacuum chamber in a region where the magnetic field is large and non-uniform.  As such,  
the tracking detector must have a low leak rate to maintain a high-quality vacuum, must be non-magnetic so as
not to perturb the magnetic field and, to minimize energy loss, must have a low 
radiation length. The performance of the tracking detector has met or surpassed the design requirements, 
with adequate electronic noise levels, an average straw hit resolution of $(110 \pm 20) \,\mu$\si{\meter}, a 
detection efficiency of 97\% or higher, and no performance degradation or signs of aging. 
The tracking detector's measurements result in an otherwise unachievable understanding of the muon's beam 
motion, particularly at early times in the experiment's measurement period when there are a significantly 
greater number of muons decaying. This is vital to the statistical power of the experiment,  as well as facilitating 
the precise extraction of several systematic corrections and uncertainties. This paper 
describes the design, construction, testing, commissioning, and performance of the tracking
detector.}

\begin{document}

% Make the title
\maketitle
\flushbottom

\section{Introduction}
\label{sec:intro}
%% A.Keshavarzi, Stevenage (UK), 03/17/21
%% A.Keshavarzi, Stevenage (UK), 03/16/21

The Muon $g-2$ Experiment (E989) at the Fermi National Accelerator
Laboratory (Fermilab) aims to measure the muon magnetic anomaly,
$a_\mu \equiv (g-2)/2$, to high precision.
Analysis of the Run-1 data has determined $a_\mu$ to 462~parts per billion (ppb)~\cite{Run1PRL,Run1BD,Run1wa,Run1field}
with a systematic uncertainty of 157~ppb and has confirmed the
previous result from the Brookhaven National
Laboratory Experiment (E821)~\cite{Bennett:2002jb,Bennett:2004pv,Bennett:2006fi}.
The Fermilab experiment's sensitivity to a muon electric dipole moment (EDM) will be 
two orders of magnitude beyond the limit set by E821~\cite{Bennett:2008dy}.

Essential to achieving these goals are the experiment's two straw
tracking detector stations. 
They provide non-destructive, time-dependent beam-profile
measurements, by extrapolating the reconstructed trajectories 
of the positrons emitted when the muons decay.
The experiment determines $a_\mu$~\cite{Run1PRL} by measuring two
quantities: the spin precession frequency $\omega_a$ of the
muons~\cite{Run1wa,Run1BD} circulating in the storage ring and the magnetic
field $\vec{B}$~\cite{Run1field} that confines the muons to the storage ring 
and causes their spin to precess relative to their momentum.  The beam-profile 
measurements from the tracking detector are integral to the precise extraction 
of both these quantities and, therefore, the precise measurement of $a_\mu$. 
They have contributed significantly to the overall understanding of the necessary systematic 
corrections and uncertainties, and allow the data from the experiment's electromagnetic calorimeters 
to be fit from an earlier time than would be otherwise possible, resulting in the ability to capitalize 
on the higher number of positrons that are detected earlier in the measurement period.  Horizontal and 
vertical betatron frequencies of the muon beam motion are extracted from the time dependence of 
beam profile measurements~\cite{Run1BD}. The combined beam profile and 
magnetic field measurements determine the average magnetic field traversed by the 
muons~\cite{Run1field}. 

The requirements of the detector to adequately perform these 
measurements are significantly defined by the experimental environment.
They must have an acceptable radiation length to avoid significant
degradation of the positron energy measurement by the downstream
calorimeters, have a low leak rate despite thin straw walls to
maintain a high-quality vacuum, and must be nonmagnetic to ensure they do not
perturb the highly uniform magnetic field. The tracking detector stations were installed in the experiment in 2017 and have
operated successfully without any significant issues since that time. 

This paper focuses on design and construction of the tracking detector. The design of the
tracking detector is described in Section~\ref{sec:design}, followed by details
of the construction in Section~\ref{sec:construction} and the
frontend electronics/data acquisition system in
Section~\ref{sec:electronicsDAQ}. Section~\ref{sec:testInstall}
describes the testing and installation of the detector, and the
subsequent commissioning and experimental performance are detailed in
Section~\ref{sec:performance}.

\section{Design}
\label{sec:design}
\subsection{Detector Principles and Location}

The tracking detector provides non-destructive, time-dependent measurements
of the stored muon beam-profile that are crucial to the experiment's 
determination of $\omega_a$ and $\vec{B}$ for the extraction of $a_\mu$, 
and paramount for its sensitivity to a muon EDM. The primary analysis outputs of 
the tracking detector are positron trajectories from the decays of a high-rate of 
muons, henceforth referred to as `tracks'. The tracks are reconstructed from positron hits in the 
tracking detector and are used to obtain associated properties such as 
momentum and position.  Hits are recorded in the detector for $\sim700~\mu$s,
 commencing 2~$\mu$s after the muon beam enters the storage ring, with the 
 beam arriving in 16 separate spills over a 1.4~s period.  All track times are measured
 relative to the injection time of the muon beam into the storage ring.

Extrapolation of tracks back to the point of azimuthal 
tangency gives an estimate the position of the parent muon at the time of decay. 
Extrapolation at the accuracy needed requires multiple measurements (at a high rate) of the particle's 
three-dimensional position along a number of points. These measurements 
define the track, from which its momentum can also be determined due 
to the presence of the magnetic field. The fundamental design objective 
of the detector is to maximize the number of reconstructed tracks 
and the precision with which the track properties are determined, 
whilst minimizing any detrimental effects on other systems.

To meet the physics goals and design constraints, the tracking detector is built from straw chambers arranged in planes oriented to optimize measurement of the positron tracks in their horizontal bending plane. The straws in the tracking detector layers detect electrons and ions produced when a decay positron passes through a gas-filled straw.\footnote{Multiple
electrons may be produced in a single collision, with the resulting
free electrons referred to as a cluster.} Each straw contains a
central wire held at high voltage (HV) acting as an anode; the grounded
straw walls acting as a cathode. Ionization electrons initially drift radially 
inward in the resulting electric field, but are deflected by the storage ring's 
magnetic field. The signal induced in the wire by drifting electrons and the avalanches they create in the high field indicates the passage of the charged particle.\footnote{Primary electrons 
produced by ionization from the traversing decay positron incur 
subsequent ionization of the gas producing secondary electrons, and so on.}
Ions produced in the avalanche also induce a delayed ($\mathcal{O}\left(\mu
s\right)$ relative to the electrons) signal in the wire which is
safely suppressed by the electronics to avoid interference.

\begin{figure}[!t]
\begin{center}
\begin{subfigure}[h]{0.55\textwidth}
\includegraphics[width=\textwidth]{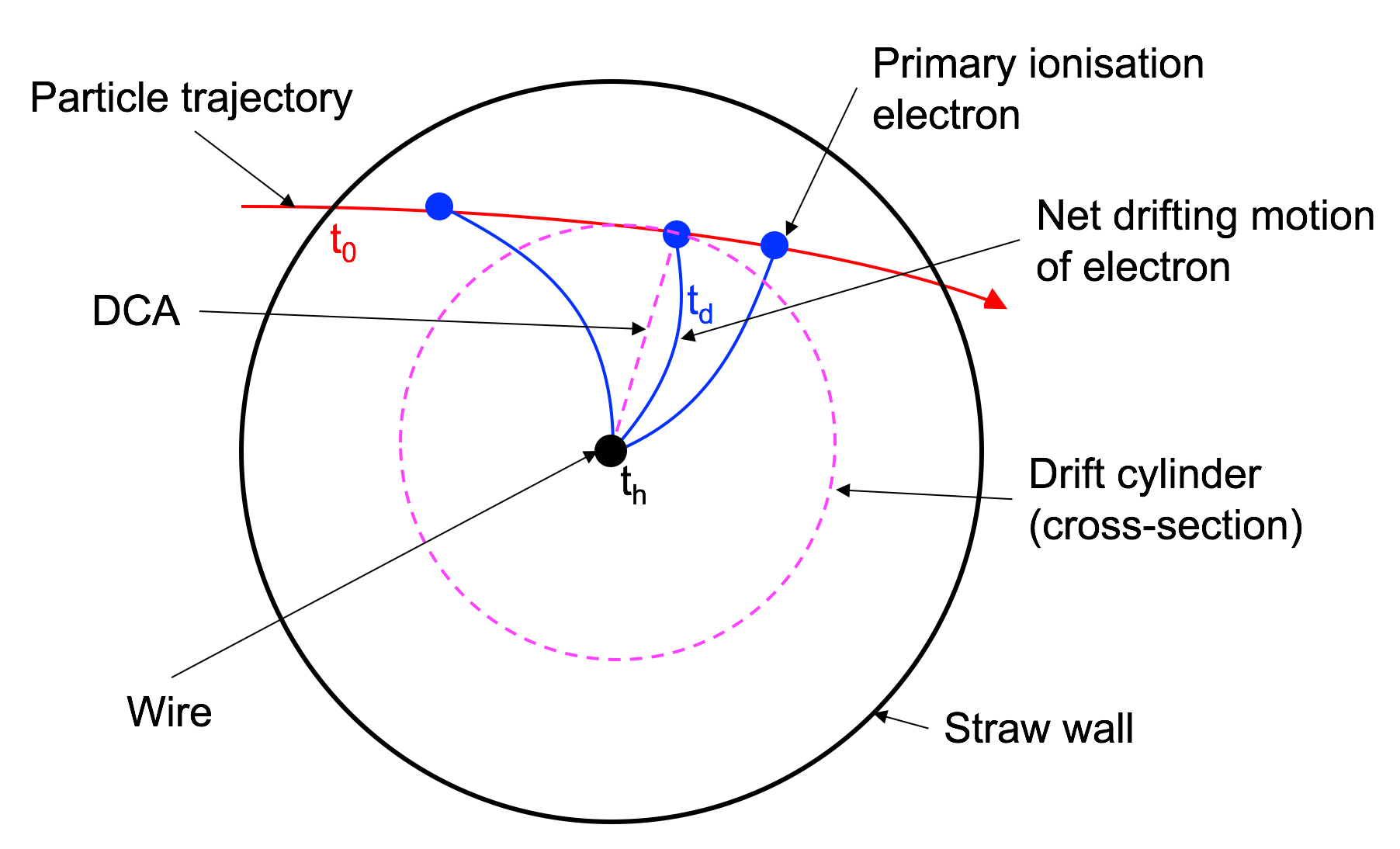}
\caption{Single drift cylinder.}
\label{fig:driftDCAsingle}
\end{subfigure}
\begin{subfigure}[h]{0.265\textwidth}
\includegraphics[width=\textwidth]{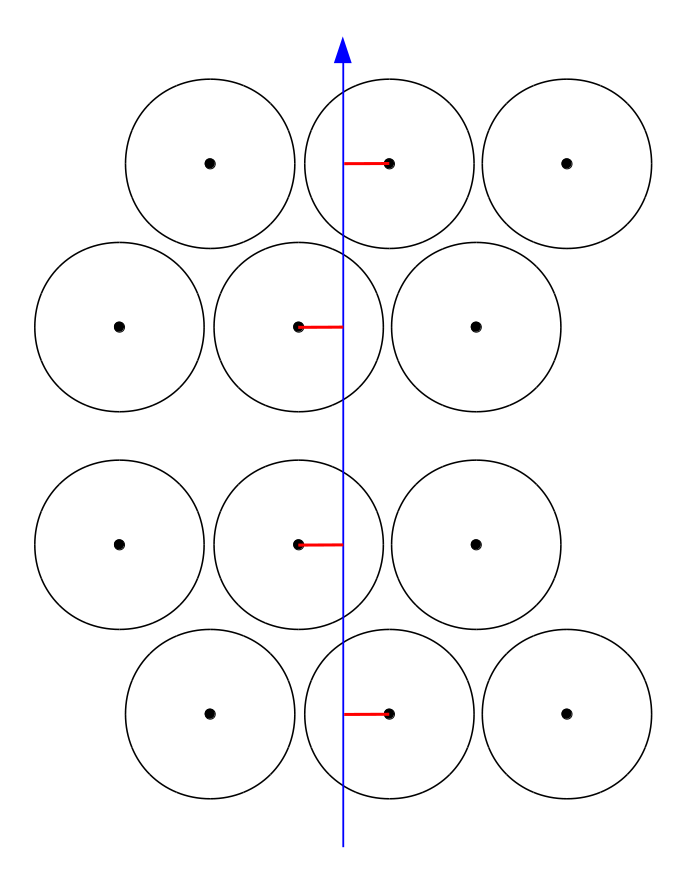}
\caption{Multiple drift cylinders.}
\label{fig:driftDCAmultiple}
\end{subfigure}
\end{center}
\caption{Top-down depiction of the drifting of primary ionization electrons to the straw wire, the DCA reconstructed from the drift time $t_d$ and the subsequent drift cylinder specified by the DCA. Also shown are the measured hit time $t_h$ and the time that the charged particle traversed the straw $t_0$.} \label{fig:driftDCA}
\end{figure}

When the shaped fast electron signal in a given straw exceeds a configurable 
threshold, the readout electronics record the hit time, $t_h$. This his time
includes both the time $t_0$ that the charged particle traversed the
straw and the drift time $t_d$ it took the primary ionization
electrons to drift to the wire, cause an avalanche and trigger the
electronics. The drift time yields the distance of closest approach
(DCA) of the charged particle to the wire, shown in
Figure~\ref{fig:driftDCA}. This distance measurement is a magnitude
with no directional information, and thus specifies a cylinder around
the wire that the traversing particle intersected at a single
point. The track reconstruction combines the drift cylinders produced by a single charged particle,
and the stereo angle between neighboring straw planes allows the determination of the full 3D
particle trajectory.

\begin{figure}[!t]
\centering
\includegraphics[width=0.51\textwidth]{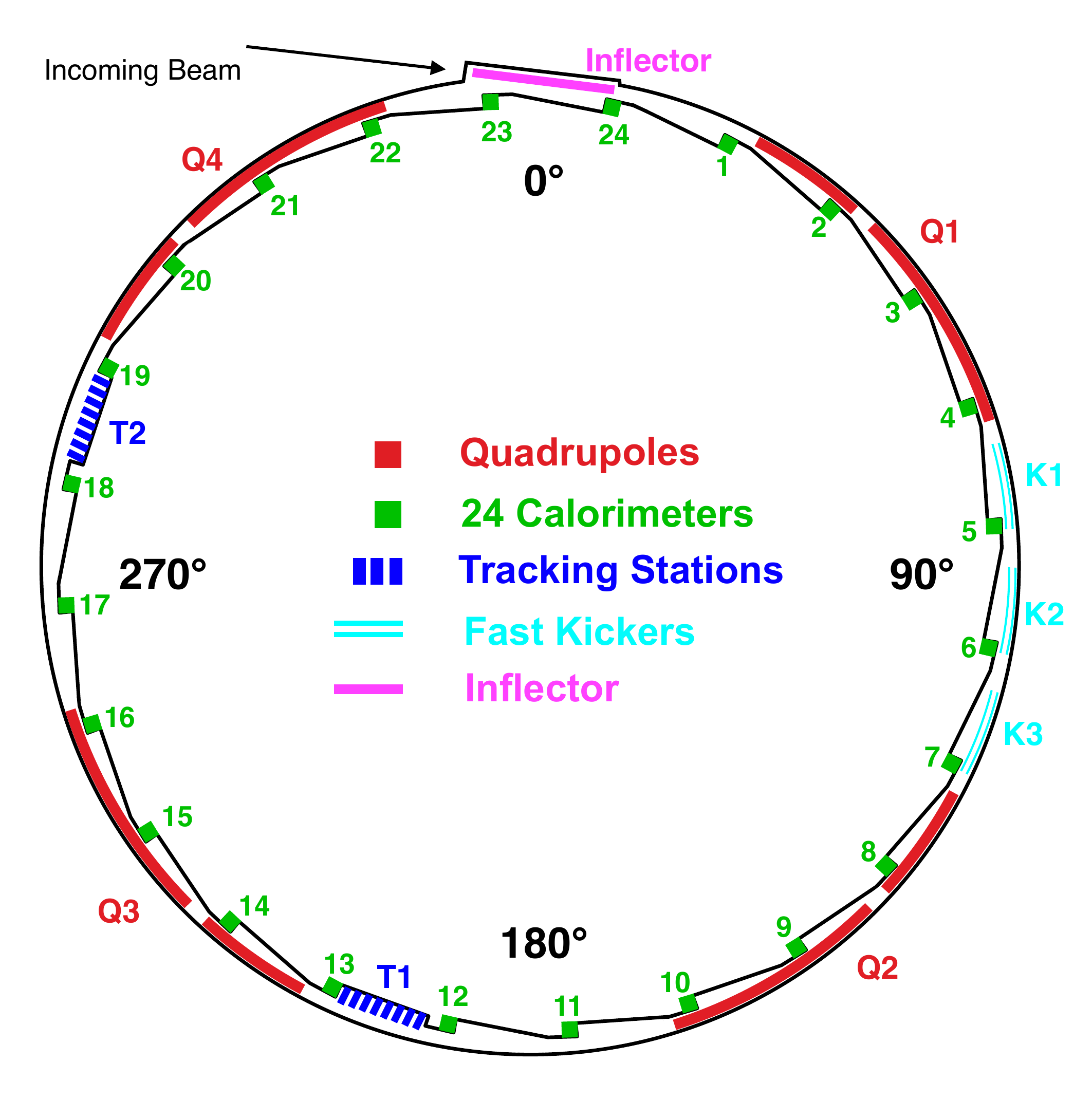}
% AL (06/02): Ring design slightly modified
% AL (05/28): Ring design slightly modified: different colors and less objects 
% - i.e., T1, T2, Q1, Q2, Q3, Q4, Kickers & Inflector -
\caption{The storage ring layout showing the 
location of the 24 calorimeters indicated by number. The beam enters via the inflector and circulates in the clockwise
direction. The tracking detector stations must be placed at locations that have an unobstructed view of the
stored muon beam, where there are no electrostatic quadrupole or fast kicker plates~\cite{Schreckenberger:2021kur,Run1BD} 
in the path of the decay positrons. The design of the vacuum chambers permits the positrons to travel in vacuum to a thick aluminum window, 
behind which the calorimeters are placed. The location of the two straw tracking detector stations (T1 and T2) 
are shown by the dark blue lines just downstream of the $180^\circ$ and $270^\circ$ positions. 
}\label{fig:ring_layout} 
\end{figure}

The tracking detector stations are positioned inside the experiment's vacuum chambers, upstream of a calorimeter~\cite{Fienberg:2014kka,Kaspar:2016ofv,Muong-2:2019hxt} within the storage ring vacuum, but outside of the region where the muons circulate.
Tracking detector stations were installed in two of the three available locations that provide an unobstructed view of the muon storage region: at 180$^\circ$ and 270$^\circ$ with respect to the muon injection point, as shown in Figure~\ref{fig:ring_layout}. The multiple tracking detector stations provide redundancy, cross-checks and complementary information for tuning and optimizing the stored muon beam.

With the tracking detector stations positioned in only two locations in the ring, they measure two azimuthal regions. Measurement of closed-orbit distortions due to azimuthal variation in the experiment's main dipole field requires at least two stations. Each tracking detector station accepts positron decays over a $25^{\circ}$ (3\,m) azimuthal range. Initial studies indicated that, to determine the beam profile, the tracking detector needed to measure the radial position of the tracks per detection plane with a resolution of $<300 ~\mu$m. The demands on vertical resolution are more relaxed as the bending is predominantly in the horizontal plane. 


\subsection{Detector Requirements}
%The overall design requirements for the Muon $g-2$ tracking detector are summarized in Table~\ref{tracker:req_table} and described in detail below. 
The decay positron momenta follow a boosted muon-decay spectrum with an end point of 3.09 GeV/$c$, the momentum of the stored muons. As such, the tracking detector must detect as many decay positrons as possible in a momentum range up to this limit 
%(with the ability to cope with initial hit rates at beam injection of $\simeq 10~{\rm kHz/cm^2}$) 
(with the ability to cope with the initial hit rates at beam injection) 
and reconstruct their tracks with enough fidelity to determine the muon beam profile at the millimeter level after extrapolation. 
%\begin{table}[!t]
%\begin{center}
%\begin{tabular}{|l|c|c|}
%\hline %\hline
%Parameter & Specification Value \qquad & Performance Value \\
%\hline
%Impact parameter resolution &\hfill $\ll 10$ mm & \hfill $\simeq 3$~mm \\
%\hline
%Vertical angular resolution &\hfill $\ll 10$ mrad & \\
%\hline
%Momentum resolution &\hfill $\ll 3.5\%$ at 1 GeV & \\
%\hline
%Vacuum load &\hfill $5 \times 10^{-5}$ Torr L/s &assumes $10^{-6}$ Torr vacuum \\
%& & and E821 pumping speed \\
%\hline
%Instantaneous rate &\hfill $10$ kHz/cm$^2$ & Extrapolated from E821 \\
%\hline
%Ideal coverage &\hfill $160 \times 200$ mm & Front face of calorimeter \\
%\hline
%Number of stations &\hfill $\ge 2$ & Required to constrain beam \\
%& & parameters \\
%\hline
%Time independent field &\hfill $<10$ ppm & Extrapolation from E821 \\
%perturbation & & \\
%\hline
%Transient ($< 1$ ms) field &\hfill $< 10 $ ppb & Invisible to NMR \\
%perturbation & & \\
%\hline
%\end{tabular}
%\end{center}
%\caption{Summary of the major requirements and environmental considerations
%for the tracking detector.\label{tracker:req_table}
%\textcolor[rgb]{1,0,0}{This table is missing a lot of performance values}} 
%\end{table}

\begin{figure}[!t]
\centering
\includegraphics[width=0.8\textwidth]{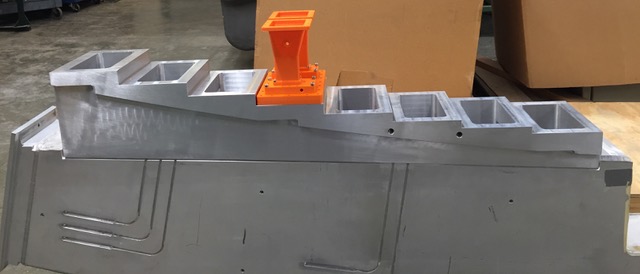}
\caption{Side-view of the modified vacuum chambers before welding. A 3D printed model of a tracker module is shown inserted into one of the step-like ports.} \label{fig:chamberModifications}
\end{figure}

To reduce the uncertainty on decay position from scattering through material, the tracking detector stations reside inside the muon vacuum chamber in scalloped sections. These sections were specifically modified to accommodate the stations immediately upstream of a calorimeter and to locate them as close as possible to the stored muon beam. A fixture welded onto the scalloped vacuum chamber section, as shown in Figure~\ref{fig:chamberModifications}, supports the detector modules in each tracking station. Being located inside the vacuum chamber imposes the requirement that the tracking detector must function in the vacuum of $<10^{-6}$ Torr. Their vacuum load on the system must fall below $5 \times 10^{-5}$ Torr L/s to avoid compromising the performance of the pulsed HV electric quadrupoles and kicker magnet~\cite{Schreckenberger:2021kur} (see~\cite{Run1BD} for further details of these systems).

\begin{figure}[!t]
\begin{center}
%\begin{subfigure}[!t]{0.49\textwidth}
%\includegraphics[width=\textwidth]{pictures/2_Design/BFieldX_y0.pdf}
%\caption{$y=0.0$ mm.}
%\label{fig:bx0}
%\end{subfigure}
\begin{subfigure}[!t]{0.49\textwidth}
\includegraphics[width=\textwidth]{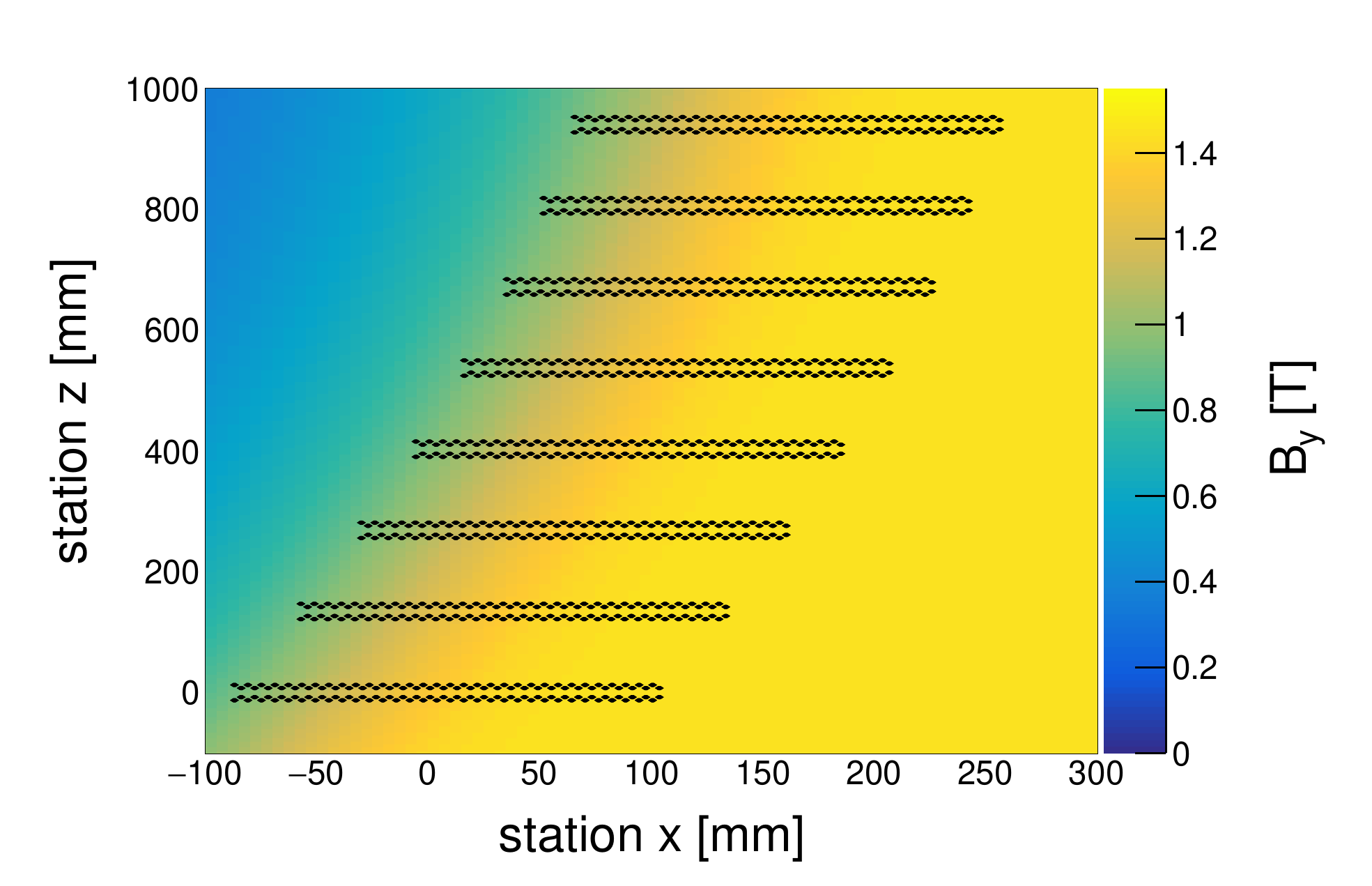}
\caption{Vertical, $y=0.0$ mm.}
\label{fig:by0}
\end{subfigure}
\begin{subfigure}[!t]{0.49\textwidth}
\includegraphics[width=\textwidth]{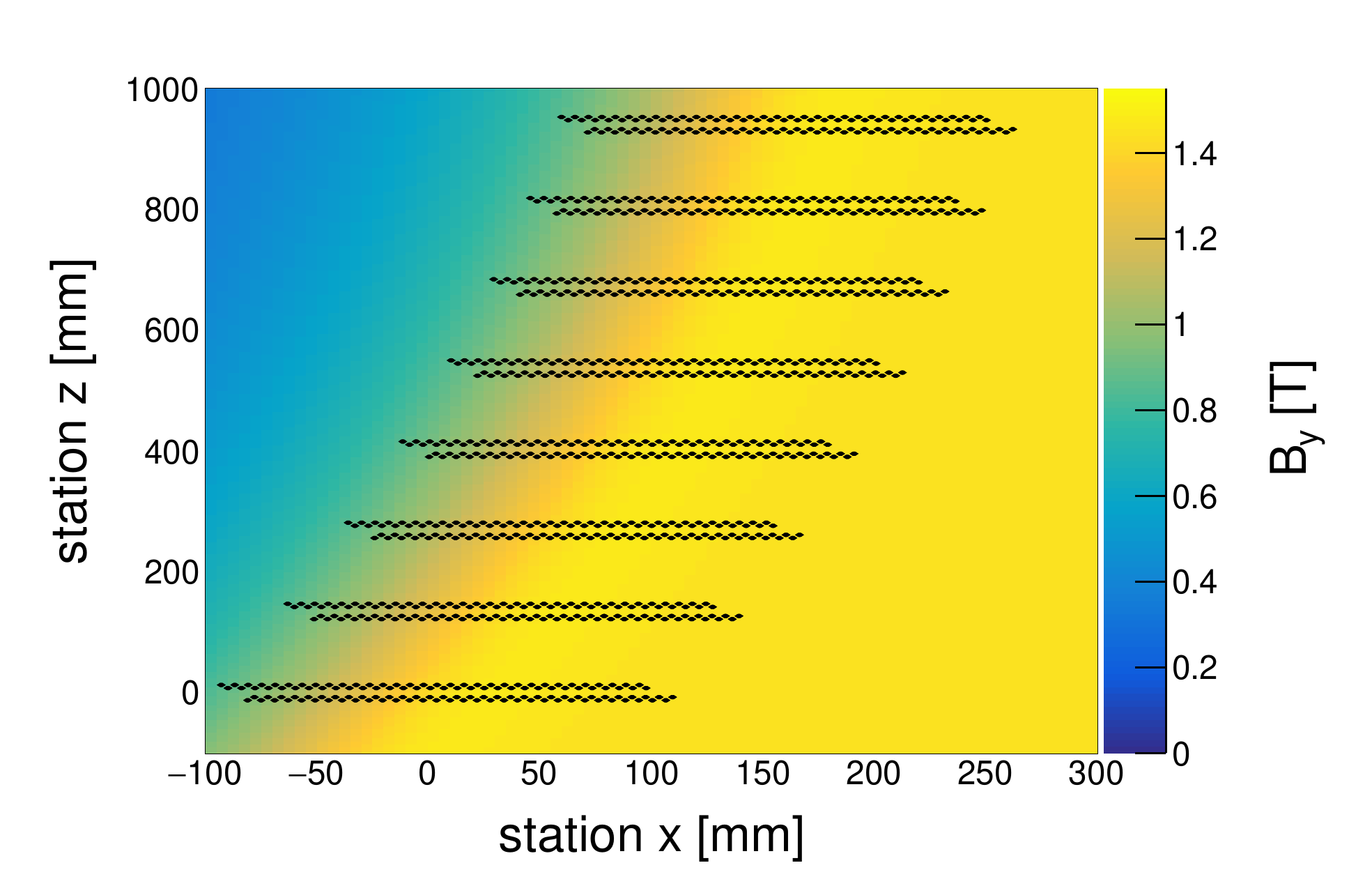}
\caption{Vertical, $y=45.0$ mm.}
\label{fig:by45}
\end{subfigure}
\begin{subfigure}[!t]{0.49\textwidth}
\includegraphics[width=\textwidth]{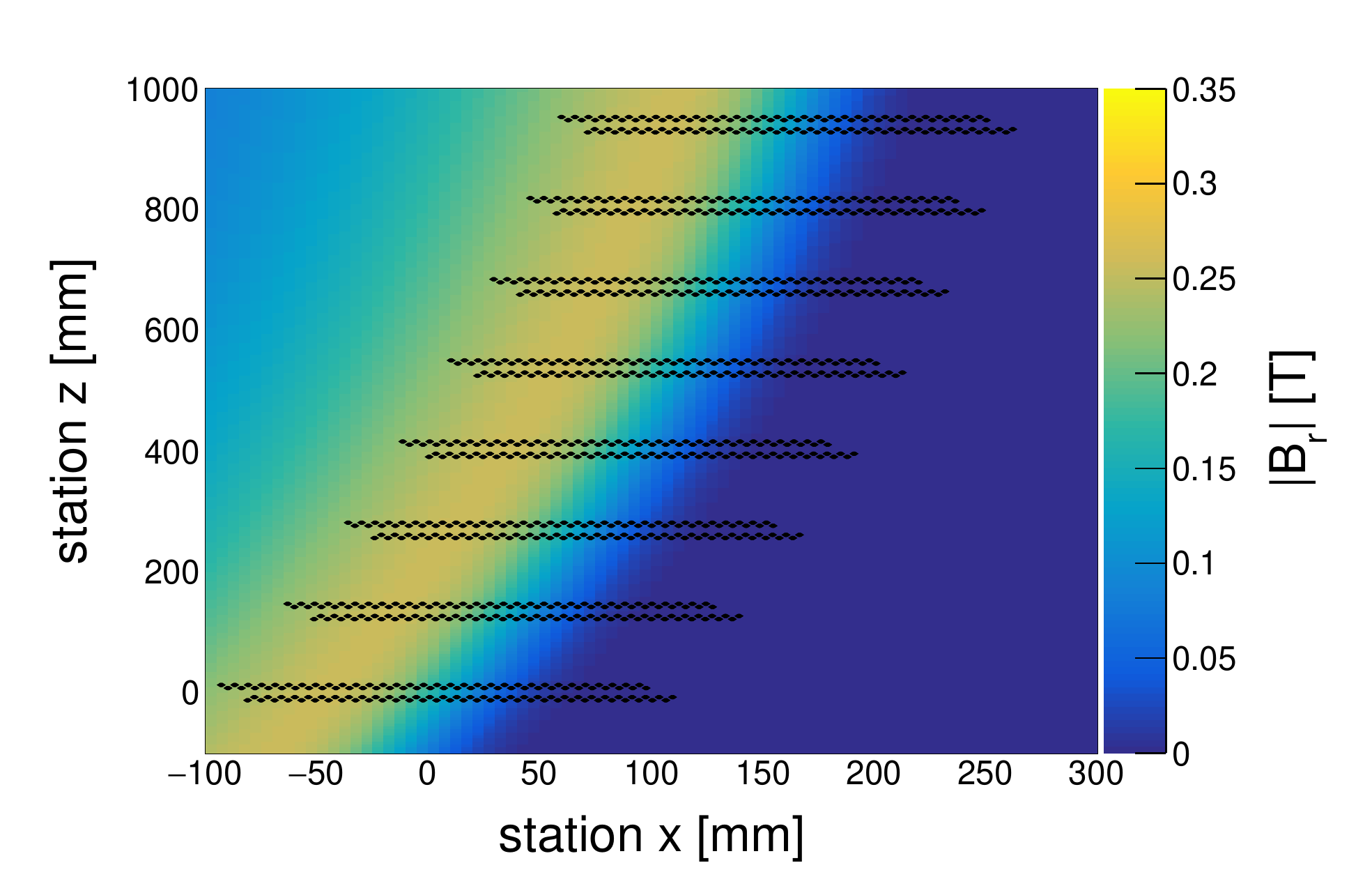}
\caption{Radial, $y=45.0$ mm.}
\label{fig:bx45}
\end{subfigure}
\end{center}
\caption{Plane view of the vertical ($B_y$) and radial ($B_r$)
magnetic field in a tracking detector station location calculated with \texttt{Opera-2D}~\cite{Opera3D}. The muon
beam is to the right of each region, and the storage ring center is
to the left. The black dots indicate the straw locations.
(a) The vertical magnetic field at the midplane of the storage ring.
(b) The vertical magnetic field at the top of tracking detector modules.
(c) The radial magnetic field at the top of the tracking detector modules, 45~mm above the midplane. Note that the scale for the magnitude of the magnetic field is smaller in the radial (c) compared to the vertical (a) and (b).
}\label{fig:fringeField} 
\end{figure}

The vacuum chambers are installed in the 180~mm gap between the magnet pole pieces and are surrounded on three sides by the ring magnet~\cite{Danby:2001eh}. 
As such, the tracking detector modules reside in the magnetic fringe field region of the storage ring as shown in Figure~\ref{fig:fringeField}, resulting in specific requirements on the design. The detector must perform in this field, the significant nonuniformity of which should also not affect the reconstruction algorithm. Additionally, perturbations that the tracking detector modules make to the magnetic field due to material or electrical currents must be below 10 ~parts per million (ppm) at the center of the storage region over an azimuthal extent of greater than 2$^\circ$. Any perturbations due to transient currents on time scales below $1$ ms must be below $10$ ppb since these cannot easily be monitored by the NMR probes. 

Two factors drive the number and positioning of the detector planes. Shorter low momentum tracks must traverse enough planes to allow their reconstruction, while higher momentum tracks require coverage over as large an azimuthal extent possible to provide the lever arm needed for accurate determination of the muon beam profile. Although positioning the tracking detector stations as close as possible radially to the storage region maximizes the overall acceptance, the introduced material must not significantly disrupt the muon beam or degrade the performance of the calorimeters. Consequently, and to minimize multiple scattering, each tracking plane had a $0.5\%$ radiation length material budget. To cover the beam aperture of $\pm 45$ mm, the tracking detector acceptance must be sensitive to 100~mm in height at the storage region. 

%\subsection{Tracking Detector Straws}

\subsection{Detector Design}

\begin{figure}[!t]
\begin{center}
\begin{subfigure}[a]{1.0\textwidth}
\includegraphics[width=\textwidth]{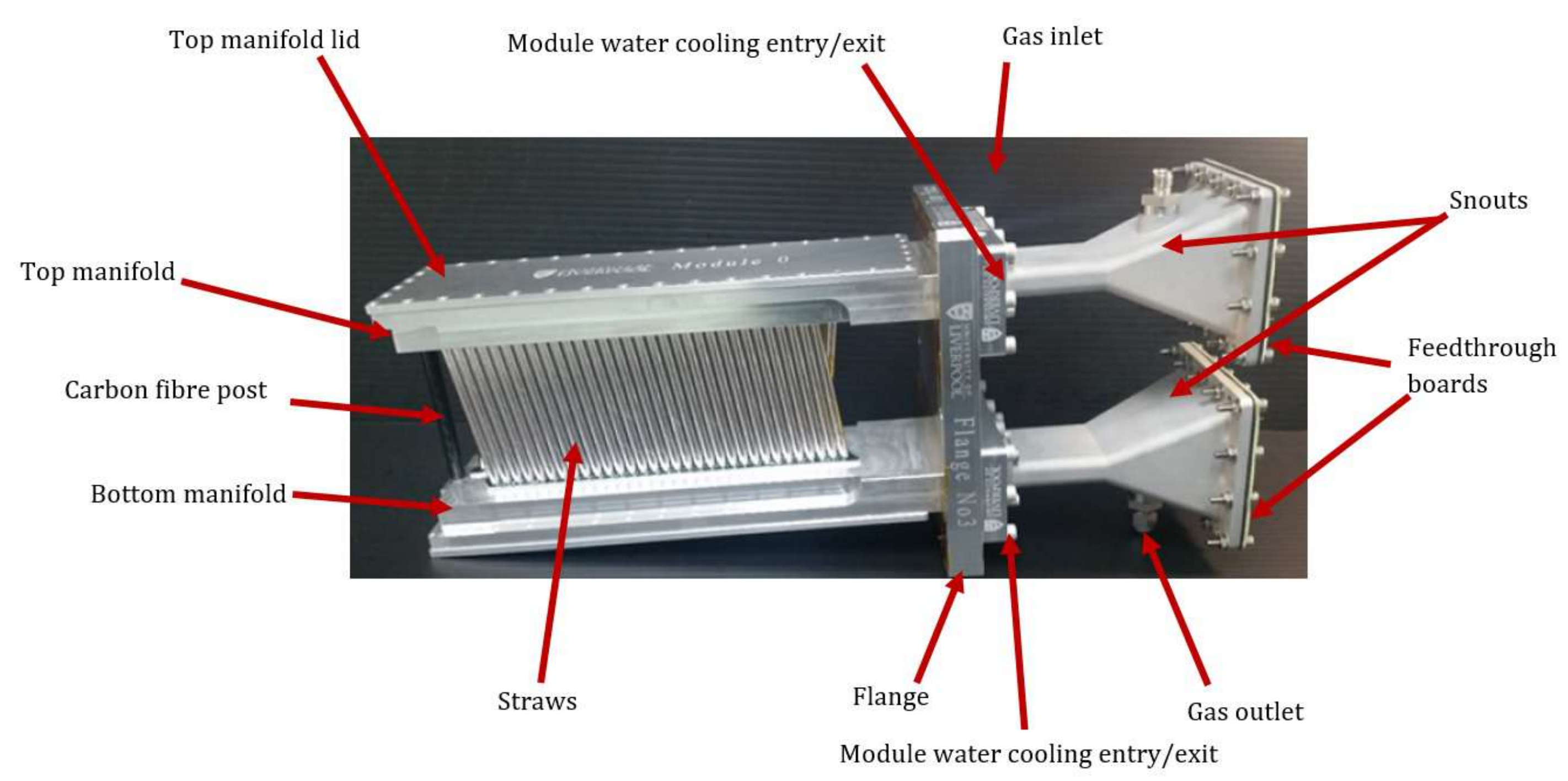}
\caption{A single, labelled tracking detector module.}
\label{fig:labelledModule}
\end{subfigure}
\begin{subfigure}[a]{0.49\textwidth}
\includegraphics[width=\textwidth]{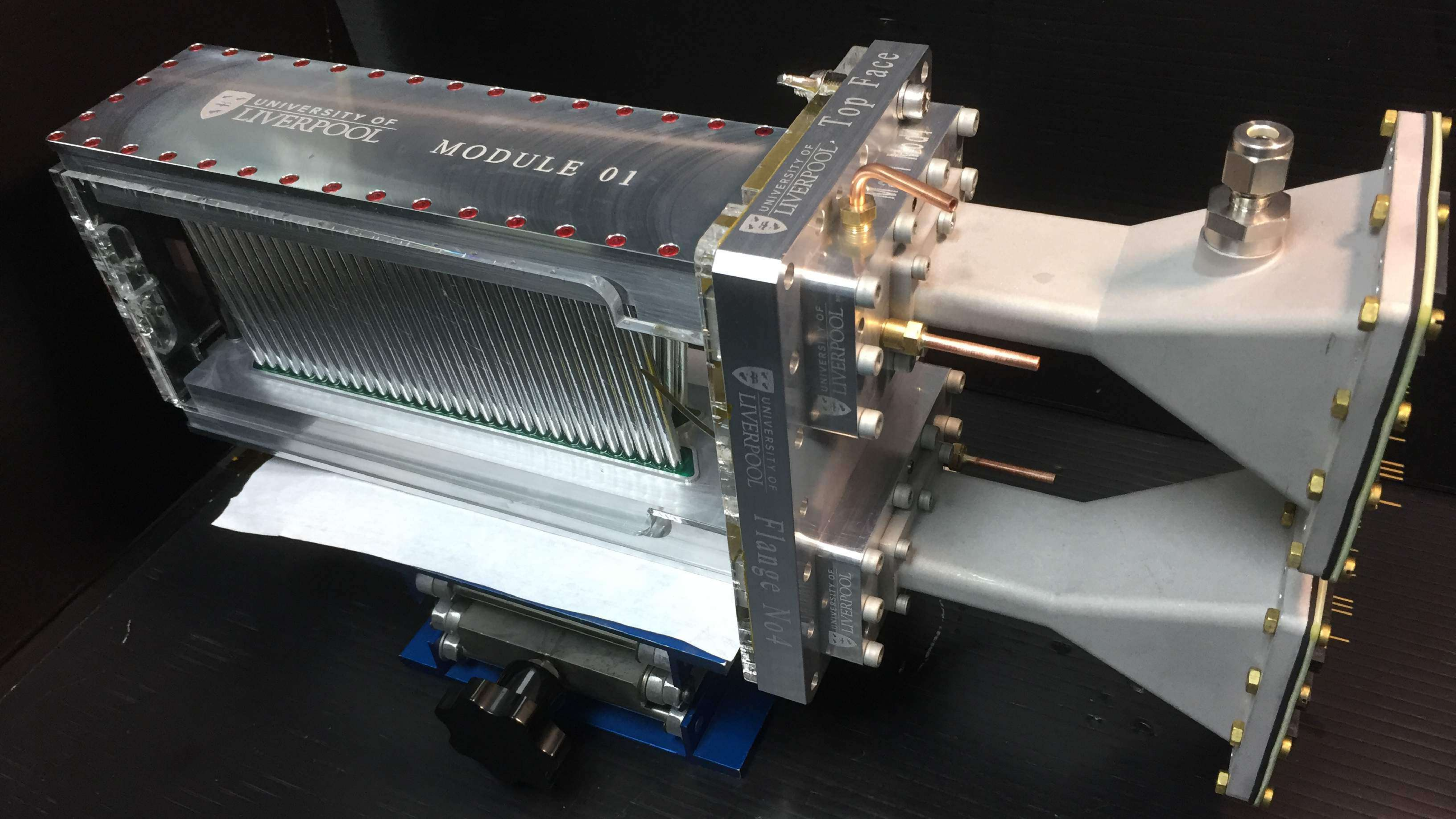}
\caption{The first completed tracking detector module.}
\label{fig:firstModule}
\end{subfigure}
\begin{subfigure}[a]{0.49\textwidth}\vspace{0.45cm}
\includegraphics[width=\textwidth]{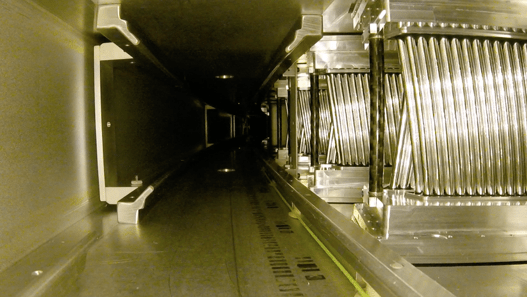}
\caption{Tracking detector modules installed in the storage ring vacuum chamber.}
\label{fig:modulesInChamber}
\end{subfigure}
\begin{subfigure}[b]{0.9\textwidth}\vspace{0.5cm}
\includegraphics[width=\textwidth]{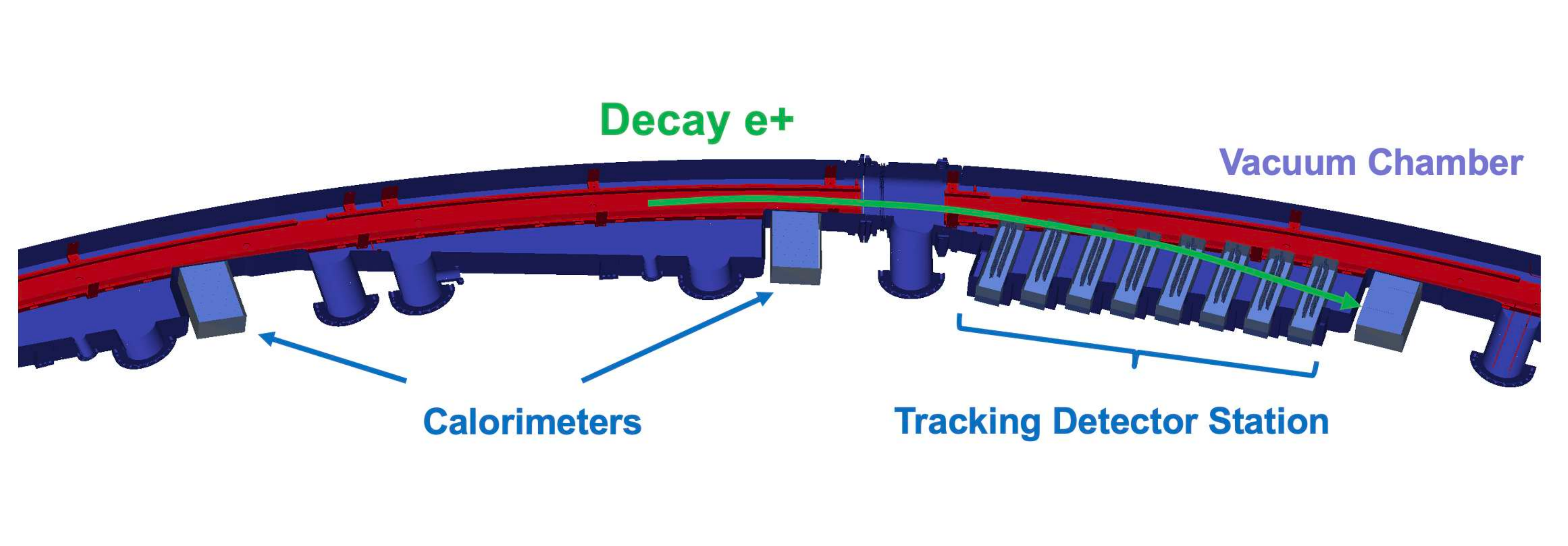}
\caption{Schematic showing the arrangement of modules in a tracking detector station.}
\label{fig:trackingSchematic}
\end{subfigure}
\end{center}
\caption{ Overview of the straw tracking detector configuration.}\label{fig:TrackerDesignOverview}
\label{fig:trackerDesign}
\end{figure}

\begin{figure}[!t]
\begin{center}
\begin{subfigure}[b]{0.5\textwidth}
\includegraphics[width=\textwidth]{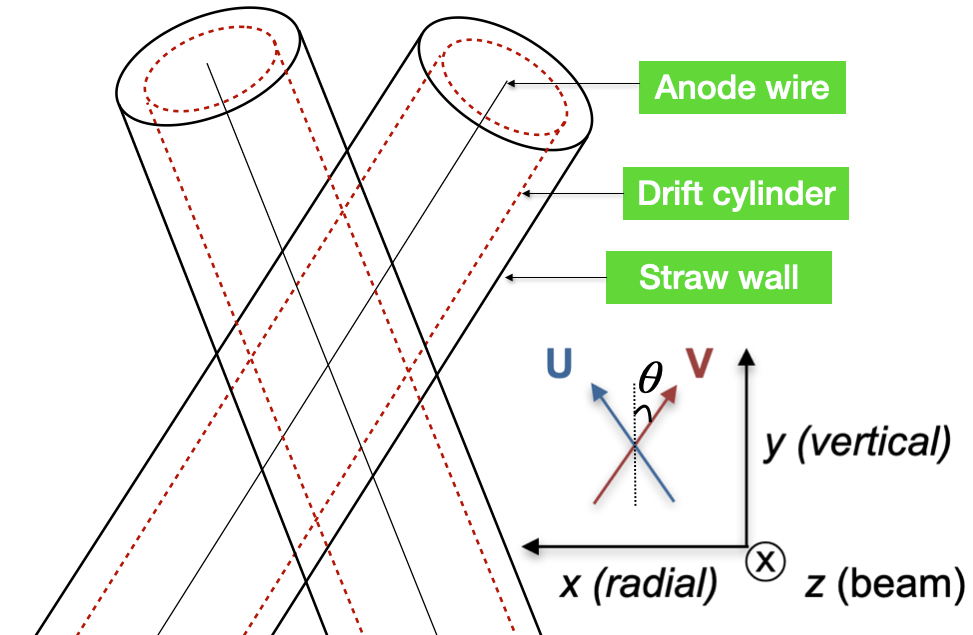}
\end{subfigure}
\end{center}
\caption{The straw coordinate system, where the straws are split into two stereo orientations: U and V, each at $\theta=7.5^\circ$ to the vertical.}\label{fig:UV}
\end{figure}

Each tracking detector station consists of 8 identical tracking `modules', one of which is shown in Figure~\ref{fig:labelledModule}. Each module consists of four layers of 32 straws, grouped into two pairs of `UV' layers. The layer pairs are staggered by half a tube diameter with respect to each other to minimize `dead' regions and to aid in the resolution of left-right ambiguities in the track pattern recognition. In one pair of layers, straws are oriented at stereo angles of $\theta=+7.5^\circ$ and $\theta=-7.5^\circ$, as shown in Figure~\ref{fig:UV}. This geometry allows the straw readout to be located above and below the path of most positrons, whilst permitting the determination of the vertical position of the tracks with sufficient precision without compromising the measurement in the bending plane. 

As shown in Figure~\ref{fig:labelledModule}, two aluminum manifolds support the straws, and house the frontend electronics and feed the gas mixture into the straws. The top and bottom manifolds are supported by an aluminum flange at one end and by a carbon fiber post at the end closest to the stored muon beam. The post avoids deflection of the manifolds when the straws are under tension. The use of carbon fiber minimizes the material in front of the calorimeters~\cite{Behnke}. The eight modules insert through holes in the wall of the vacuum chamber, with the installed modules shown in Figure~\ref{fig:modulesInChamber}. The step-like structure shown in Figure~\ref{fig:chamberModifications} and Figure~\ref{fig:trackingSchematic}, enable the use of identical modules at each position, minimizing cost and manufacturing time. 

The readout electronics and HV connect directly to the wire ends. To cool the electronics boards within the manifolds, each manifold includes a cylindrical cooling channel with a concentric internal cylinder, that allows cooling water to be pumped in and extracted within the same channel. Copper bars and silicone-based thermal pads form a thermal connection between the water-cooled manifold and the electronics~\cite{LuoThesis}. Digital readout signals on flexicables and HV cables exit the manifolds via channels through the flange. Beyond the flange, these openings expand as aluminum “snouts” (see Figure~\ref{fig:labelledModule}) until they terminate in a feedthrough electronics board to which the downstream electronics connect. Section~\ref{sec:electronicsDAQ} discusses the electronics in detail. 

The snouts also contain the gas connections, with the gas entering one manifold through its snout, flowing through the straws, and exiting the opposite manifold and its snout. The snouts extend the connections radially inwards away from the storage ring to clear the magnet pole pieces whilst maintaining the gas seal and providing an unbroken Faraday cage for the electronics, thereby shielding them against RF noise. The straw gas, a 50:50 mixture of argon-ethane (Ar:C$_2$H$_6$ 50:50), provides good resolution performance and a minimal leak rate. Argon provides a relatively inexpensive but effective ionization medium. Ethane, the quencher gas, provides good absorption of ionization photons that could otherwise trigger new avalanches. Together, Ar:C$_2$H$_6$ 50:50 provides a saturated drift velocity. Ar:CO$_{2}$ in an 80:20 ratio mixture, an inflammable alternative tested with detector prototypes, exhibited a weaker gain plateau and an unacceptable leak rate compared with the permeation rate of Ar:C$_2$H$_6$ 50:50 through aluminized Mylar. Rate-of-rise tests in a vacuum chamber (see Section~\ref{subsec:gas}) confirmed the reduced leak rate of Ar:C$_2$H$_6$ 50:50.

\section{Construction}
\label{sec:construction}
\begin{figure}[!t]
\centering
\includegraphics[width=0.99\textwidth]{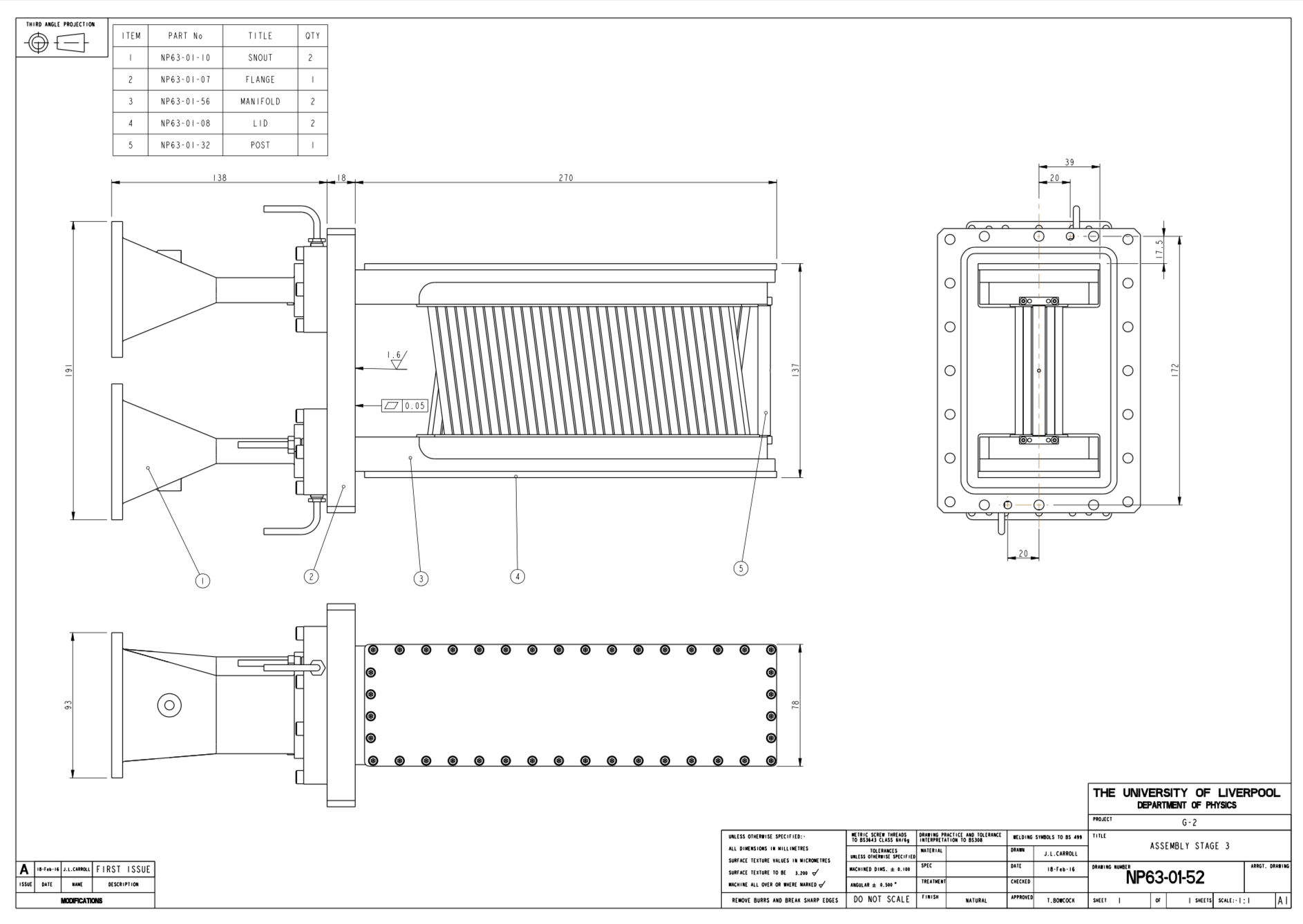}
\caption{The technical design of a tracking detector module~\cite{SaskiaThesis}.}\label{fig:trackerCAD}
\end{figure}

The technical design of a tracking detector module is shown in Figure~\ref{fig:trackerCAD}. All stages of module construction were performed in ISO 5 (Class 100 and Class 10000) clean rooms, with all staff fully equipped and trained for the process. 

\subsection{Machining of Module Components}\label{sec:Machine1}

The initial stage of construction involved the machining of the low-magnetization aluminum manifolds, flanges, manifold lids and snouts form Grade 6061 aluminum. Treatment with no-color Alocrom 1000 (Alodine) prevented oxidization without adding or impeding the function of the material. The manifolds sit in a flange that bolts into the vacuum chamber. Two snouts on the other end of the flange connect the manifolds to the crate that houses the backend readout electronics. 

\begin{figure}[!t]
\centering
\begin{subfigure}[t]{0.4\textwidth}
\includegraphics[width=\textwidth]{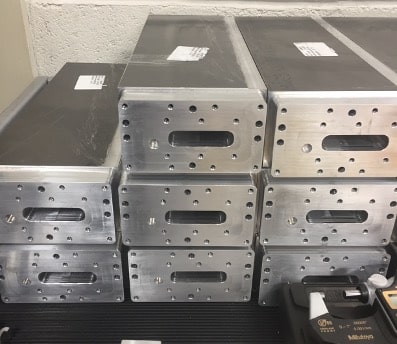}
\caption{Stock aluminum with one face machined and alignment holes added.}
\label{fig:machine1}
\end{subfigure}\hspace{1cm}
\begin{subfigure}[t]{0.4\textwidth}
\includegraphics[width=\textwidth]{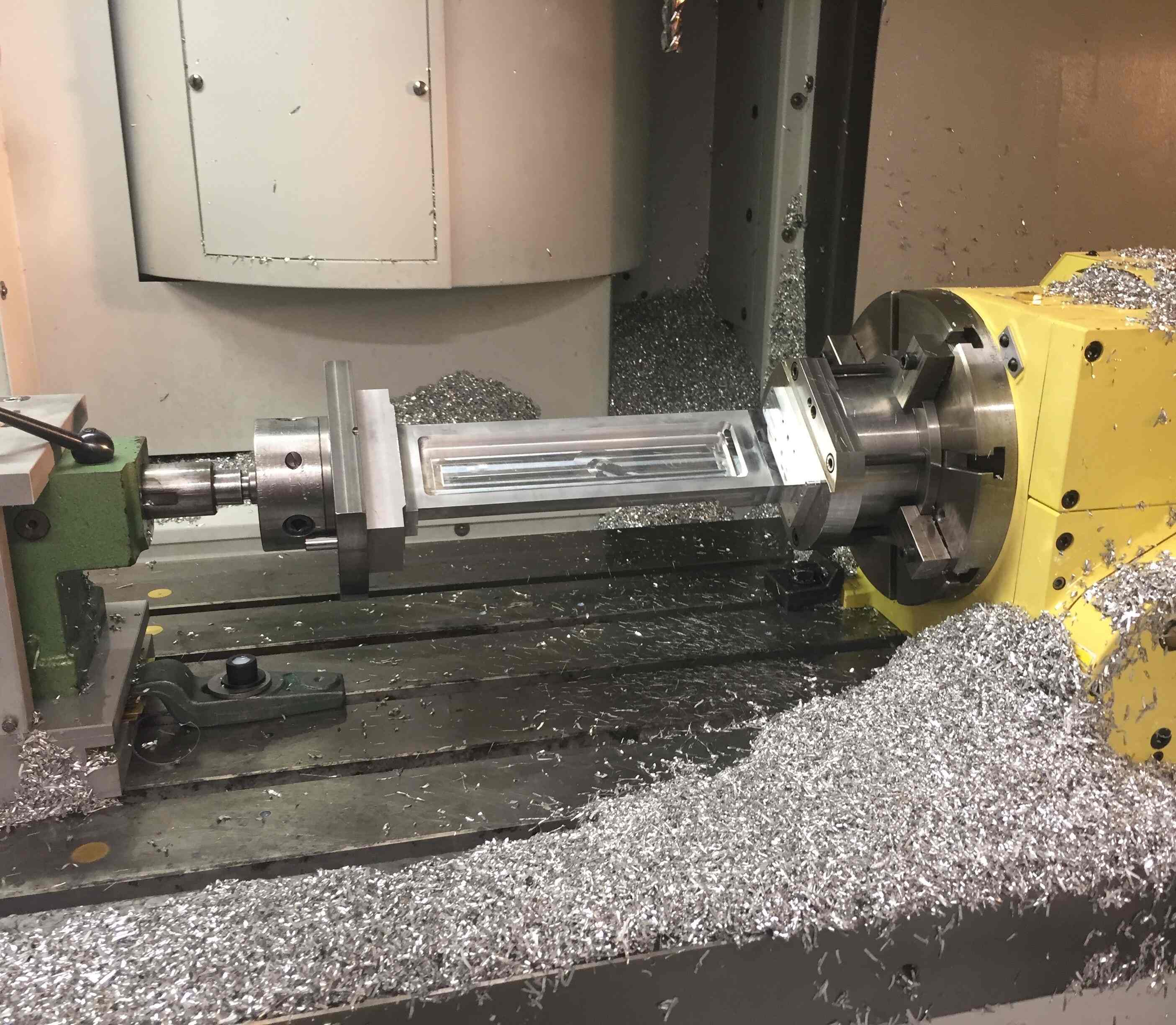}
\caption{Pocket for electronics roughed out on 4-axis CNC.}
\label{fig:machine2}
\end{subfigure}\hfill
\begin{subfigure}[t]{0.4\textwidth}
\includegraphics[width=\textwidth]{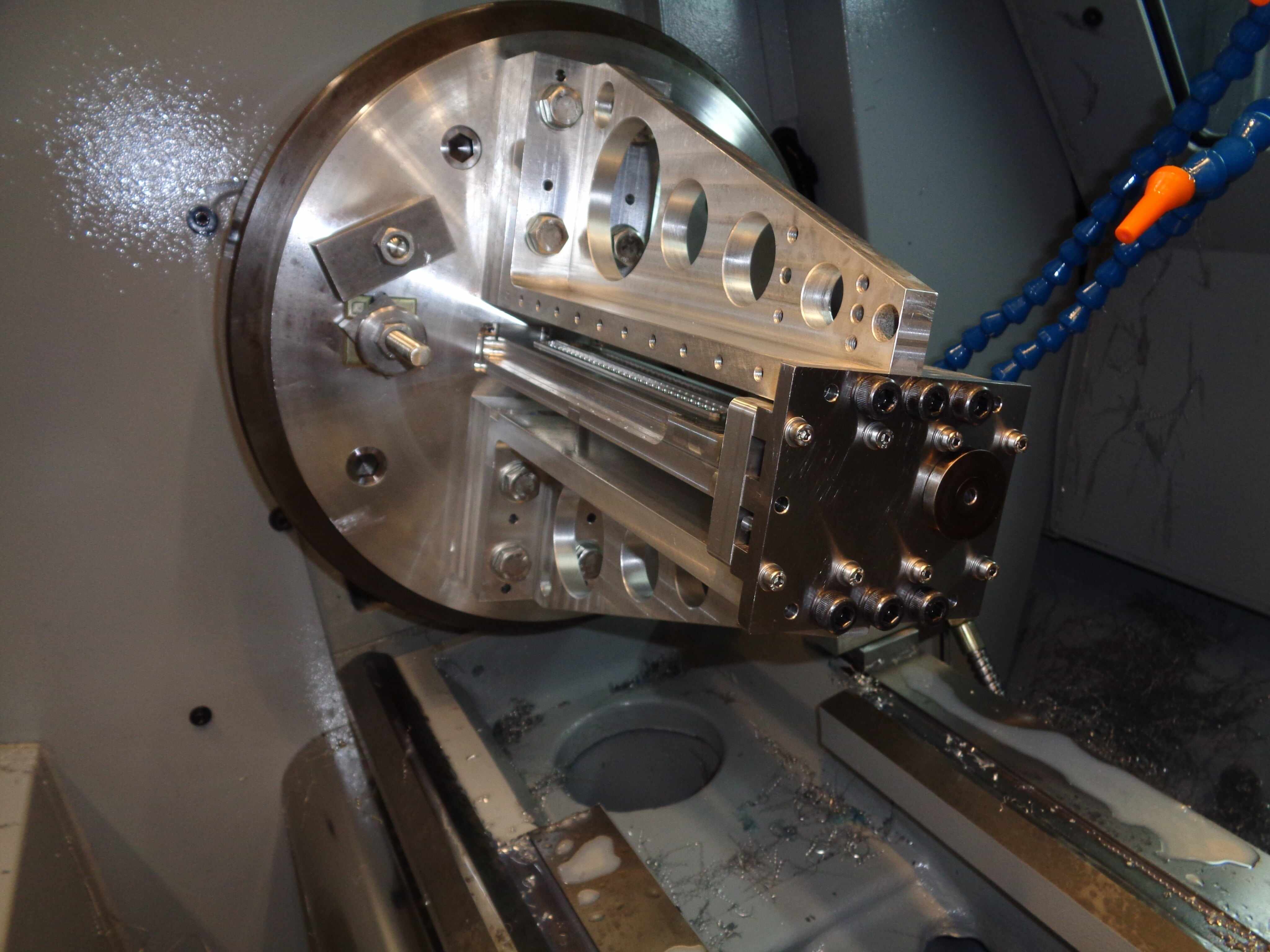}
\caption{Drilling of the $5\times260$\,mm deep cooling hole.}
\label{fig:machine3}
\end{subfigure}\hspace{1cm}
\begin{subfigure}[t]{0.4\textwidth}
\includegraphics[width=\textwidth]{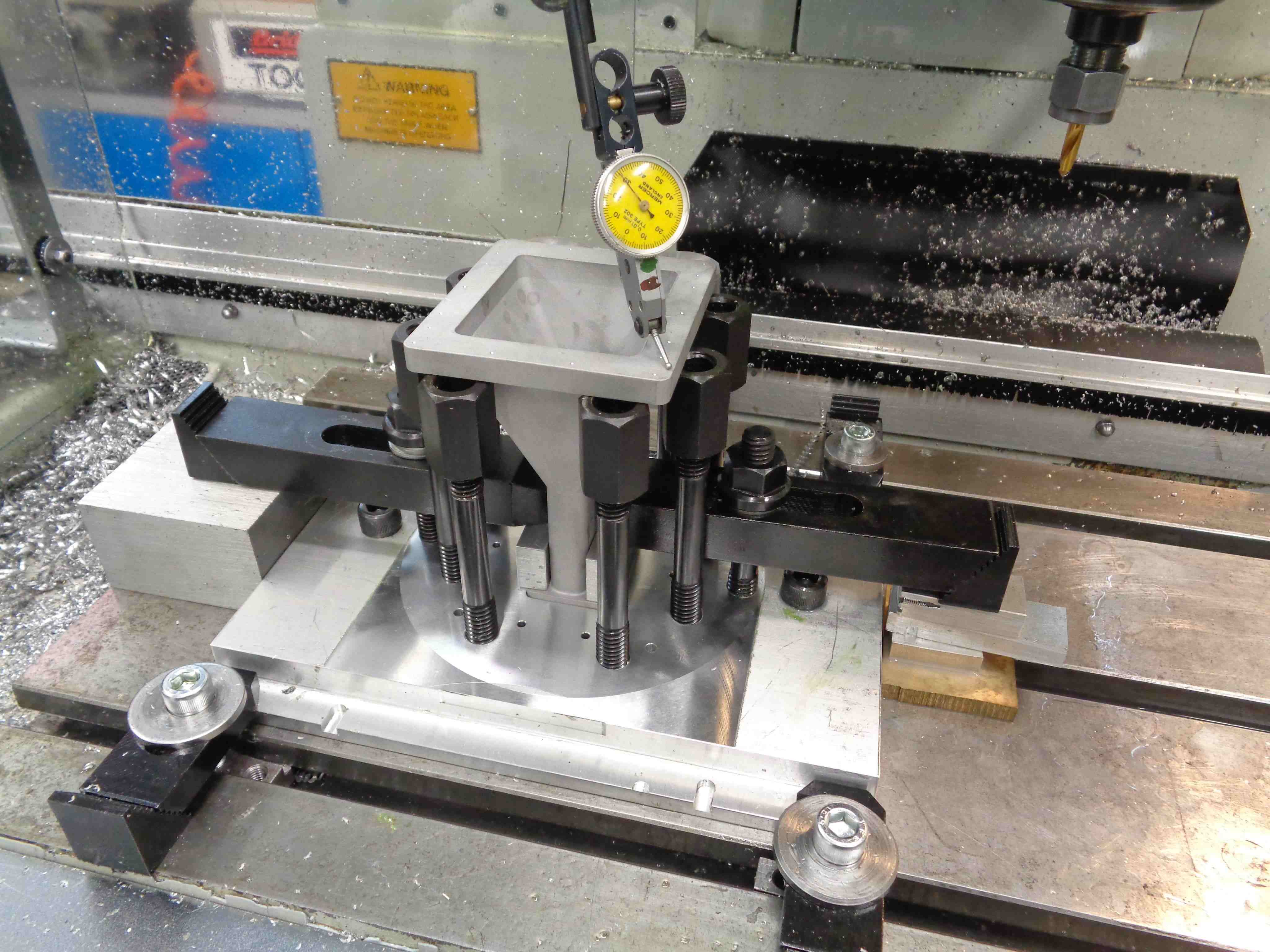}
\caption{Machining of the snout.}
\label{fig:machine4}
\end{subfigure}\hfill 
\begin{subfigure}[t]{0.4\textwidth}
\includegraphics[width=\textwidth]{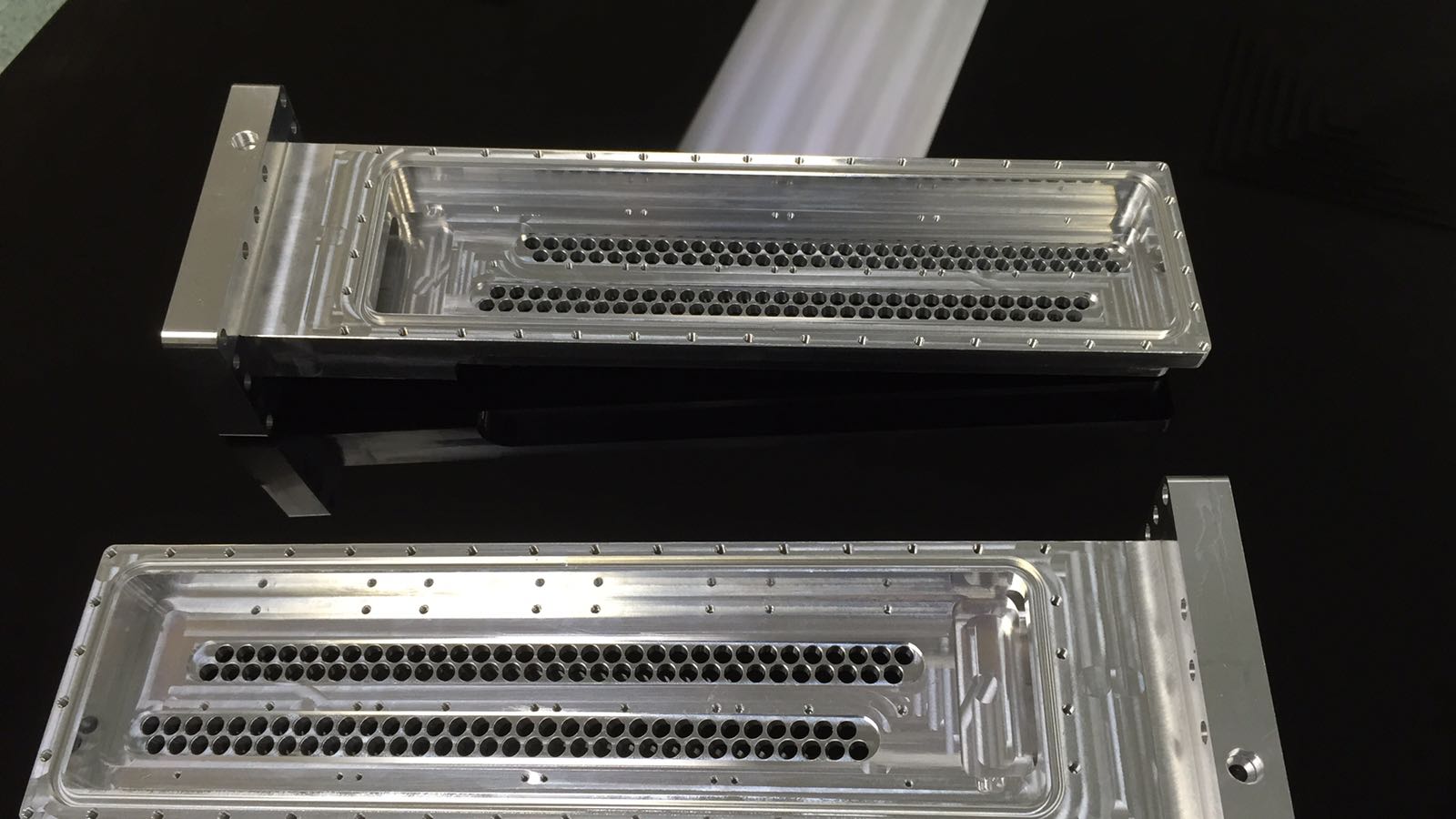}
\caption{Fully machined pair of manifolds.}
\label{fig:machine5}
\end{subfigure}\hspace{1cm}
\begin{subfigure}[t]{0.4\textwidth}
\includegraphics[width=\textwidth]{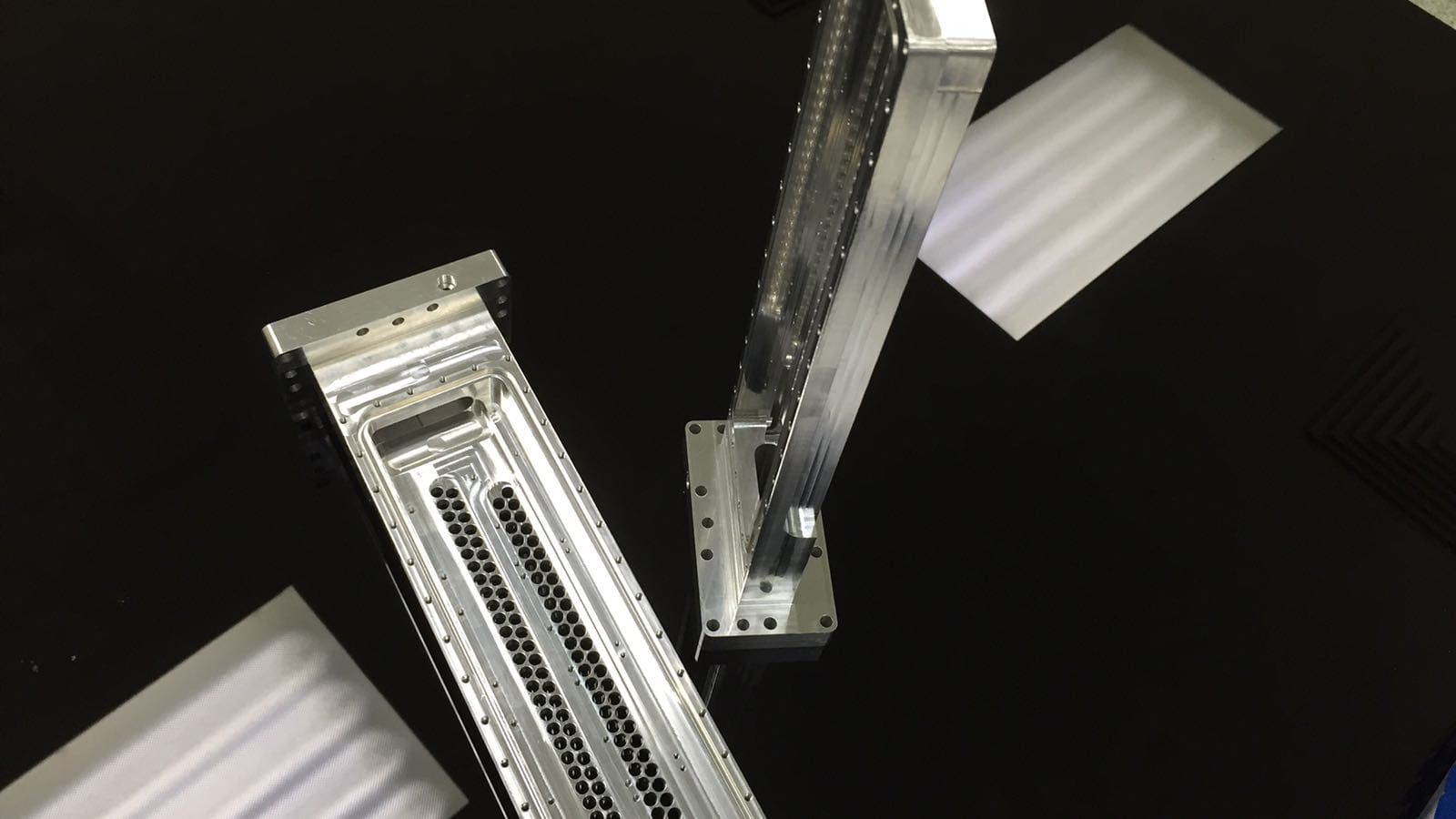}
\caption{Fully machined pair of manifolds.}
\label{fig:machine6}
\end{subfigure}\hfill 
\begin{subfigure}[t]{0.4\textwidth}
\includegraphics[width=\textwidth]{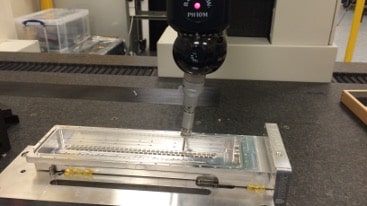}
\caption{Manifold undergoing metrology.}
\label{fig:machine7}
\end{subfigure}\hspace{1cm}
\begin{subfigure}[t]{0.4\textwidth}
\includegraphics[width=\textwidth]{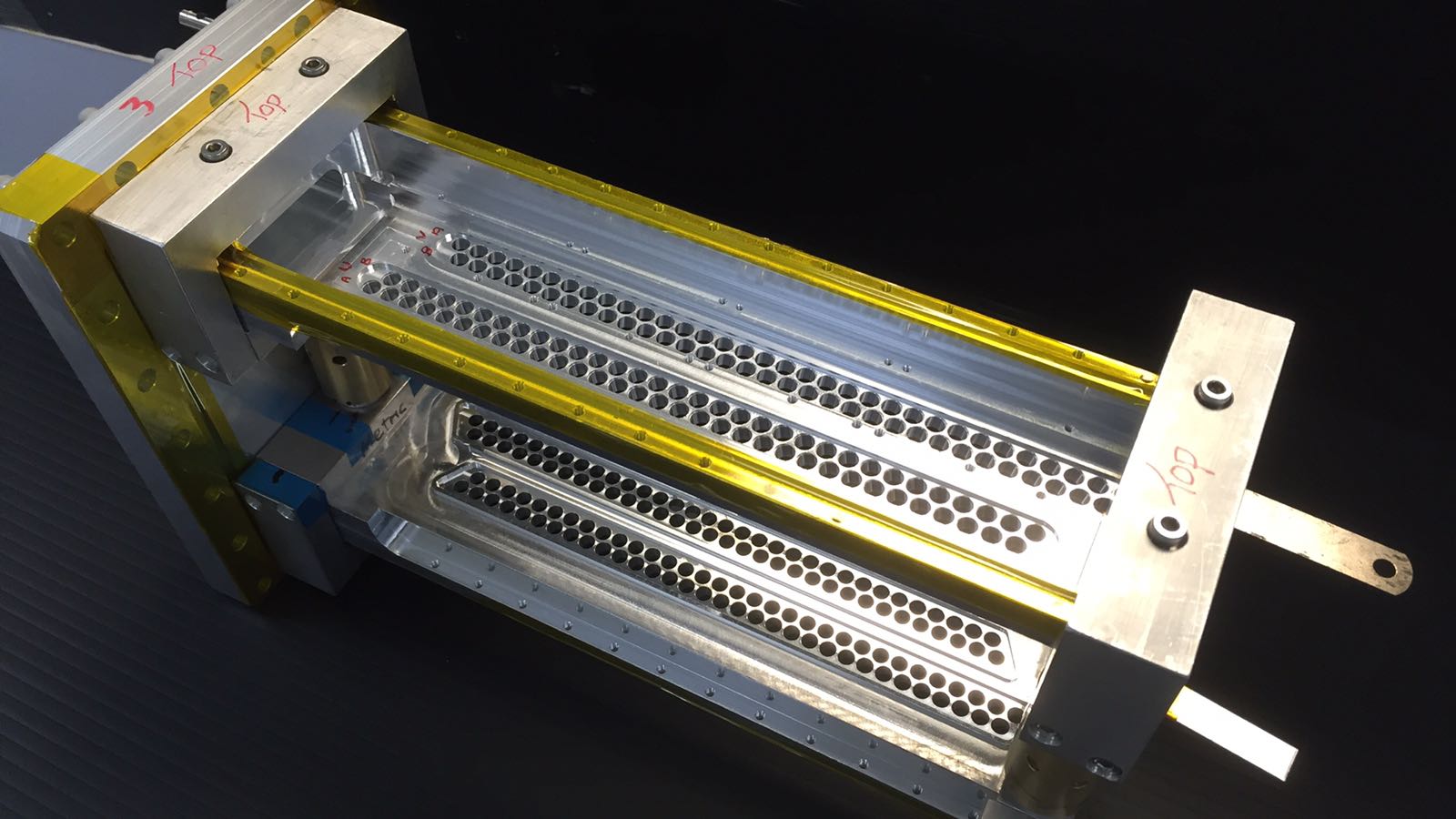}
\caption{Manifolds and flange paired with jacks.}
\label{fig:machine8}
\end{subfigure}\hfill
\caption{Stages of manifold production and module construction~\cite{TurnerThesis}.}\label{fig:machine}
\end{figure}

Figure~\ref{fig:machine} shows the stages of machining. The manifolds in Figure~\ref{fig:machine1} were first machined on a 5-axis computer numerical control (CNC) machine to manufacture the flange end, create a datum face on both ends, provide precise placement details such as alignment holes for dowel pins, and enable mounting for subsequent machining cycles. Letterbox features to link to an electronics pocket allow the necessary electronics to pass through to the flange face. Shown in Figure~\ref{fig:machine2}, the large electronics pocket itself was then rough-machined on a 4-axis CNC. A break in machining between this stage and the next allowed time for the material to relax before further critical machining was carried out. The precise finishing of the electronics pocket and all angled straw holes (helically bored to a tolerance of 10$\mu$m and spaced so each straw is 6 mm apart any neighboring straw) were then completed on the 5-axis CNC, including a shelf for cooling within the pocket, two troughs for glue to well into around the straws, and an O-ring groove in the top face for the lid to seal. Separate operations on the 5-axis CNC finished the bottom with wells for glue, as well as the flange end and tail end.


A lathe drilled a $5\times260$\,mm deep cooling hole down the length of the manifold block, providing space for a plastic pipe to transport cooling water along the length of the manifold to remove heat generated by the internal electronics.\footnote{Original attempts to use copper piping resulted in galvanic corrosion. Plastic tubes, the alternative, had a lower flow rate than copper, but still adequate for the required heat transfer.} In this stage, shown in Figure~\ref{fig:machine3}, the drilling runout could not exceed 0.5mm as this would break through into the pocket housing the electronics. Final details were added by hand, including polishing and deburring the straw holes to remove any imperfections that could damage the straws on insertion. The snouts, shown in Figure~\ref{fig:machine4}, were cast and then machine finished. Full details of the manifold production are documented in~\cite{TurnerThesis}.

%% AK 03/22/21 - The info on metrology is copied/paraphrased/constructed from Tabitha's thesis 
All machined pieces of the tracking detector modules underwent a metrology survey (using CAD models and a contact probe with a coordinate-measuring machine) before assembly to ascertain how accurately the parts had been machined and assess if they lie within acceptable tolerances. All manifold faces, o-ring planes, dowel holes, straws holes, etc. were surveyed as part of this process. Manifolds were paired with a flange based on the offsets measured from the flange surveys and any alterations to the machined pieces of equipment based on the acquired data were carried out. The metrology process is depicted in Figure~\ref{fig:machine7}. Full details of the metrology survey can be found in~\cite{TabithaThesis}.

\subsection{Straw Testing and Assembly}

The straws never left the clean room environment during assembly. They are made from two layers of spiral-wound aluminized Mylar with an outer diameter of 5 mm. The straw wall is made of two layers of 6\,$\mu$\si{\meter} Mylar with a 3\,$\mu$\si{\meter} layer of adhesive between. The inner surface has 500\,\si{\angstrom} of aluminum overlaid with 200\,\si{\angstrom} of gold to act as a cathode. The outer surface has 500\,\si{\angstrom} of aluminum to reduce the permeation rate of gas through the straw walls and provides increased electrostatic shielding. The straws have the same specification as the 1.3\,{m} straws to be used by the Mu2e experiment~\cite{Bartoszek:2014mya} and arrived from the vendor, DuPont~\cite{DuPont}, cut to this length specification with a supportive layer of paper inside. Quality control procedures were performed prior to handling on all acquired straws. Two different batches of straws were used in the production of the detectors. The first batch had an average wall thickness of 15\,$\mu$\si{\meter} and diameter of 5.07 mm, whilst the second batch had an average wall thickness of 12\,$\mu$\si{\meter} and diameter of 4.92 mm. The average resistance was measured from one end of each uncut straw to the other. The resistance of the second batch was found to be lower than the first batch (before and after paper removal), but acceptable overall. The average resistance was found to be $155 \, \Omega$m$^{-1}$, with 98\% of all straws found to have an adequate resistance of $\sim 200\, \Omega$. 

\begin{figure}[!t]
\centering
\begin{subfigure}[t]{0.4\textwidth}
\includegraphics[width=\textwidth]{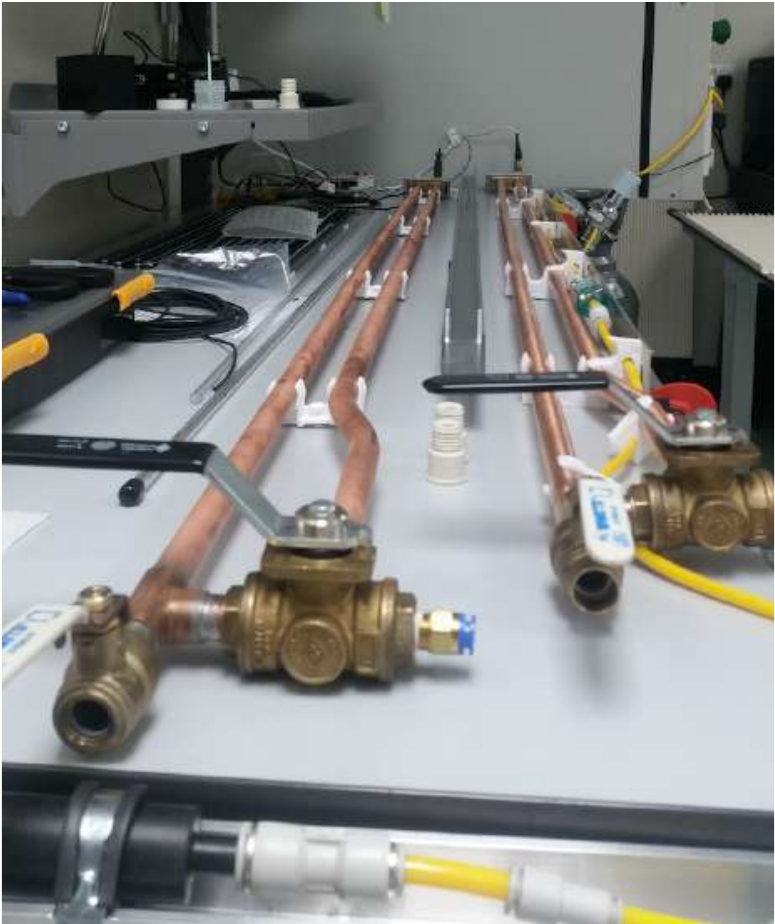}
\caption{Photograph of the straw leak testing apparatus. }
\label{fig:leak1}
\end{subfigure}\hfill
\begin{subfigure}[t]{0.6\textwidth}
\includegraphics[width=\textwidth]{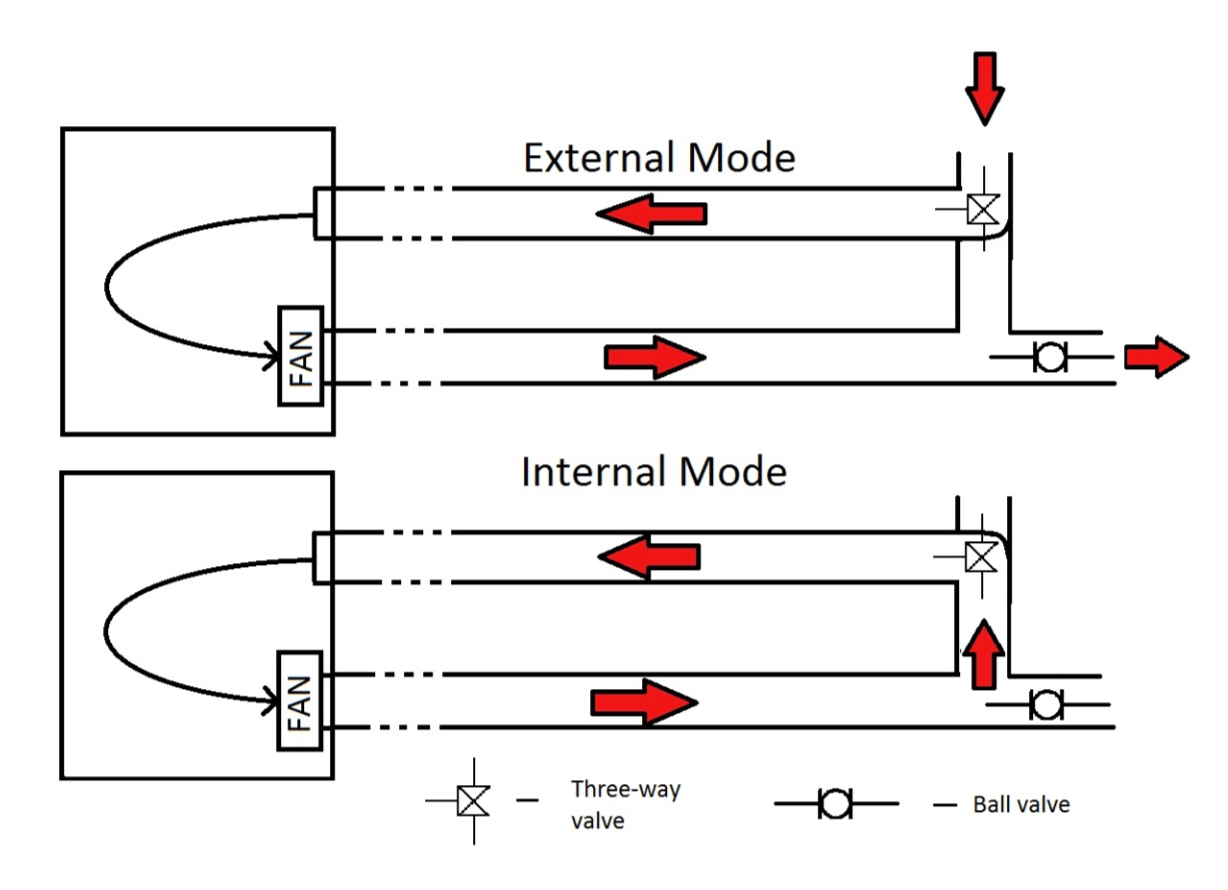}
\caption{Diagram of the straw leak testing apparatus. }
\label{fig:leak2}
\end{subfigure}\hfill
\caption{The straw leak testing apparatus.}
\label{fig:strawLeak}
\end{figure}

%% AK 03/22/21 - This info and figure on the leak test is copied/paraphrased/constructed from Saskia's thesis
Prior to being used in the tracking detector modules, the straws underwent a rigorous leak-testing procedure.\footnote{This procedure was developed by the Mu2e experiment, since straws of identical design and manufacture will be used in both experiments.} For optimal vacuum performance, it was required that each module had a leak rate of $< 5.6 \times 10^{-6}$ Torr L/s, with a worst-case threshold of $1.5 \times 10^{-5}$ Torr L/s. The leak testing chamber, shown in Figure~\ref{fig:leak1} and Figure~\ref{fig:leak2}, consists of two copper tubes of approximately 1.5 m in length, the ends of which are connected to either a small bridging tube or electronics box. A fan installed in the electronics box ensures gas circulation. There are two bidirectional valves in the chamber. One connects the chamber to the gas line, and the other is used to switch the chamber between the ‘open’ configuration (where the gas flows from the inlet around both pipes of the chamber and is then exhausted after a single circuit) and the ‘closed’ configuration (where the chamber is sealed at both ends and the fan blows the gas around the entire volume).

The leak rate of every straw was measured in the custom-built leak-testing chamber before being cut to size.  The purpose of these tests was to measure the permeation rate of the individual straws over two hours to ensure that they would not surpass the maximum permeation rate of an entire module.  During testing,  any ``real'' leak or hole in a straw that was not caught during an initial visual inspection stage was immediately identifiable as the leak testing chamber would immediately fill with CO$_{2}$.  In these cases, the test would be stopped immediately and the straw discarded.  

Approximately 5~cm of Viton tubing was attached to an injection-molded straw end-piece, one of which was inserted into either end of the straw to be leak tested.  The diameter of the end-piece was slightly smaller than the straw diameter in order for it to be inserted into the straw without damaging it.  The end-pieces were glued into the straws using a generous amount of rapid Araldite to ensure a good gas seal.  The straw was then left in a preparation tray for up to 3 hours.  Once the glue had cured, the straw was carefully transferred into a metal leak-testing tray.  The tray had higher walls than the straw height in order to protect it during its time in the leak testing chamber,  and was flat at either end where the viton tubing rested.  A CO$_{2}$ gas line was connected to one end of the straw and the straw was flushed with CO$_{2}$ for 30~s.  After flushing, a barbed plug was used to seal the other end of the straw and stop the flow of CO$_{2}$  

The straws in the storage ring vacuum experience a pressure of 15 psi, and so were leak tested at a relative pressure of 15 psi above atmospheric pressure (30 psi absolute pressure). However,  the straws must be able to withstand up to 25 psi (relative) to ensure they would not be damaged in the event of a problem with the gas supply (a relief valve in the experimental setup directs the gas away from the straws if the pressure in the system exceeds 25 psi).  To this end, the pressure of the CO$_{2}$ inside the straw was increased to 25 psi and held for 1 minute before being reduced to 15 psi for the leak test.  When the pressure reached 15 psi, the Viton tubing was clamped to isolate the straw from the gas line, and then cut between the clamp and the gas line.  A second barbed plug was inserted into the open end of the tubing and the clamp was released.  The straw was secured in its metal tray using Kapton tape and was then inserted into the leak testing chamber,  which continued to be flushed with nitrogen at a rate of 2.5 L/min.  The nitrogen flush was continued for 2 minutes after the straw was inserted to remove any residual air that was introduced with the straw and tray.  The chamber was then switched to the `closed’ configuration and the nitrogen line removed before running the leak test. 

\begin{figure}[!t]
\centering
\includegraphics[width=0.7\textwidth]{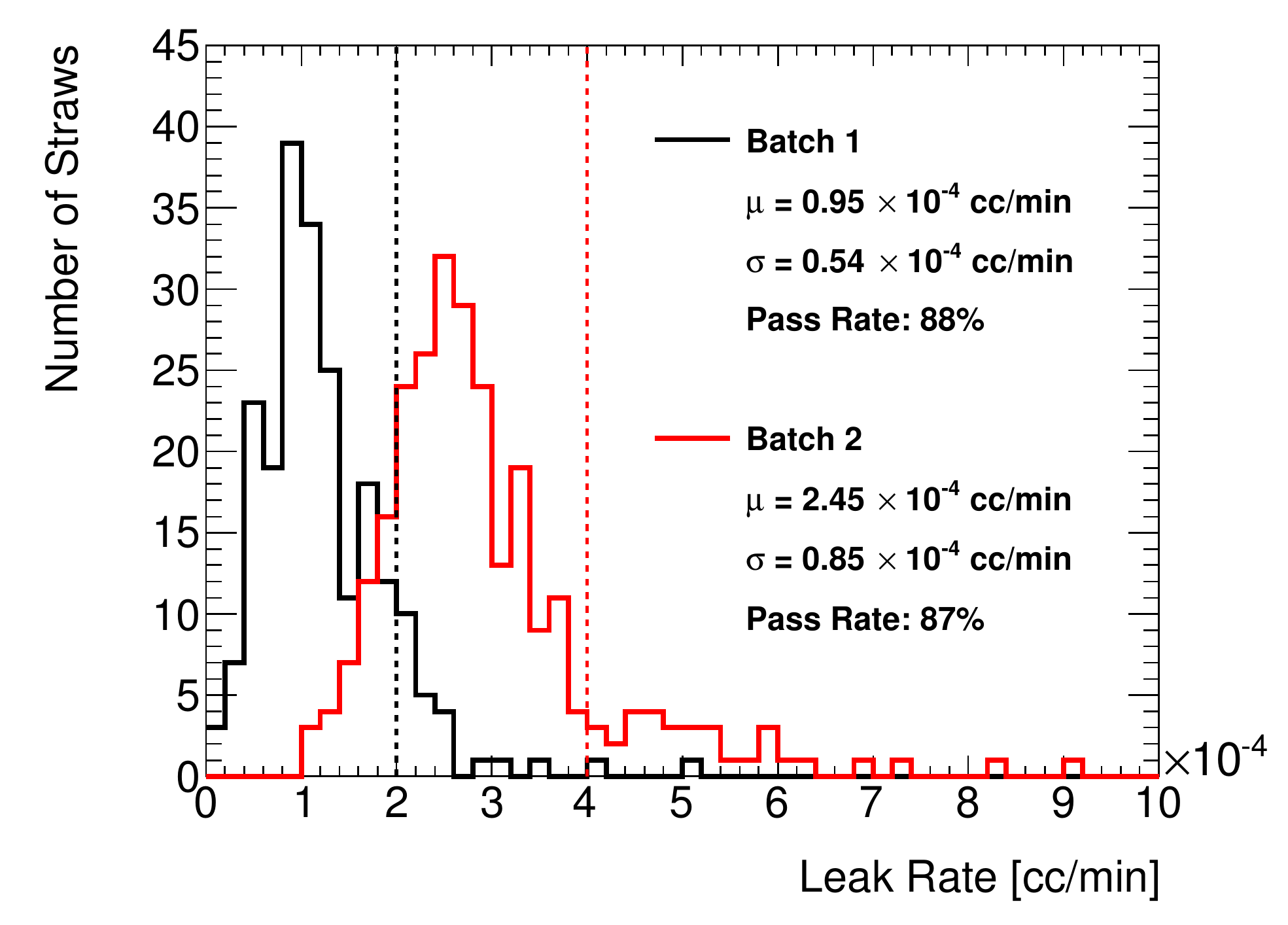}
\caption{Histograms of leak rates for all Batch 1 (black) and Batch 2 (red) straws. All leak rates are quoted for Ar:CO$_2$ 50:50 gas. The mean leak rate for batch 1 straws was $0.95 \times 10^{-4}$~cc/min for Ar:CO$_2$ 50:50 gas; the pass rate with the batch 1 threshold (dotted black) of $2 \times 10^{-4}$ cc/min was $88\%$. The mean leak rate for batch 2 straws was $2.45 \times 10^{-4}$ ccm; the pass rate with the batch 2 threshold (dotted red) of $4 \times 10^{-4}$ ccm was $87\%$.}
\label{fig:strawLeakResults}
\end{figure}

Each straw was filled with Ar:CO$_2$ 50:50, and the leak rate measured using a CO$_{2}$ detector inside the testing chamber. Although the gas used in the experiment is Ar:C$_{2}$H$_{6}$ 50:50, it was acceptable to test the leak rates using CO$_{2}$ since it is known to permeate through Mylar at a greater rate than ethane. The exact conversion between the leak rates of the two gases was determined on the first fully completed module after production by measuring the rate-of-rise using the different gases, with Ar:C$_{2}$H$_{6}$ 50:50 being found to have a $\sim 20$ times smaller leak rate than 100\% CO$_{2}$.\footnote{An original threshold for the straw leak rate was set prior to this. A less stringent threshold was safely established later defined by the average leak rate of the straws per tracking detector station (see Figure~\ref{fig:strawLeakResults}).} Due to the thinner walls, the leak rates for the second batch of straws were significantly higher than the first, as shown in Figure~\ref{fig:strawLeakResults}, but both batches met the specifications for the required vacuum performance. The pass rate was $\sim88\%$ for the first batch of straws at the original threshold and $\sim87\%$ for the second batch at the relaxed threshold. The full leak testing procedure is documented in~\cite{SaskiaThesis}. %Importantly, all straws for a single tracking detector station (8 modules) came from the same batch to ensure that all straws in a station had the same geometry. 

In addition to the testing in the chamber, each straw was visually inspected to look for manufacturing imperfections. Other QA procedures were performed on a subset of spare straws, prior to the production procedure. Repeated pressure tests were performed, where the pressure on an individual straw was increased and decreased repeatedly to demonstrate robustness of the straws under cycles of the vacuum in the experiment. Over-pressure tests were performed to test that the straws would survive any sudden failures in the gas supply system that could cause the straws to experience a relative pressure of 30 psi. The straws were leak-tested before and after having tension applied to them to demonstrate that the leak rates would be unaffected by the tension applied to the manifolds. Finally, tests were performed to measure the hygroscopic properties of the straw, by measuring the mass before and after being stored in a humid environment. These tests revealed that the straws should be stored in a dry nitrogen environment. 

\begin{figure}[!t]
\centering
\includegraphics[width=0.9\textwidth]{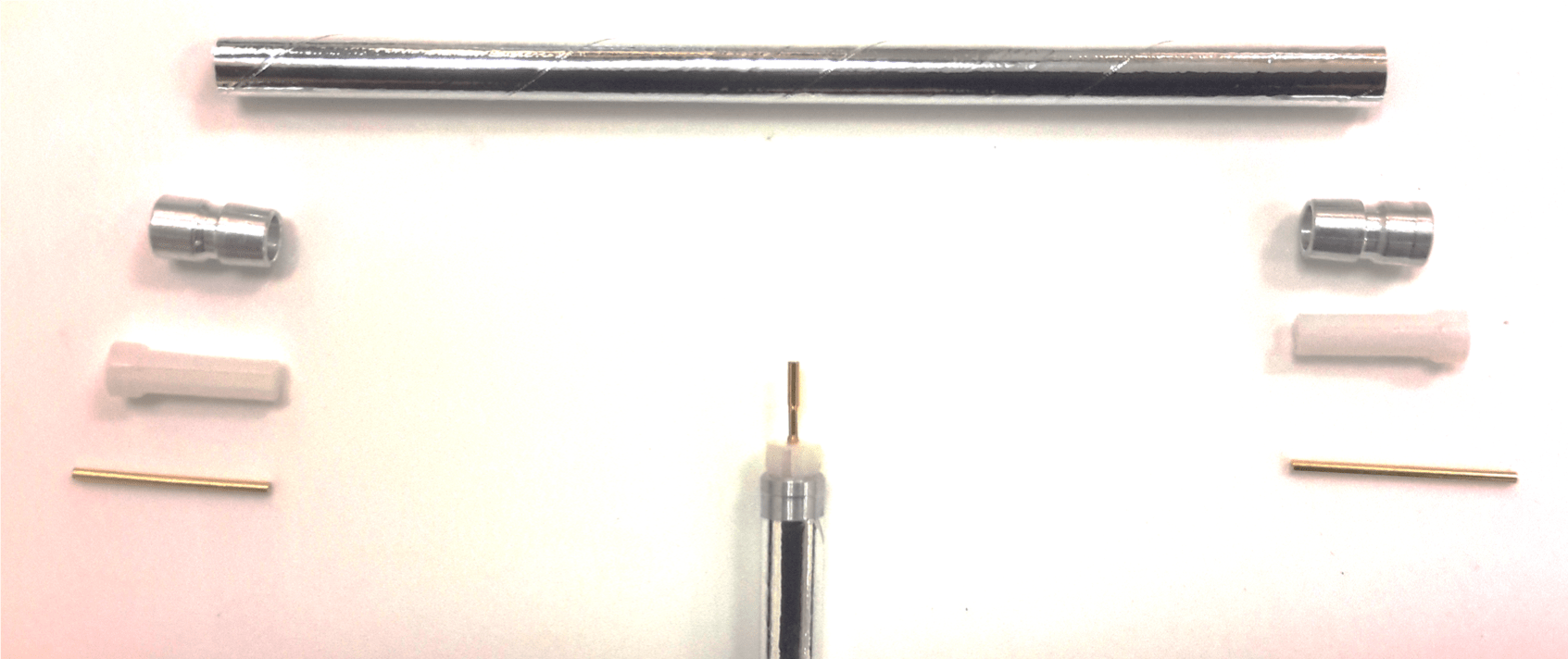}
\caption{Components making up the
straw assembly. The 5 mm diameter straw is equipped with aluminum end
pieces, ABS plastic inserts and gold-plated copper pins of two different
types. The longer pin and `top hat' endpiece shown on the right is connected
to the frontend electronics.} \label{StrawComponents}
\end{figure}

The component parts of a straw can be seen in Figure~\ref{StrawComponents}. The straws were cut to a length of 90.8~mm using a custom designed and built guillotine. They were glued and sealed into the top and bottom aluminum manifolds with green dyed Araldite 2020 (for its radiation hardness and high viscosity properties) and with conductive silver epoxy to ensure a good ground connection between the straw and the module. An additional thick layer of structural epoxy was also applied to provide a gas seal between the outside of the straw and the manifold. A gold-plated tungsten-rhenium sense wire, with a diameter of 25~$\mu$m, is held at the center of the straw by two gold crimp pins (one at each end). Each straw is readout from only one end. The pins at the end that is seated into the electronics cards are slightly longer than the pins at the other end. The pins were inserted into injection-molded plastic inserts that were themselves inserted into the straw aluminum end-pieces. On either side of the plastic insert is a channel to enable the flow of gas through the straw. The inserts were not glued into the aluminum end-pieces allowing an individual wire to be removed if it breaks. 

\subsection{Module Assembly}

\begin{figure}[!t]
\centering
\begin{subfigure}[t]{0.47\textwidth}
\includegraphics[width=\textwidth]{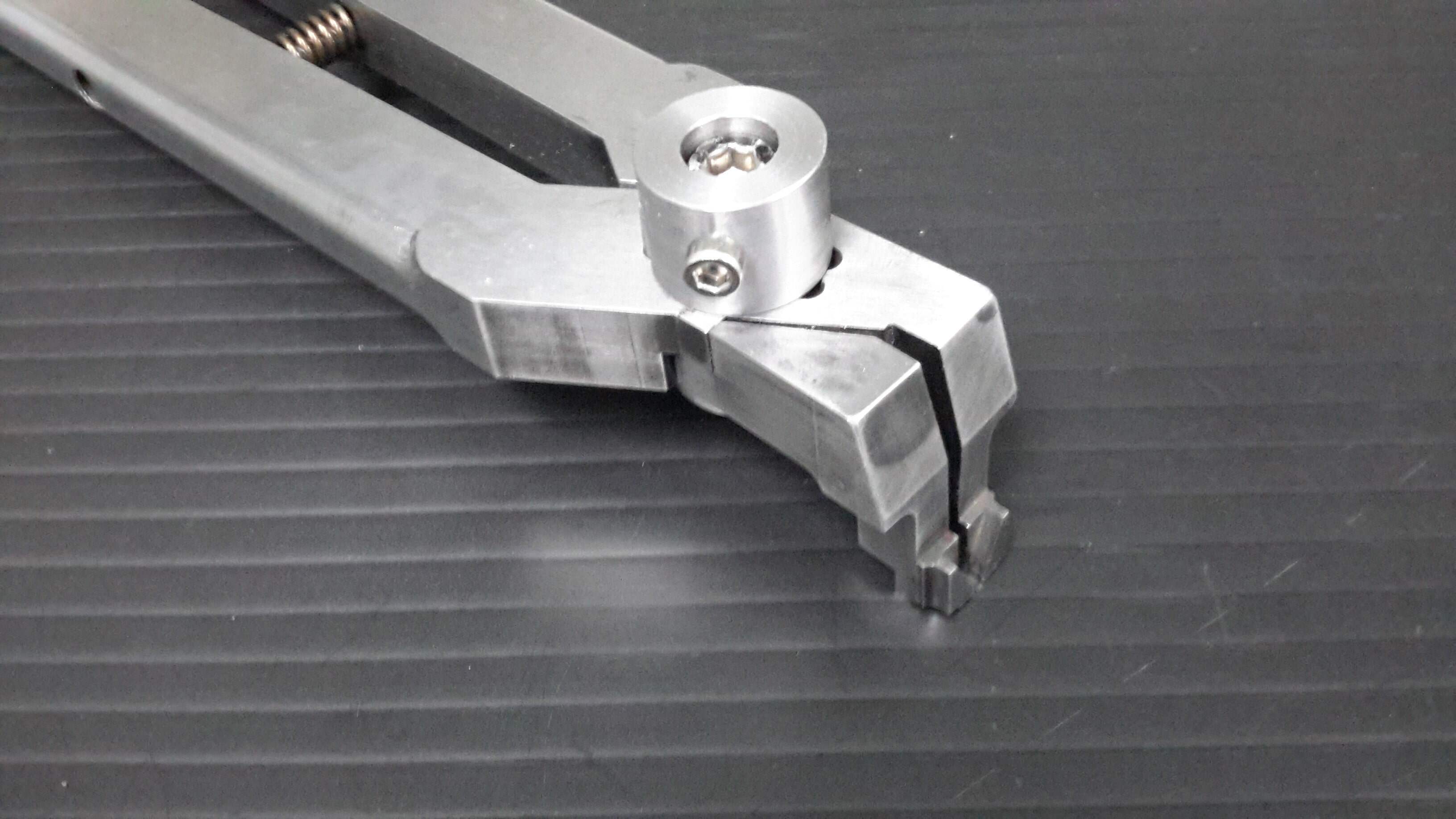}
\caption{The custom designed and built machine crimping tool for the ‘long’ pins.}
\label{fig:crimpTool1}
\end{subfigure}\hfill
\begin{subfigure}[t]{0.47\textwidth}
\includegraphics[width=\textwidth]{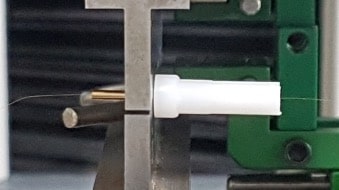}
\caption{Machine crimping the ‘long’ pins.}
\label{fig:crimpTool2}
\end{subfigure}\hfill
\begin{subfigure}[t]{0.47\textwidth}
\includegraphics[width=\textwidth]{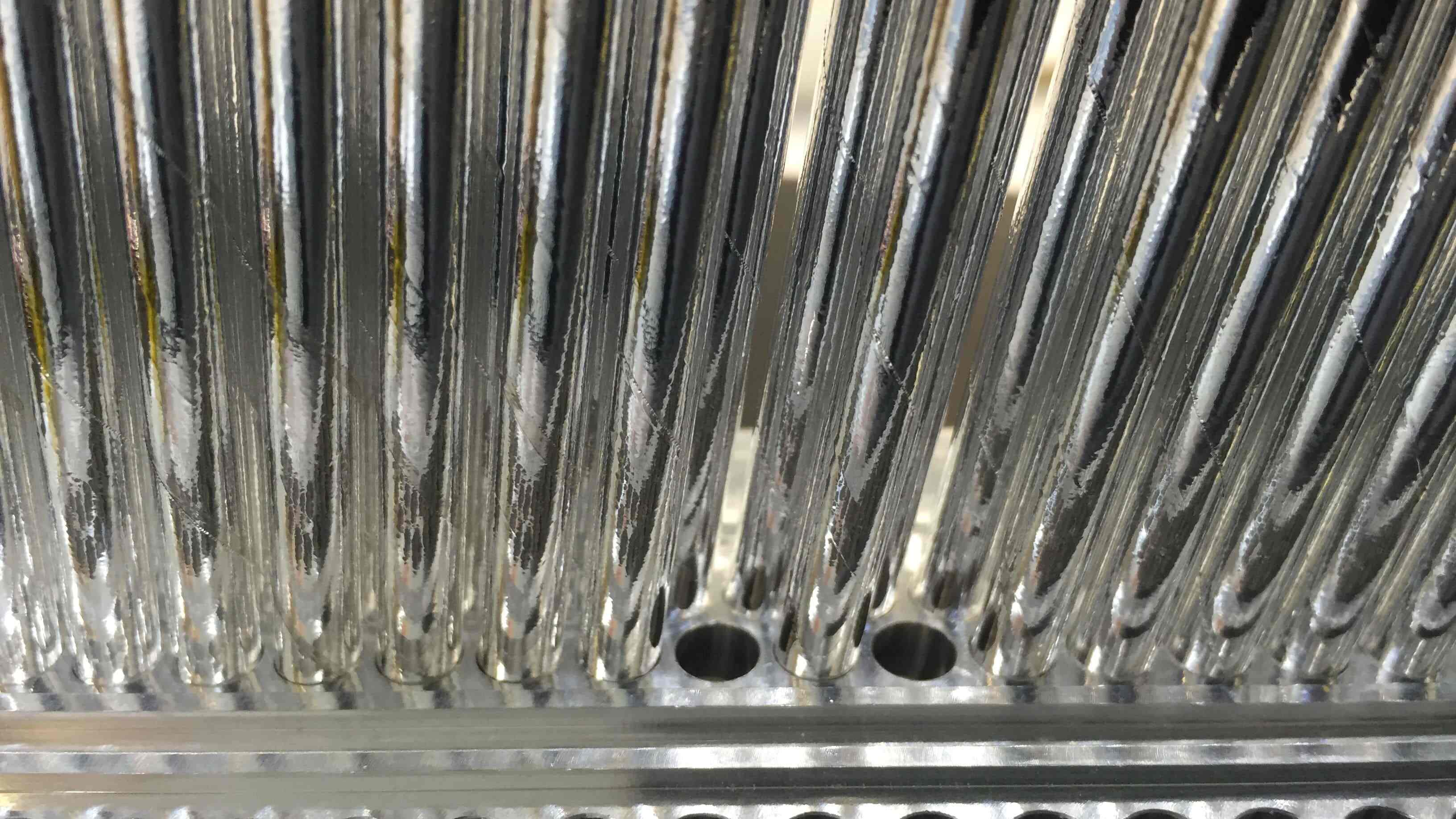}
\caption{Straws inserted into manifold.}
\label{fig:insertedStraws}
\end{subfigure}\hfill
\begin{subfigure}[t]{0.47\textwidth}
\includegraphics[width=\textwidth]{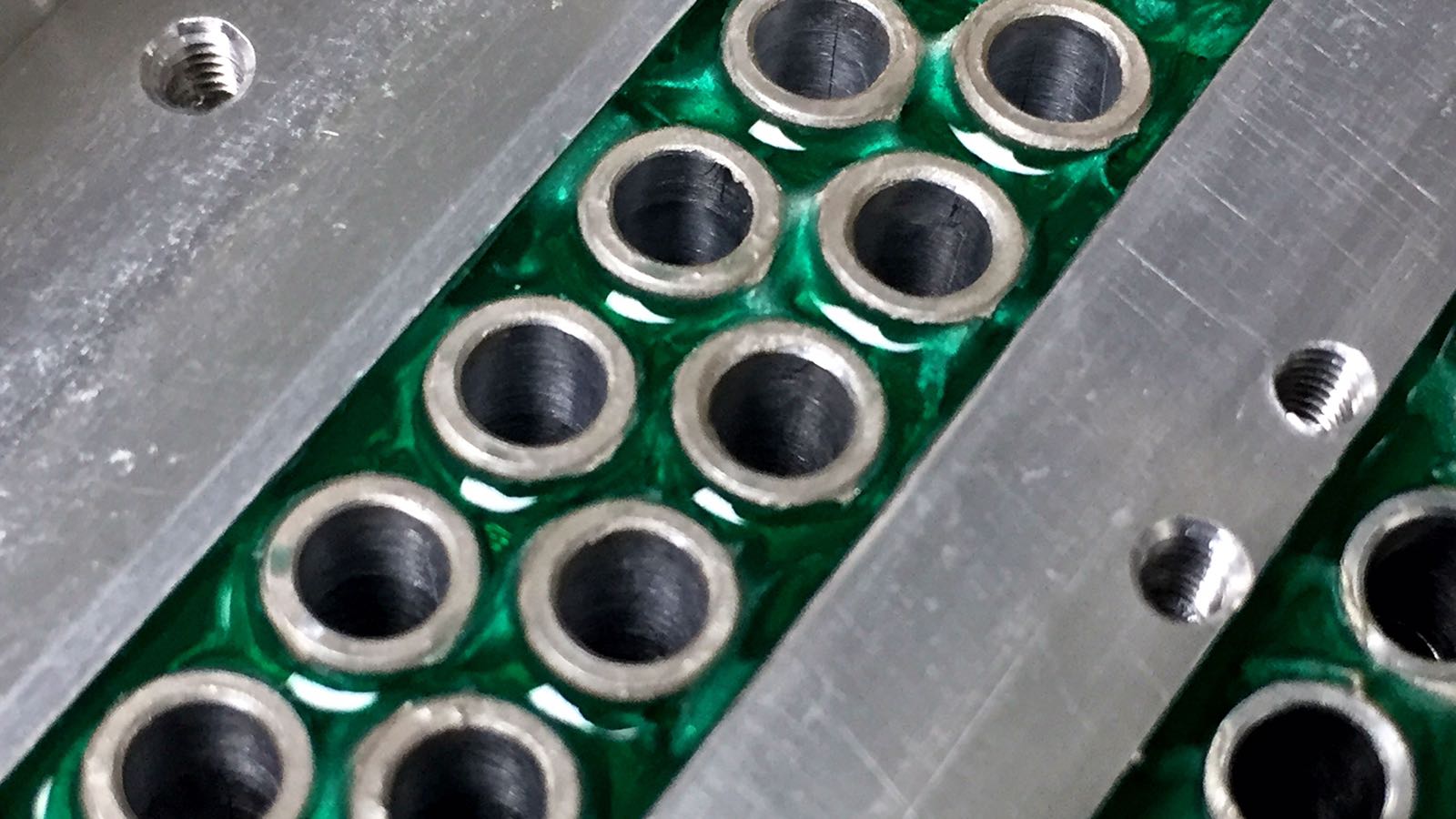}
\caption{Straws glued into the module.}
\label{fig:gluedStraws}
\end{subfigure}\hfill
\begin{subfigure}[t]{0.47\textwidth}
\includegraphics[width=\textwidth]{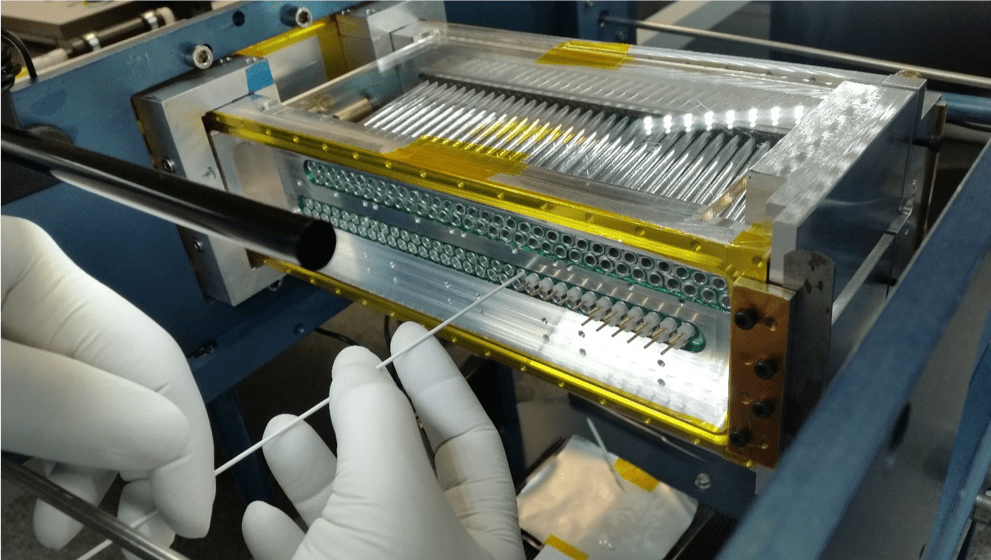}
\caption{A wire being threaded into a straw on the stringing jig.}
\label{fig:crimpJig}
\end{subfigure}\hfill
\begin{subfigure}[t]{0.47\textwidth}
\includegraphics[width=\textwidth]{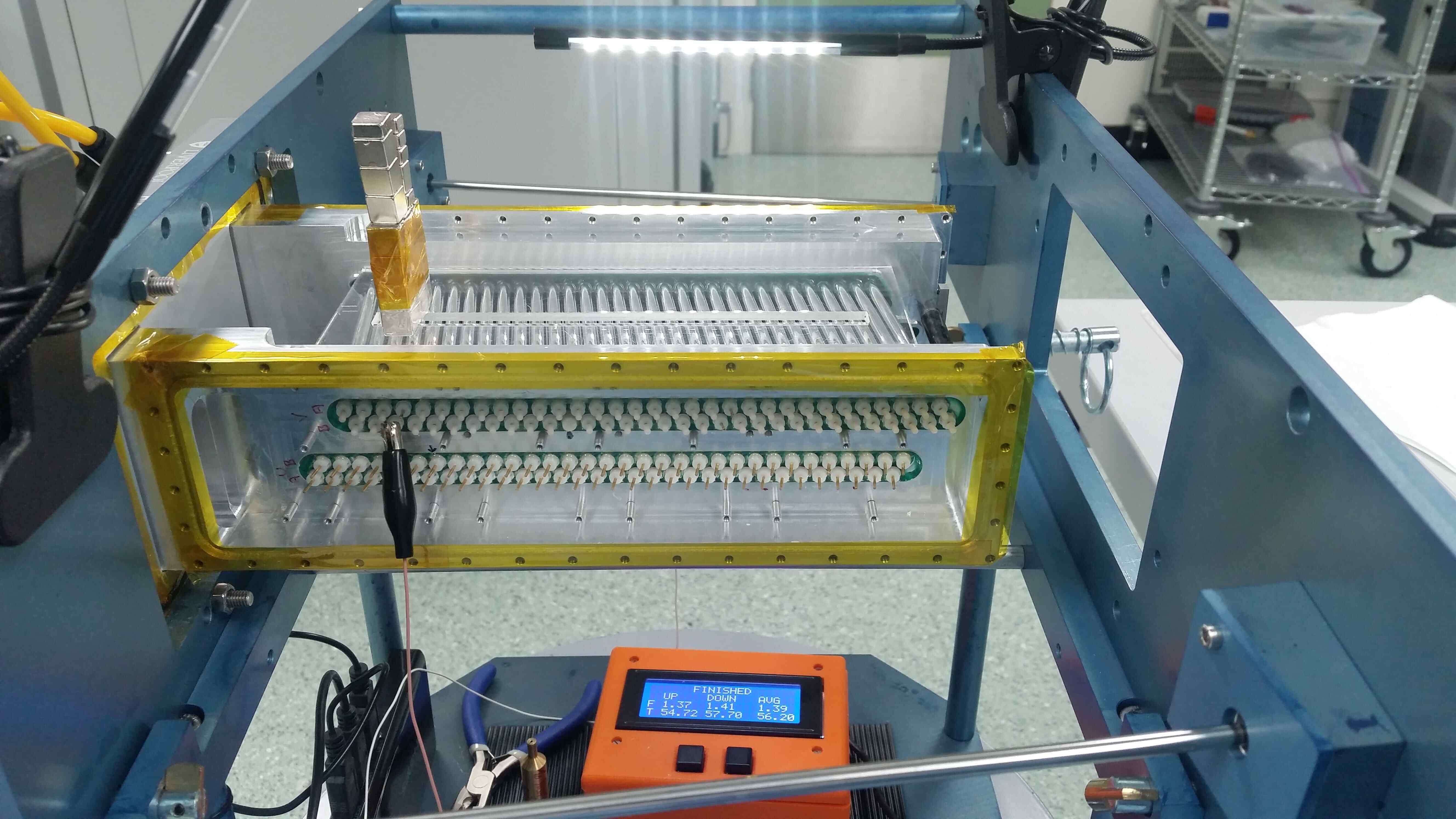}
\caption{A tension test being carried out on a wire.}
\label{fig:tension}
\end{subfigure}\hfill
\caption{Photographs of wire procedures~\cite{SaskiaThesis,TurnerThesis,TabithaThesis}. The 25~$\mu$m wire is threaded into the ‘long’ pin prior to insertion in the module. The pin is crimped onto the wire by applying a force of 1~kN to the pin which is seated in a gap between two plates. The plates are machined such that the gap is a fixed depth, ensuring uniform crimping across all pins. The wire is then threaded into the module and held in place by a short pin that does not connect to any electronics. This pin is crimped in place using a custom-built crimp tool while the module is mounted in a stringing jig.}\label{fig:crimp}
\end{figure}

The straws inserted and glued into the modules are shown in Figures~\ref{fig:insertedStraws}-\ref{fig:crimpJig}. All straws are separated such that the straw centers are spaced 6 mm from each other. Before being inserted into the module, the `long' pins were crimped onto the wire using a custom tool constituting a set of jaws attached to a tension machine. Images of this procedure are shown in Figure~\ref{fig:crimp}. A force of 1~kN was applied between the jaws to crimp the pin securely onto the wire, without snapping or damaging the wire inside the pin. The wires were then threaded down the straws and the plastic insert seated in place in one end of the straw.  The wires were strung and crimped with an applied tension of 0.3--0.5~N (30~g) to remove any kinks or sharp bends that could be discharge hazards. While under tension, the plastic insert was carefully placed in the aluminum end-piece, and then the pin was pushed into the insert and secured with a small amount of epoxy. The crimping method allows good control over the location of the wire in the straw, and each pin could be worked on independently without risk of damage to neighboring pins. After the pin was crimped, the excess wire was cut off and a small amount of glue (50\% epoxy, 50\% rapid Araldite) was applied to the end of the pin to smooth over the cut piece of wire. 

Before the electronics were installed, the manifolds were pulled apart by 50~$\mu m$ using a jack in order to ensure an overall 50~g tension was applied equally to the straws and wires. This removed any kinks or creases in the straw walls such that every individual drift tube has a constant radius, and there are no sharp points which could cause discharges, and allowed for compensation for expansion under vacuum during experimental operation. The tension had to be applied very carefully in order not to damage the straws or introduce small holes from which gas could leak. As part of a series of quality assurance tests, a resistance test was carried out on each strung wire under tension. Each should be within the range of $10$–$13$~$\Omega$. If wires were outside this range, they would be removed, re-strung and re-tested. At the edge of the straw module furthest from the flange is a 5 mm-diameter carbon fiber post, which supports the manifolds and ensures that they remain a fixed distance apart for the entire length of the module, such that the tension across every straw and wire is constant. The post is made from carbon fiber as this has a comparatively high strength-to-density ratio, reducing the risk of multiple scattering while enabling sufficient support to the module. 

After the module had been tensioned and the carbon fiber support post had been installed, custom-built tension-testing devices were used to survey the straws and wires to ensure none were damaged during the tensioning process.  The wire tensions were measured by placing a magnet above the wire, sending a current through the wire and measuring the amplitude of the wire oscillations. The current was changed until the resonant frequency was found, and the tension deduced from that resonant frequency.  Assuming no wires needed to be replaced, the module was then inserted into a vacuum chamber, which was pumped down to a value of $< 10^{-6}$ Torr and left in the chamber for 24 hours to ensure it could hold the required pressure. Rate-of-rise tests were done to check for slow leaks. During the early stages of construction, six of the straws were found to leak. As it was not possible to repair these \textit{in situ}, the end caps were plugged and were deemed inactive. As an additional check, a light-transmission testing stage was added to the process, after which no more straws required plugging. Electrical performance tests were also performed to ensure the module could hold HV for a sustained period and that the straws could detect hits either from cosmic rays or a radioactive source. No wires were broken during transport, testing or installation. Further details on the procedures and tests described here are documented in~\cite{TurnerThesis}. The testing of fully constructed tracking detector modules is discussed in Section~\ref{sec:testInstall}.

For the purpose of handling the completed modules, custom Perspex shields were constructed to reduce any risk of potential damage, e.g., inadvertently changing a module's resistance. The Perspex was custom cut to fit in the gap beneath the manifolds, with one piece placed on either side. This allowed, among other things, for safe transport of the modules to Fermilab and the ability to clean the modules with isopropyl alcohol without damaging the straws.

\section{Electronics and Data Acquisition}
\label{sec:electronicsDAQ}
\subsection{Frontend Readout Electronics}

\begin{figure}[!t]
\centering
\includegraphics[width=.49\textwidth]{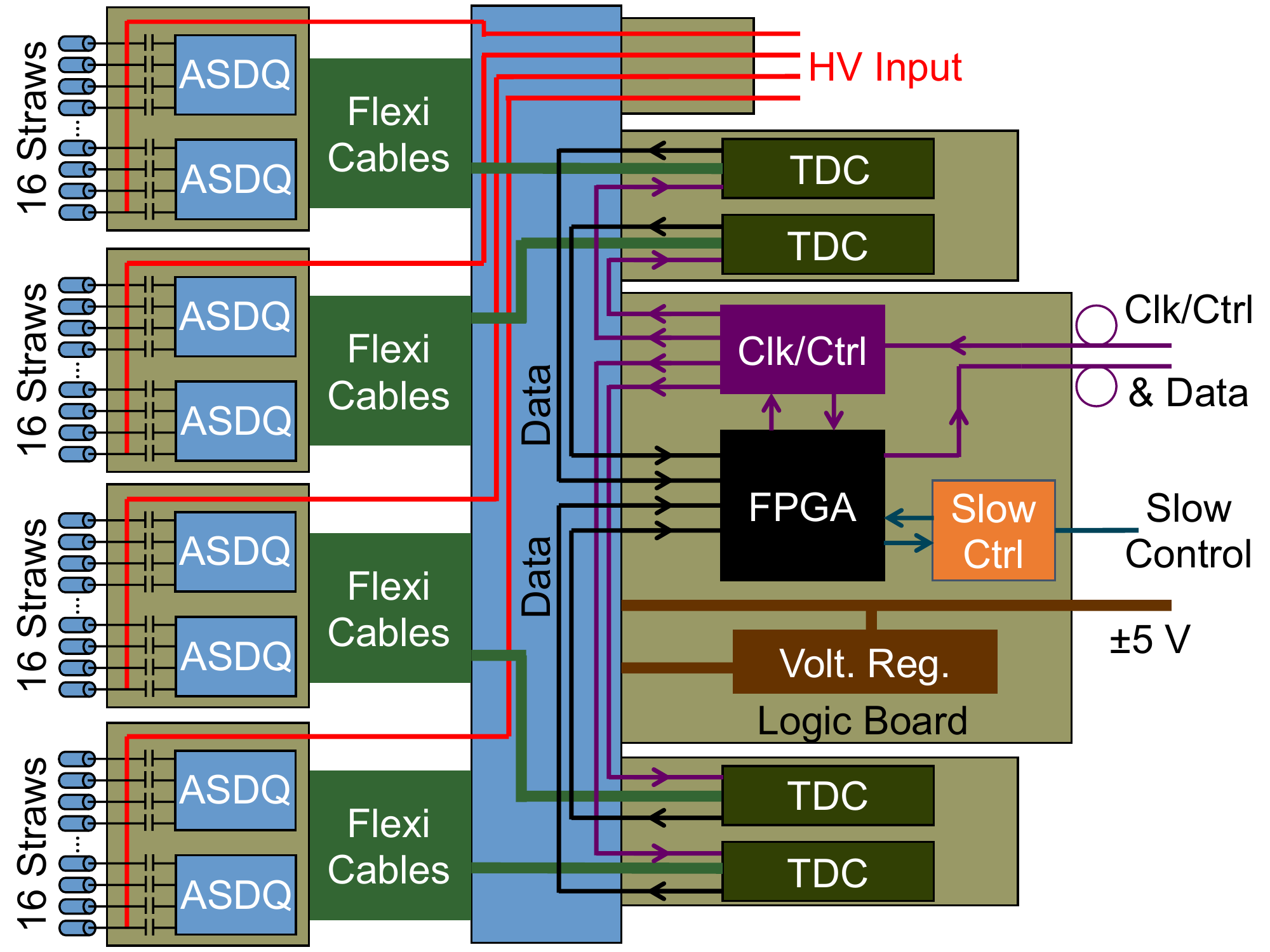}
\caption{ A schematic diagram of the frontend tracking detector readout electronics for
64 straws corresponding to two layers. The straws connect to the ASDQs on
the left and the data are passed through to the right.} 
\label{fig:FrontendSchematic} 
\end{figure}

\begin{figure}
\begin{subfigure}{0.2\textwidth}
\includegraphics[width=\textwidth]{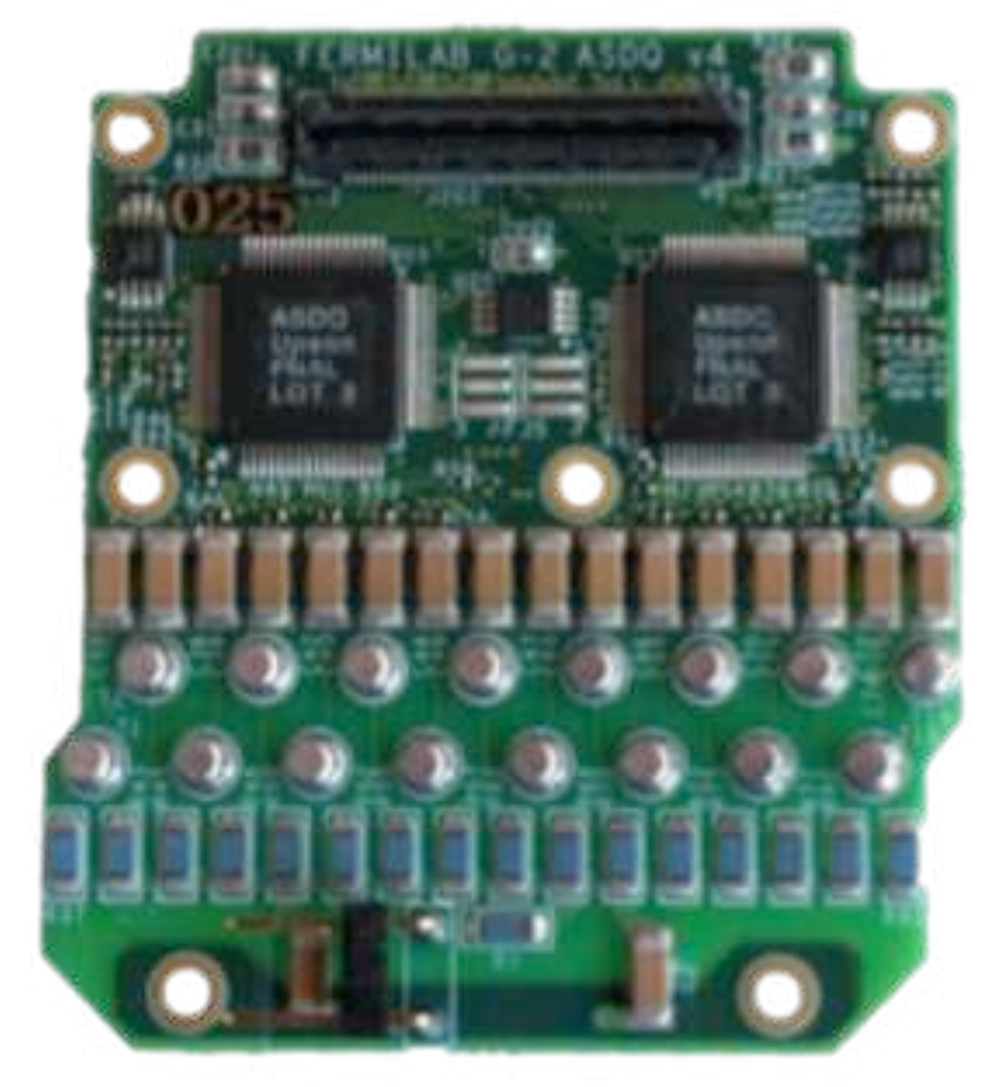}
\caption{ ASDQ}
\label{fig:ASDQ}
\end{subfigure}
\begin{subfigure}{0.2\textwidth}
\includegraphics[width=\textwidth]{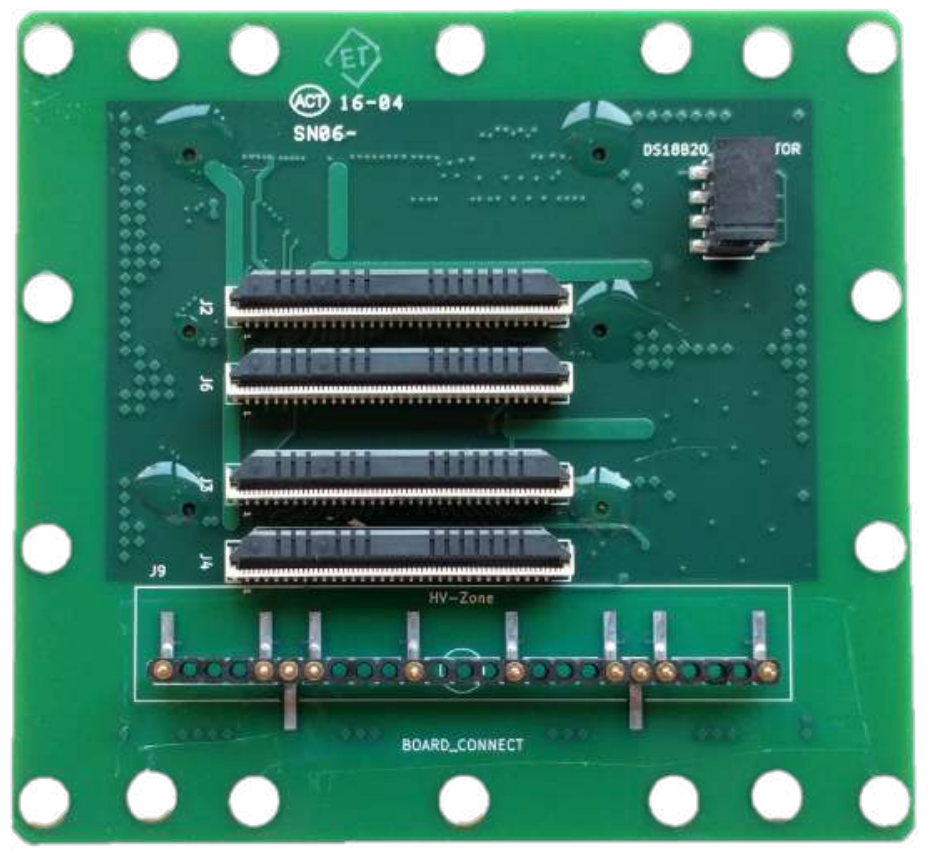}
\caption{ Feedthrough}
\label{fig:Feedthru}
\end{subfigure}
\begin{subfigure}{0.25\textwidth}
\includegraphics[width=\textwidth]{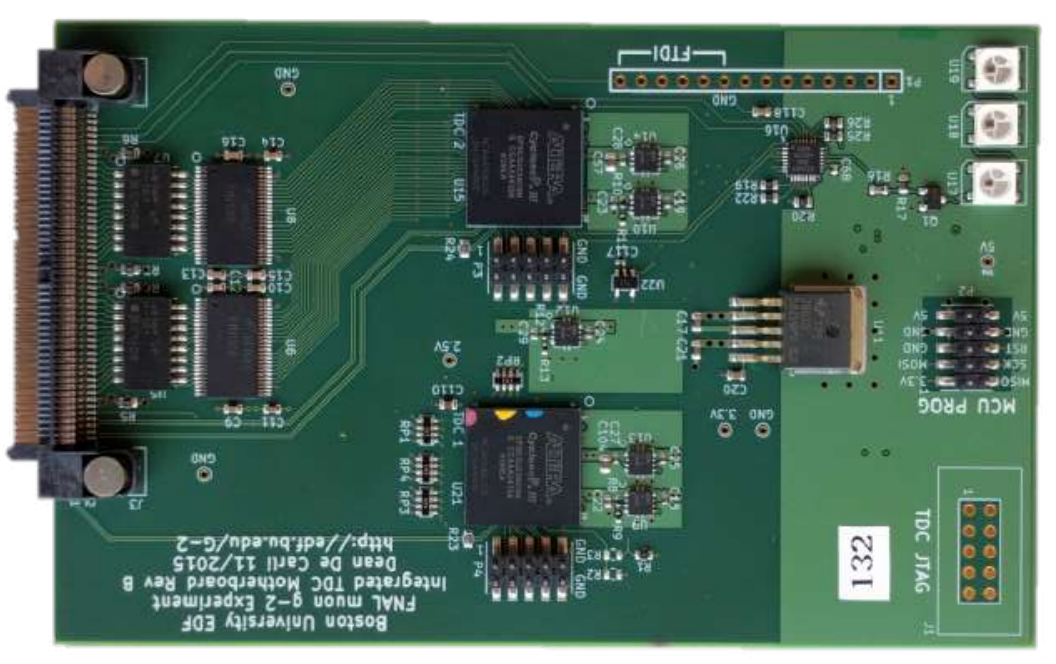}
\caption{ TDC}
\label{fig:TDC}
\end{subfigure}
\begin{subfigure}{0.28\textwidth}
\includegraphics[width=\textwidth]{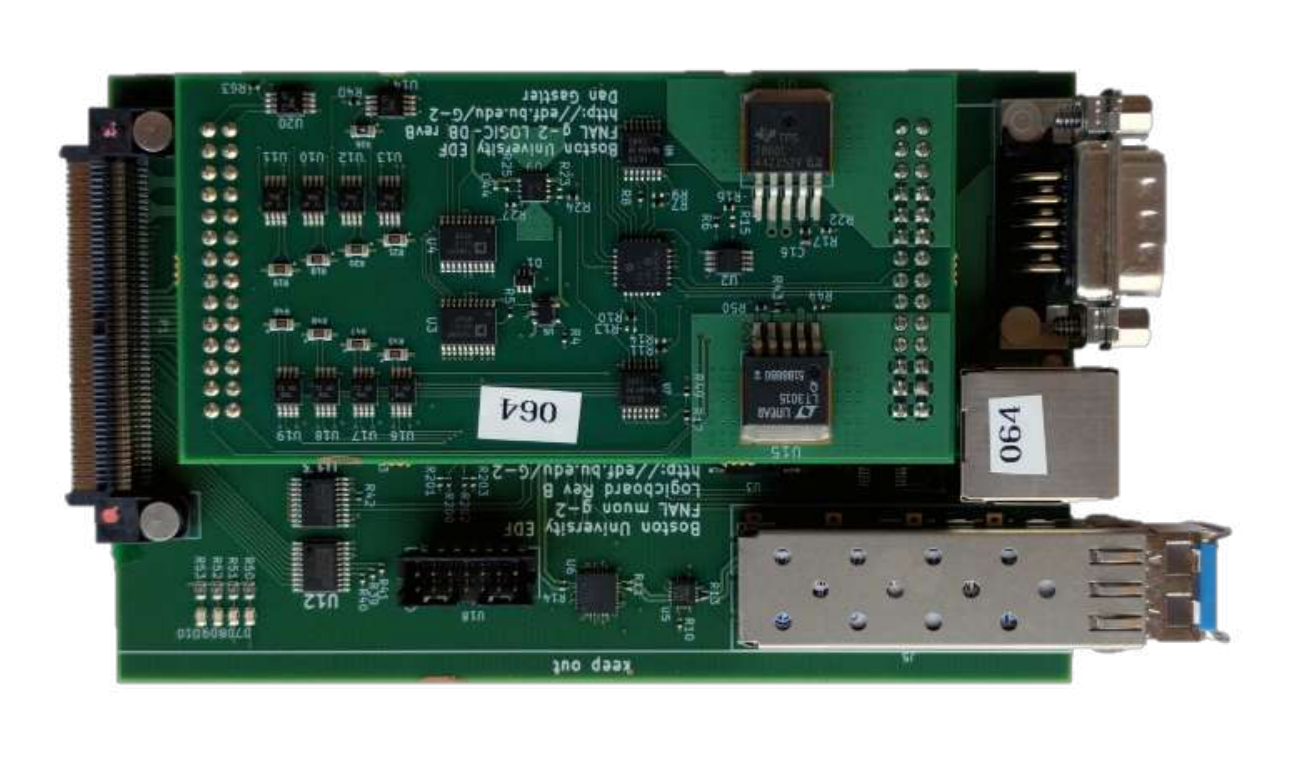}
\caption{ Logic Board}
\label{fig:Logic}
\end{subfigure}
\caption{ Four of the different types of frontend electronics boards. The functionality of each board is described in the text.}
\label{fig:FrontendBoards}
\end{figure}

\begin{figure}[!t]
\centering
\includegraphics[width=0.6\textwidth]{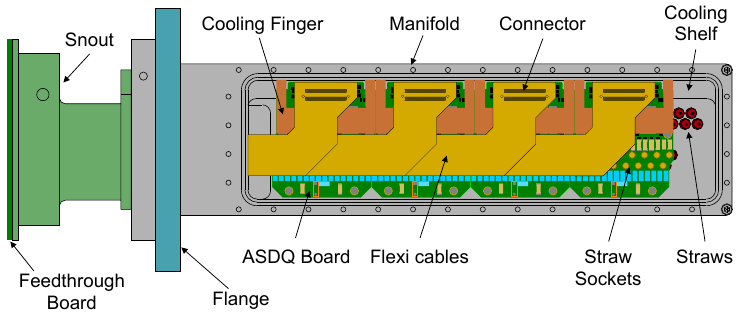}\qquad
\includegraphics[width=0.24\textwidth]{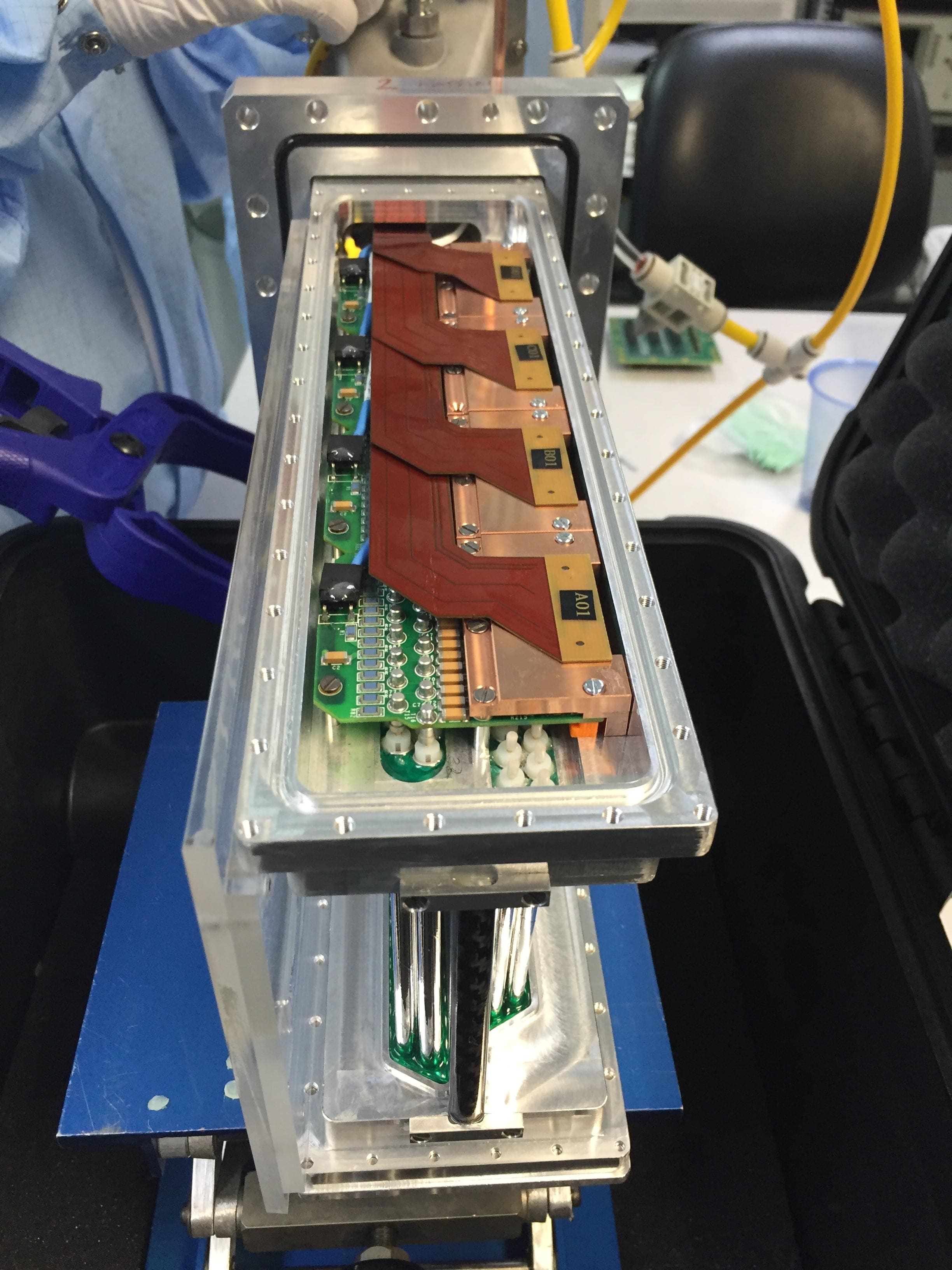}
\caption{ Schematic diagram and photograph of a tracking detector module with the electronics installed inside the manifold.} 
\label{fig:moduleElectronics} 
\end{figure}

The frontend readout electronics consist of six different board types with
the schematic design shown in Figure~\ref{fig:FrontendSchematic}. The ASDQ (amplifier shaper discriminator with charge (Q)) boards, shown in Figure~\ref{fig:FrontendBoards}(a), are connected directly to a group of 16 straws and
housed in the manifold of the tracking detector module, as shown in Figure~\ref{fig:moduleElectronics}. Each board contains two
eight-channel ASDQ ASICs (application-specific integrated circuit) which were originally developed for the CDF Central
Outer Tracking Detector~\cite{Bokhari:1998asdq,Newcomer:1996asdq}. These ASICs provide amplification, shaping, a discriminator and
charge measurement. They meet the design requirements for the tracking detector with
fast charge collection (8 ns), good double pulse resolution (30 ns), low
power (40 mW/ch) and low threshold (2 fC). They also provide baseline
restoration and ion tail compensation using a pole-zero cancellation
technique. The discriminator threshold for the leading edge can be adjusted.  In addition, 
the width of the signal can be altered by changing the drain current into the integrating 
capacitors in the discriminator circuit. It is also possible to inject realistic input pulses 
using the calibration circuits, allowing the circuit delay time and pulse width-to-input charge
relationship to be determined on a channel-by-channel basis. The boards are
cooled by a silicone-based thermal pad and copper cooling finger attached to the
water cooling line described in Section~\ref{sec:construction}. After
assembly, the ASDQ chip heights were measured and only boards within a tolerance
of 30\,\si{\micro\meter} were used to ensure good contact with the cooling finger.
The temperature on each board is monitored with a Maxim Integrated DS18B20
digital thermometer~\cite{Maxim}. The digitization and readout were found to
increase the noise level, so measurements are only performed in between muon 
fills when no data are stored.

The output of each ASDQ chip is eight LVDS (Low Voltage Differential Signal) pairs, where the leading edge represents
the time the threshold was crossed, and the width is proportional to input
charge. These are transferred through flexicables through a feedthrough board
(Figure~\ref{fig:FrontendBoards}(b)) to one of two TDC boards
(Figure~\ref{fig:FrontendBoards}(c)). The flexicables also carry power, reference voltages
and calibration signals to the ASDQs. The feedthrough board provides a gas
seal, allows for signals to be passed in and out of the gas volume and acts
as a backplane for all other frontend boards. The TDC implementation stamps 
transitions with a coarse 400~MHz counter and is extended to fine precision (625~ps~LSB)
by a hand placed 4-phase sampler circuit.  Each TDC board has two Altera
Cyclone III Field-Programmable Gate Arrays (FPGA)~\cite{Altera} which digitize 
up to 2016 hits per trigger to the accuracy of 625 ps.  It receives an external 10~MHz clock which
is multiplied up internally to 400~MHz for time measurements and internal
operation. After receiving a trigger, the TDC starts a readout cycle for a given time window
according to programmable delay and width parameters.


The data from the two TDC boards are transferred to a logic board (Figure~\ref{fig:FrontendBoards}(d)) 
which assembles the event and sends it to the backend electronics in 8b/10b format~\cite{EightBTenB} at 125 Mbits/sec over optical fiber. The logic
board contains a Xilinx Spartan-6 FPGA~\cite{XilinxSpartan6} which can buffer up to five spills
from each TDC and can be remotely reprogrammed. It receives an external 10
MHz clock with control and trigger signals encoded using the C5 clock-command
combined carrier coding~\cite{CFive} which it multiplies and distributes to the TDCs. It has a
mezzanine board that regulates the external power and supplies the reference voltages
required by the ASDQs, with temperature, voltage and current sensors which
can be monitored. 

There are also two HV boards for each tracking detector module. Each of these boards filters an external
supply of four HV channels, passing them via the feedthrough board to the ASDQ boards along a
separate cable. The ASDQ boards further filter the HV and provide it to each
straw through a 100~$k\Omega$ current-limit resistor. Each straw has an 0.49~nF blocking capacitor
to isolate the HV from its ASIC input.

\subsection{Backend Readout Electronics}

The backend readout uses FC7 advanced mezzanine cards (AMC), originally developed
for the CMS experiment~\cite{Pesaresi:2015qfa}, to provide the clock and control signals to the
logic boards and receive the hit data back. They are housed in a $\mu$TCA (micro Telecommunications Computing Architecture) crate, with up
to 12 slots available, and connected to an AMC13 controller module~\cite{AMC13} via the
backplane. The crate also has a commercial $\mu$TCA Carrier Hub (MCH) controller which provides the
Intelligent Platform Management Interface (IPMI) functionality and Gb ethernet. The FC7 board has
a Xilinx Kintex-7 FPGA~\cite{XilinxKintex7} and can accommodate two FPGA Mezzanine Cards (FMC).

The clock, control and readout data are transmitted between the FC7s and
the logic boards via fiber optic cables which has the benefit of decoupling
the clock/control/data-readout signals from the low voltage (LV). One FC7 is sufficient to
readout a single tracking detector station since each of the 16 logic boards will communicate with the FC7
with a single fiber pair (via a small form-factor pluggable (SFP) transceiver) and each of the two FMCs on the FC7 can
accommodate eight fiber pairs. The FC7s required to readout the entire tracker detector
can therefore be comfortably housed within one $\mu$TCA crate, with space for spares to
allow for quick recovery should an FC7 fail. The FMCs containing the 16 SFPs
is the CERN EDA-02707~\cite{EDA-02707} which is also used for the Trigger, Control and
Distribution System (TCDS) at CMS. The configurations for the
AMC13 and FC7 boards are sent over the ethernet connection using the \texttt{IPBus}
communication tools developed by CMS~\cite{Larrea:2015wra}. 

The clock and control signals are sent over the backplane from the AMC13 to
the FC7s where they are encoded into C5 and transmitted to the logic boards
on a 10 MHz clock. The 8b/10b data from the logic boards are formatted in the
FC7 to match the requirements of the AMC13. The FC7 has 30 MB of clock
RAM which is more than an order of magnitude larger than the data volume
expected from one spill in a single tracking detector module. 

\subsection{Data Acquisition System (DAQ)}

The DAQ system is triggerless. Hits in the straw tracking detector are recorded 
for approximately 700~$\mu$s commencing 2~$\mu$s after muon beam injection. 
The beam arrives in 16 separate spills over a 1.4~s period. There are 8 spills separated 
by 10~ms, a 197~ms gap,  a further 8 spills separated by 10~ms, and then a 1063~ms gap.

The data collected by the AMC13 are transferred to the PC over \texttt{IPBus}~\cite{Larrea:2015wra} in 32~kB
blocks, which is sufficient given the low data rate from the
tracking detector. Functionality to transfer the data using 10GbE optical fiber links also exists.
The DAQ software is based on the \texttt{MIDAS} package~\cite{MIDAS}
that was developed at the Paul Scherrer Institute and is now maintained at
TRIUMF. It consists of function templates and library routines for processes
that handle the communication with the backend electronics, event building and data
logging. The DAQ frontend code packages the detector data into a \texttt{MIDAS} bank and passes it
to the event builder where the banks from all other detector systems are
combined into a single \texttt{MIDAS} event. 

The experimental configurations are stored in the \texttt{MIDAS} online database (ODB). 
The ODB can be accessed from the web page to adjust the parameters as required
and can be read from or written to via the frontend code. This contains
configurations for each individual board, as well as global settings, with
the most important parameters displayed on a custom web page. There are also
buttons for turning the low and HV supplies on and off which can be
used by nonexperts.

In addition to the fast frontend collecting the data, there are two slow
control frontends, asynchronous from the event builder. These run
periodically to monitor information about the boards and low and HV supplies such as temperatures and currents, storing the values in both the
ODB and a PostgreSQL database. Alarms can be thrown based on the values of
the parameters stored in the ODB. 

The data are also passed to an online analysis layer using an {\em art}-based
analyzer~\cite{Green:2012gv} together with JavaScript to publish plots of the data in real time
to a web server. As the analysis process is slower than the data rate, it is
designed such that it always processes the most recent event. The plots show both basic information (number of hits per straw, total data size, found errors) and complex results, such as the reconstructed straw drift time distributions, which have been found to be highly sensitive to issues with the HV, gas or detector timing. In addition, full track reconstruction is performed on a smaller subset of the data and the
reconstructed beam profile can be monitored. Alarms are thrown to the \texttt{MIDAS} DAQ system in case
any errors are found.

\section{Testing and Installation}
\label{sec:testInstall}
\subsection{Test Stand and Quality Assurance}\label{sec:testStand}

\begin{figure}[!t]
\centering
\includegraphics[width=1.0\textwidth]{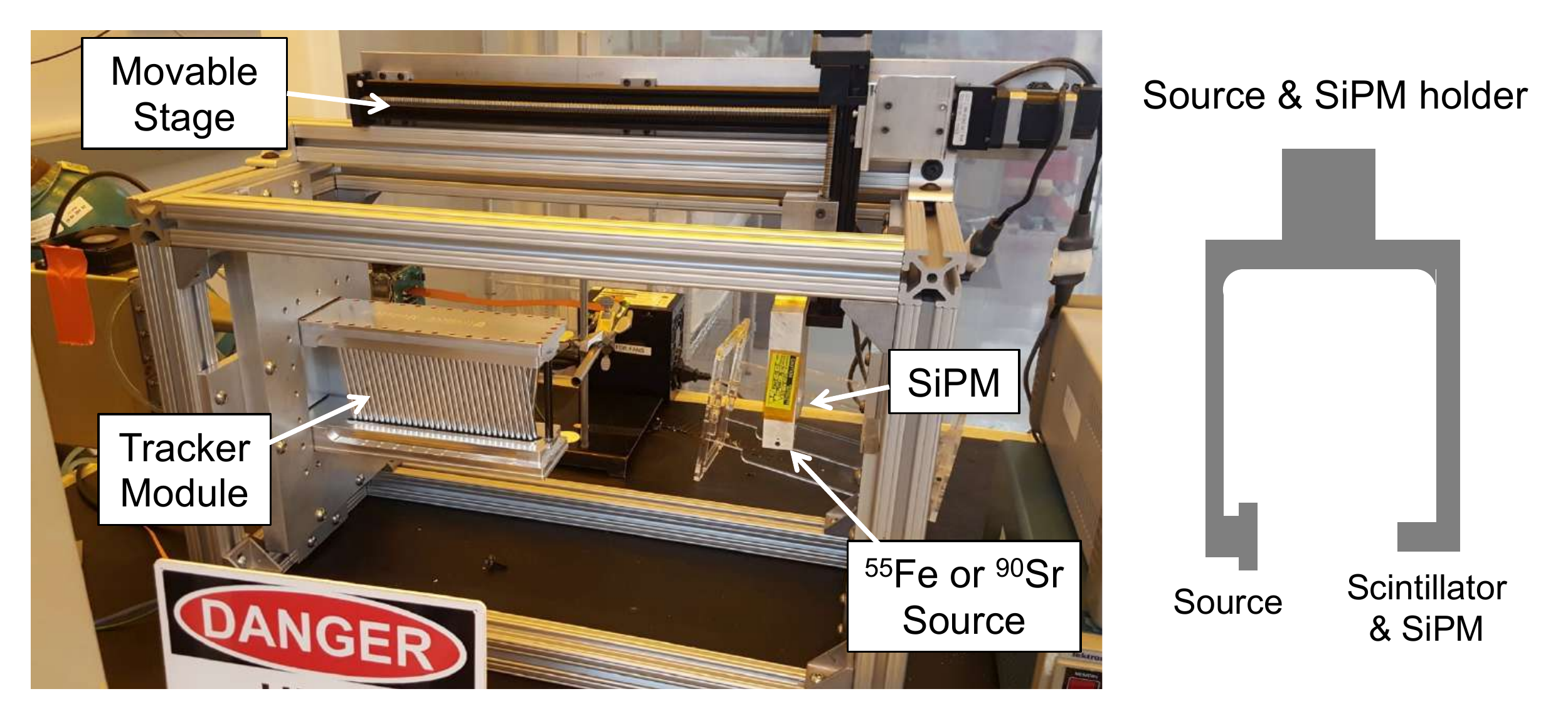}
\caption{ A photograph of a tracking detector module installed in the test stand and a diagram showing the source and detector holder. The movable stages are at the top of the frame. The radioactive source and a scintillator trigger are mounted on an inverted ``U''-shaped aluminum bracket attached to the vertical movable stage. A detailed description of the test stand design is given in~\cite{Behnke}.}
\label{fig:teststand}
\end{figure}

As part of a dedicated quality control and quality assurance plan, a radioactive source test stand was designed, constructed and installed at Fermilab~\cite{Behnke}. The purpose of the completed stand, shown in Figure~\ref{fig:teststand}, was to evaluate the performance of each tracking detector module: straw response, gain, and position of each straw and wire. The $60 \times 30 \times 40\ {\rm cm}^3$ frame was constructed from extruded aluminum, with a tracking detector module mounted into an aluminum flange identical to the mounting flanges on the storage ring vacuum chambers. Horizontal and vertical motion is achieved with a Velmex XSlide traverse system controlled with VXM motor controllers~\cite{Velmex}, which was measured to have a positional accuracy of $\sim 10~\mu$m~\cite{Epps}. The two-dimensional motion system mounts to an inverted ``U''-shaped aluminum bracket that contains a radioactive source holder on one side of the module and an Eljen EJ200 plastic scintillator~\cite{Eljen} read out by a SenSL MicroFJ-SMPTA-30035 silicon photomultiplier (SiPM)~\cite{OnSemiconductor} on the other side. The source is highly collimated and mounting pins enabled the orientation to be shifted $\pm7.5^{\circ}$ from the vertical to match the straw stereo angle. Extensive optimization was performed on the test stand's physical dimensions, collimator size, and event rate~\cite{Epps}.

This arrangement permitted scans of the entire active area of each module with either an $^{55}$Fe $\gamma$-ray or a $^{90}$Sr $\beta$-ray source, as well as data quality assurance tests of hits registered from cosmic muons. The $^{55}$Fe source provides a nearly-monoenergetic $\gamma$ ray that deposits a fixed amount of energy in the straw. This source can be used to measure the gain of each straw, which is defined as the ratio of charge produced at the sense wire to the initial charge at ionization. Once the modules are constructed, there is no direct way of measuring the charge on the sense wire. Instead, the ASDQ discriminator threshold was used as a proxy. Independent measurements using the discharge of a small capacitor determined the response of the electronics to a given input charge deposited in a few nanoseconds, providing the conversion from discriminator threshold to Coulombs. A change in threshold of $1$\,mV is equivalent to $0.035$\,fC.

\begin{figure}[!t]
\centering

\begin{subfigure}{0.49\textwidth}
\includegraphics[width=\textwidth]{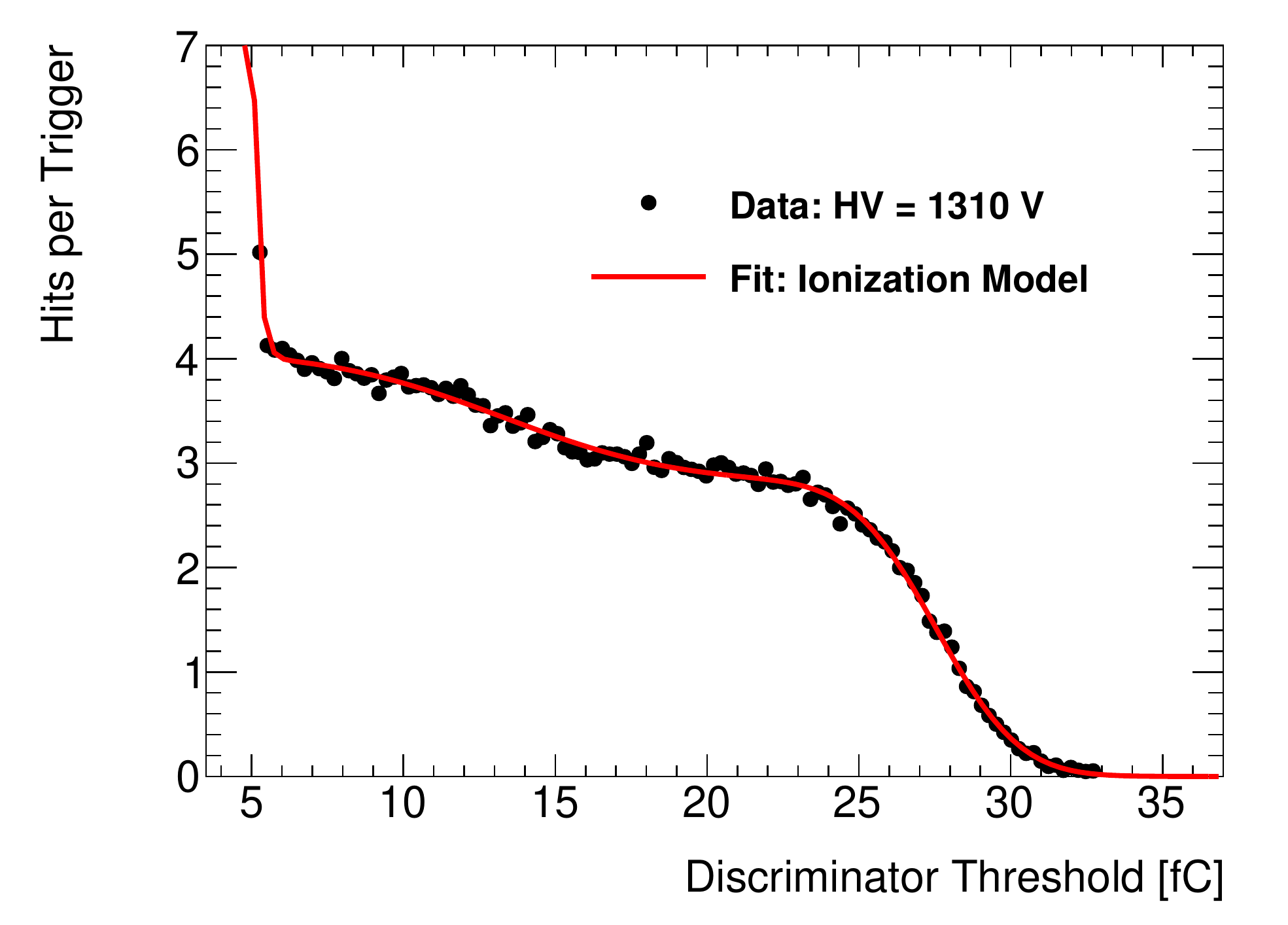}
\caption{}
\label{fig:TestStandGainScan_a}
\end{subfigure}
\begin{subfigure}{0.49\textwidth}
\includegraphics[width=\textwidth]{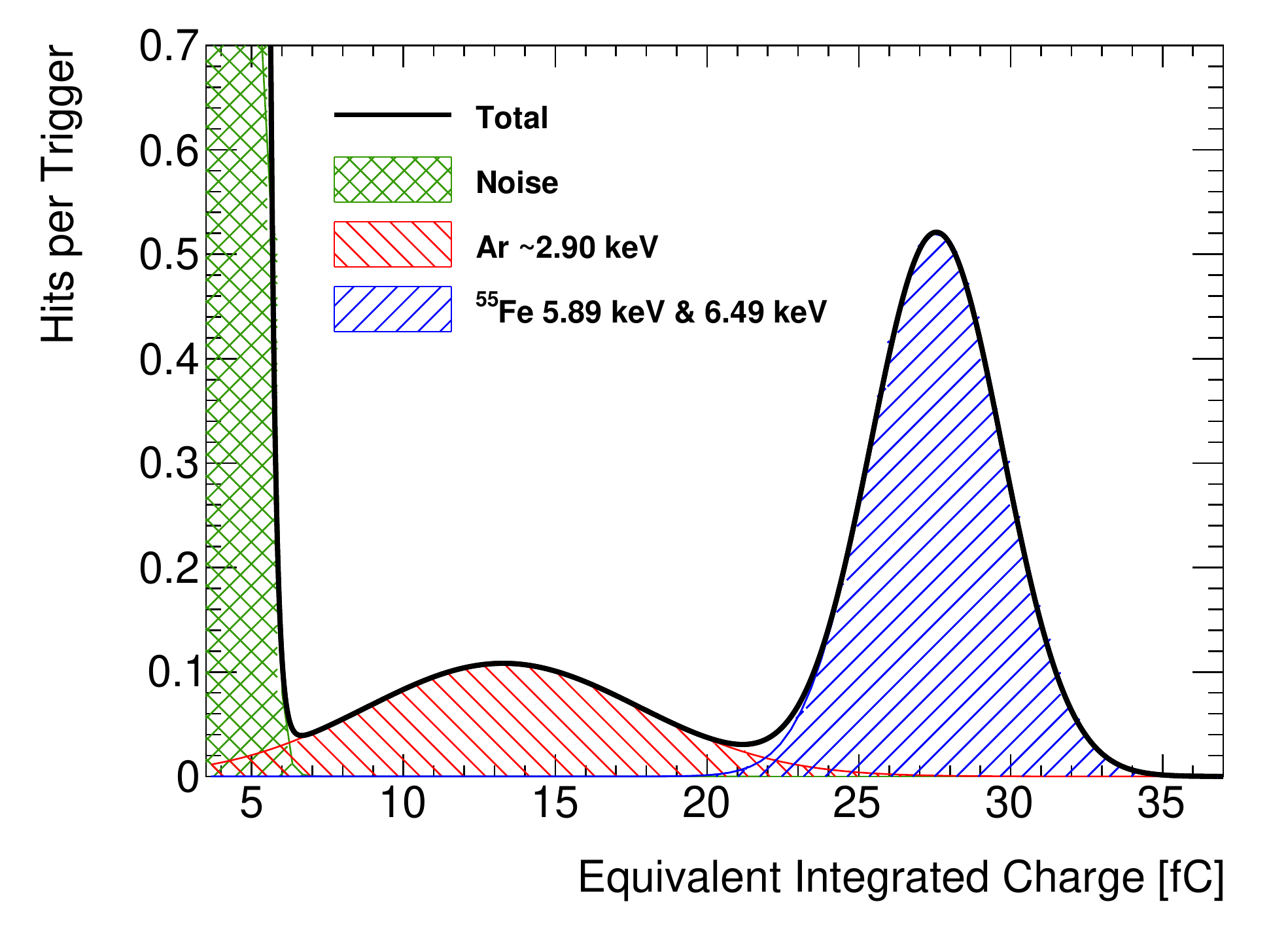}
\caption{}
\label{fig:TestStandGainScan_b}
\end{subfigure}
\caption{ (a) Event rate versus discriminator threshold converted to an equivalent charge for a single straw in the presence of an $^{55}$Fe source.
The fit function is the sum of an exponential noise term and two complementary error functions. (b) The Gaussian and exponential equivalents to the functions shown in (a). The fit function is obtained by integrating these functions above threshold. The two Gaussian distributions correspond to an argon escape peak at 2.9~keV and two unresolved $^{55}$Fe X-rays at 5.89~keV and 6.49~keV~\cite{Ahmed:2013ela}.}
\label{fig:TestStandGainScan}
\end{figure}

The $^{55}$Fe source was placed in the source holder at a constant distance away from the straw and remained at this distance for all measurements. The low energy photons from this source are fully absorbed by the argon and produce the initial ionization. There are three photon energies produced, corresponding to
$5.89$~keV $\mathrm{Mn ~ K_{\alpha}}$ X-rays, $6.49$~keV $\mathrm{Mn ~ K_{\beta}}$ X-rays and the $2.9$~keV escape peak~\cite{Ahmed:2013ela}. In practice, only two peaks are observed as the two X-rays are only resolved as a single peak. The discriminator threshold was gradually increased starting from the noise floor and the number of hits per trigger recorded until no hits were observed. 
A typical scan is shown in Figure~\ref{fig:TestStandGainScan_a}. The noise at low threshold energy was fit with an exponential, and two complementary error functions (the integral of a Gaussian) were used to model the response to the different initial charges from the X-ray and escape peaks. 
The equivalent Gaussian and exponential functions are shown for clarity in Figure~\ref{fig:TestStandGainScan_b}.

\begin{figure}[!t]
\centering
\begin{subfigure}{0.49\textwidth}
\includegraphics[width=\textwidth]{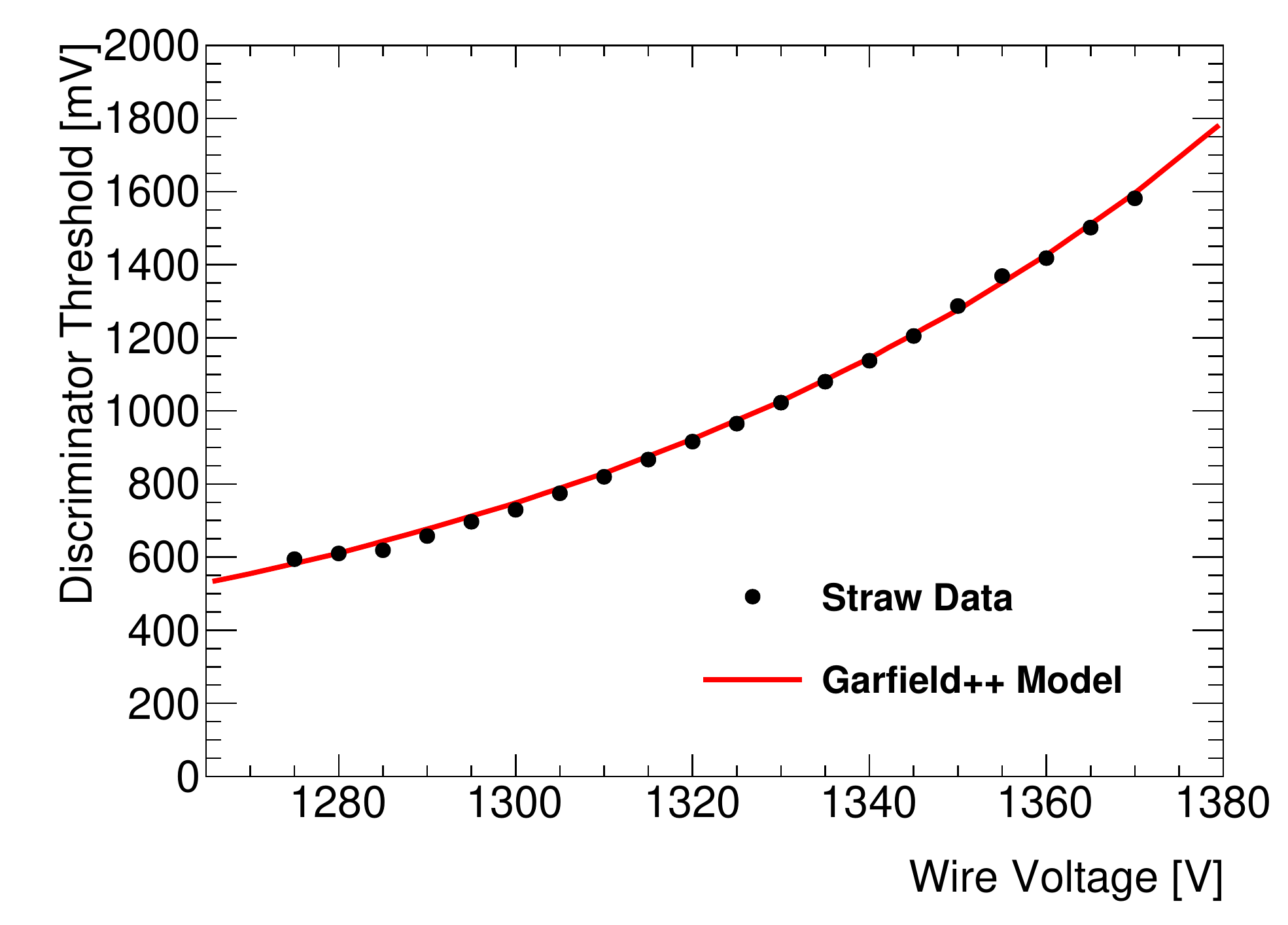}
\caption{}
\label{fig:GarfieldPlateau_a}
\end{subfigure}
\begin{subfigure}{0.49\textwidth}
\includegraphics[width=\textwidth]{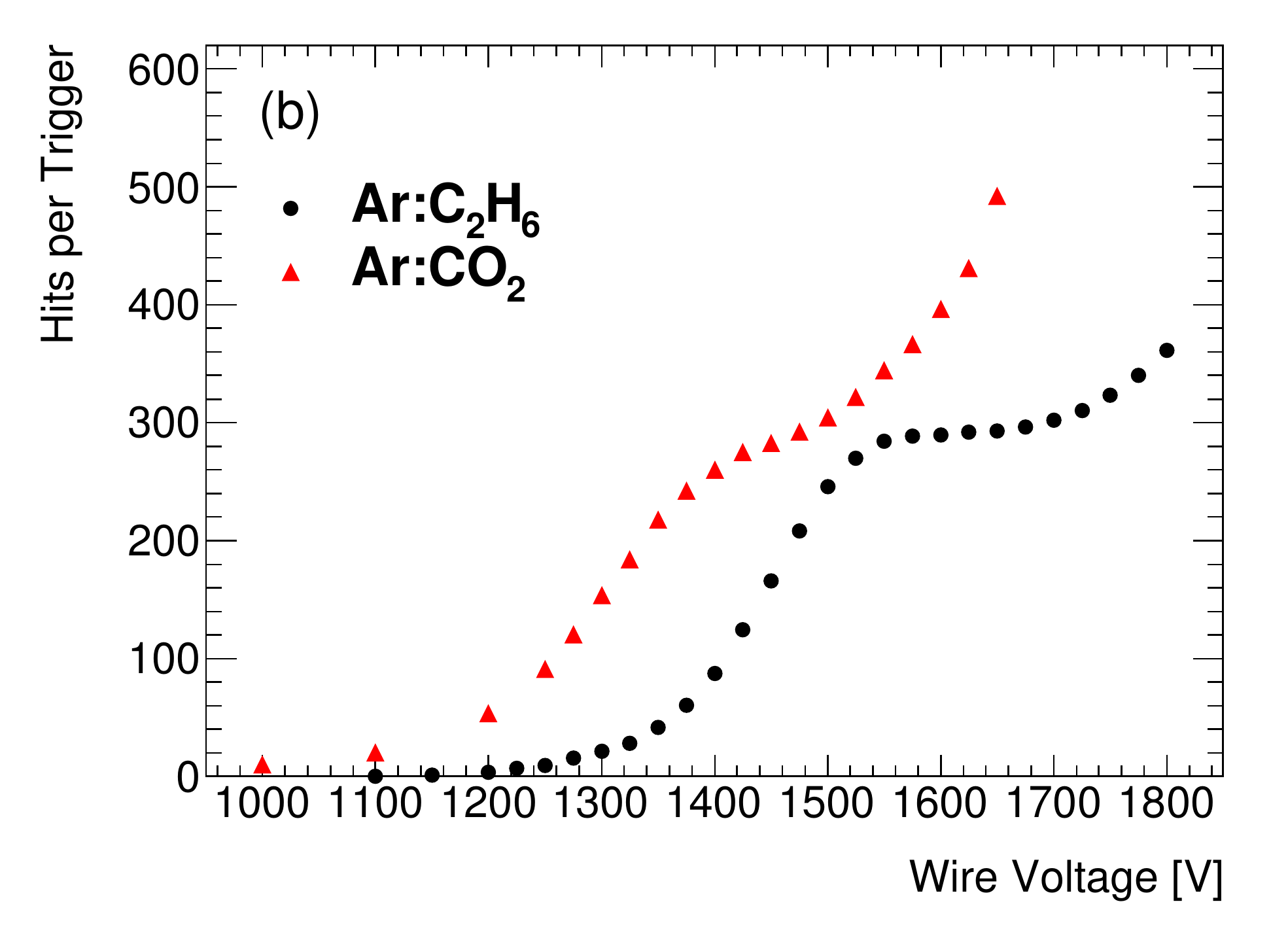}
\caption{}
\label{fig:GarfieldPlateau_b}
\end{subfigure}
\caption{ (a) Comparison of data taken with an $^{55}$Fe source and values calculated with a \texttt{Garfield++} model of
a straw and readout electronics.
Measurements of the ASDQ discriminator voltage setting corresponding to the mean of the X-ray signal peaks
are displayed for different HV settings. A single tuning parameter has been applied to the \texttt{Garfield++} model.
(b) Event rate versus voltage for a straw with Ar:CO$_2$ 80:20 and Ar:C$_2$H$_6$ 50:50 gas mixes.
The rate rises until it reaches an extended plateau before the straws begin to refire.
The tracking detector is operated in the plateau region at 1650~V with Ar:C$_2$H$_6$ 50:50. }
\label{fig:GarfieldPlateau}
\end{figure}

The number of electrons in a single ionization cluster from $^{55}$Fe X-rays is significantly larger than that expected from an ionization from a typical signal positron. In order to maintain a reasonable signal amplitude, the measurements are made at a lower HV setting than normal running, ranging from $1275 - 1370$~V in steps of $5$~V. The mean of the higher energy peak was compared with that expected from a \texttt{Garfield++} simulation~\cite{Garfield++} for each voltage setting. As shown in Figure~\ref{fig:GarfieldPlateau_a}, this was found to be in excellent agreement (following normalization with a single scale parameter). The simulation can be used to extrapolate the gain to the operating voltage.

The $^{90}$Sr/Y source releases a $\beta$-ray with an end-point energy of $\sim2.2$\,MeV and was used to measure the functionality of each straw and its readout electronics. Measurements, shown in Figure~\ref{fig:GarfieldPlateau_b} for Ar:CO$_2$ 80:20 and Ar:C$_2$H$_6$ 50:50 mixtures, were carried out to find the optimal operational wire voltage for the modules. Measuring the event rate versus the applied wire voltage produces plateau curves, indicating that in plateau regions all beta particles traveling through the straw are detected by the wire as hits. The plateau for the intended Ar:C$_2$H$_6$ 50:50 mixture resulted in a chosen operating wire voltage of 1650~V. The gas gain at the nominal straw conditions during experimental operations is estimated to be $\sim 370,000$.

\begin{figure}[!t]
\centering
\begin{subfigure}{0.49\textwidth}
\includegraphics[width=\textwidth]{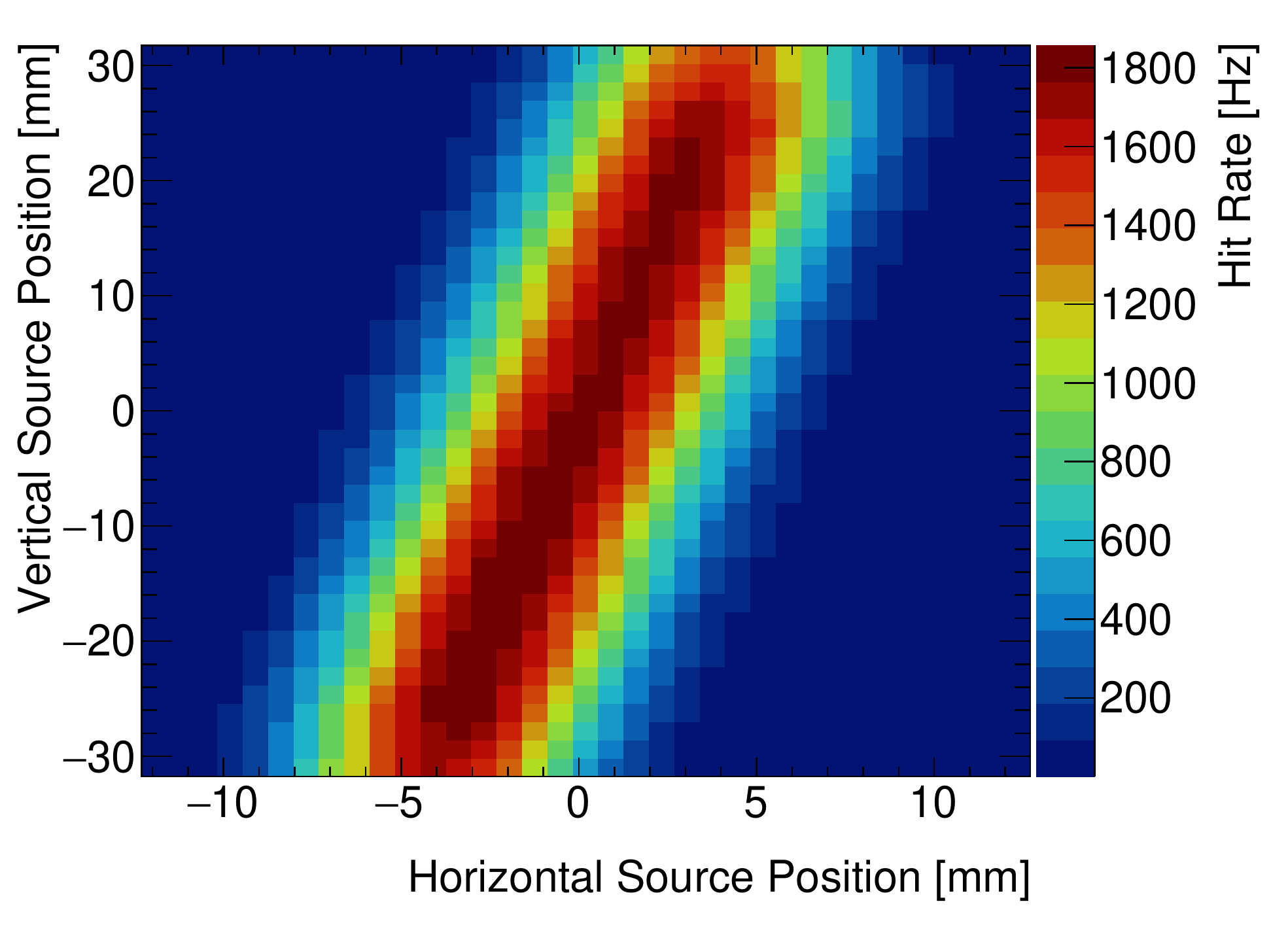}
\caption{}
\label{fig:2d-dist-straw-pos_a}
\end{subfigure}
\begin{subfigure}{0.49\textwidth}
\includegraphics[width=\textwidth]{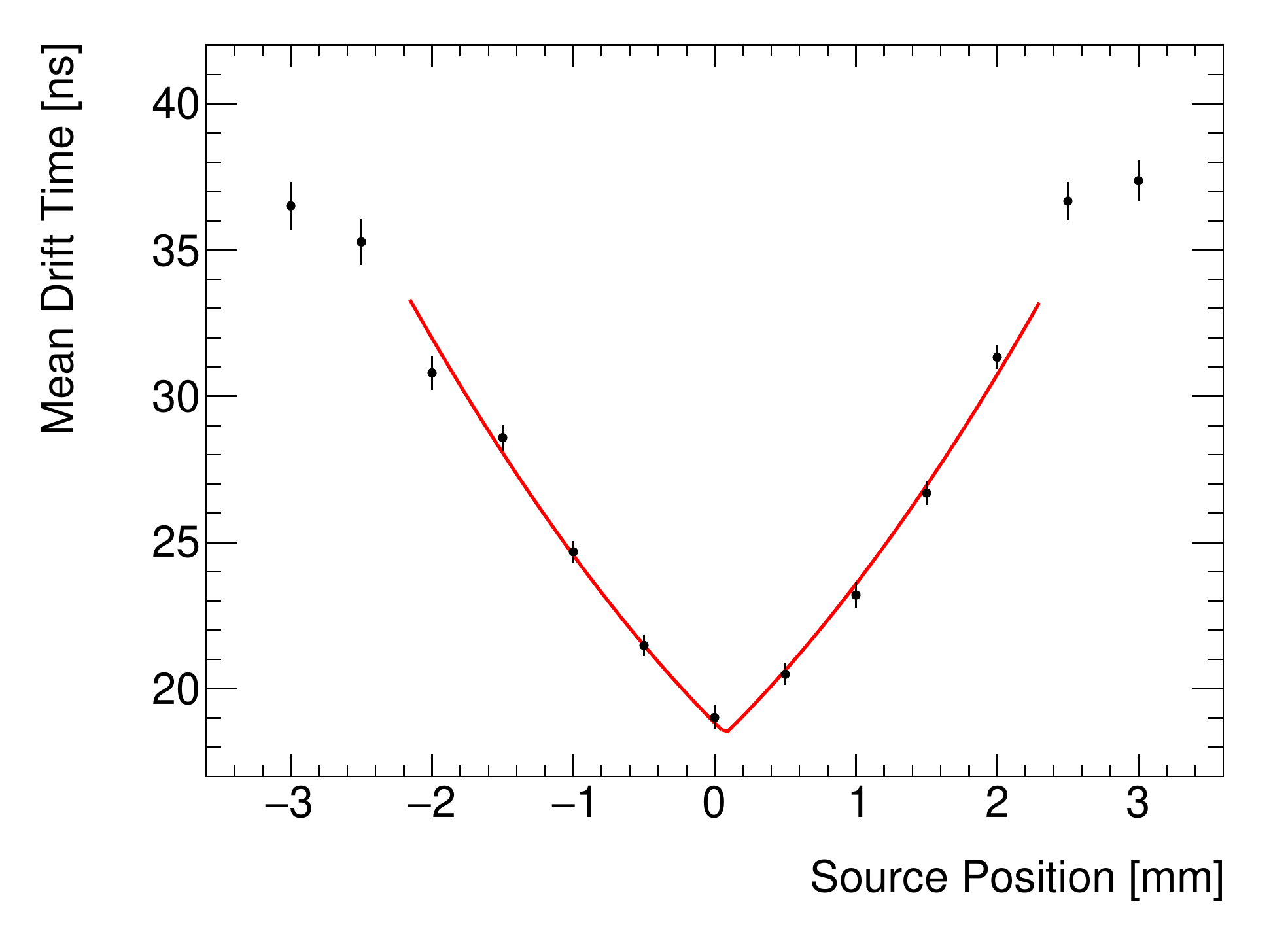}
\caption{}
\label{fig:2d-dist-straw-pos_b}
\end{subfigure}
\caption{ (a) Two-dimensional distribution of event rate versus horizontal and
vertical position for a single straw. All straws except the straw
of interest have been disabled.
(b) Average drift time versus $^{90}$Sr source position for a single
straw. The slope of the fit (red line) is the time to drift distance
relationship and the minimum of the fit is the wire position.}
\label{fig:2d-dist-straw-pos}

\end{figure}

In addition to measuring the plateau voltage, a scan across the chamber provided information on the position and angle of each straw. Figure~\ref{fig:2d-dist-straw-pos_a} shows a two-dimensional distribution of event rate for a single straw. Measurements with different horizontal and vertical source positions allow the confirmation of the straw stereo angle. Using the $^{90}$Sr $\beta^-$ source with a trigger from the plastic scintillator and silicon photomultiplier provides additional timing information. In this way, both the wire position and the time to drift distance relationship can be measured. A scan of average drift time versus horizontal source position is shown in Figure~\ref{fig:2d-dist-straw-pos_b}. The wire position can be extracted from a fit to these data and multiple scans at different heights allow the measurement of the wire angle.  All measurements were consistent with the nominal wire position within a resolution $<100~\mu$m.

\subsection{Gas System}\label{subsec:gas}

The $g-2$ experimental hall is an enclosed space containing a large cryogenic magnet and the tracking detector gas system, which is supplied in standard high-pressure industrial cylinders. The tracking detector gas is Ar:C$_2$H$_6$ 50:50. As such, the gas mixing system for the $g-2$ tracking detector is designed to flow 50\% ethane and 50\% argon by volume. To mitigate and reduce the corresponding flammable gas and oxygen deficiency hazard (ODH) risks inside the Muon $g-2$ experimental hall, a gas storage shelter containing the supply system (and where almost all the gas inventory is located) was constructed outdoors external to the experimental hall, with lockout and appropriate flammable hazard warnings. The shelter contains two ethane gas canisters, each containing 410 standard cubic feet (SFC) of ethane and a manifold of four canisters of argon, each containing 300 SCF of argon. The ethane is two-phase depending on the outdoor temperature and the pressure inside the cylinder. The ethane canisters are connected in parallel with two isolation valves controlled by programmable logic controllers (PLC). 

Due to the flammable nature of ethane, multiple safety
mechanisms were implemented to store and distribute the tracking detector
gas. The system was designed to provide a continuous flow to the tracking detector
during gas canister replacement, and permit the isolation of single points of
failure such as a burst straw. When pressure readings from the ethane supply canister in use drops
below 25 psig, PLC logic automatically switches both valves to open the
second ethane gas canister while closing the empty canister. Spent ethane canisters can be replaced individually by the gas supplier without interruption of experiment operation, as the control system only permits flow from a single cylinder at a time. When the PLC senses low pressure on the current ethane canister in operation, manual valve controls allow the technician to vent any excess air from the line and install a new ethane canister, Using the low volume flow rate of 720~std. cc/min allows for
canister replacement every 4 to 5 weeks of experiment operation. When the bank of four argon cylinders is emptied, a backup cylinder is automatically used allowing the four empty cylinders to be exchanged. 

Both the ethane and argon
supply lines penetrate the experimental hall for mixing, routing, and
delivery. Mass flow controllers each supply 50\% of the flow by volume of argon gas and ethane, with the PLC automatically adjusting the flow set point to the modules to supply 30 standard cubic centimeters per minute of Ar:C$_2$H$_6$.
The argon and ethane are combined in a swirling in-line mixer that
proceeds through a control volume buffer along with both carbon and membrane
filters. Sample valves are available at this location
to test the quality
and ratios of mixed Ar:C$_2$H$_6$. The dry supply is then passed through a
chilled 
isopropyl alcohol bubbler at a temperature of $5^{\circ}$ C. The supply line then splits, sending the gas mixture around the
outside of the ring to the
two separate tracking detector stations. If none of the stations are configured to
receive gas, then there is routing to directly vent from the supply to the
exterior of the building. An integrated pressure control valve and
flow indicator monitors the gas flow through the entire tracking detector
station. At the station, a manifold breaks the single supply
into the eight separate
lines for the individual tracking detector modules. Each module can be isolated
from the manifold through PLC solenoids that must be energized
to stay open. Control logic from vacuum monitors and pressure transducers may
shut the solenoids to prevent unwanted contamination in the vacuum chamber
due to a burst straw or other tracking detector module failure. 

Manual flow metering valves are used to monitor and control the gas
flow to individual tracking detector modules. The gas flows
into the top module manifold, over the electronics, through the Mylar straws
and out past the electronics in the bottom manifold. Pressure transducers
positioned after the flow to the module further monitor significant pressure drops (e.g., a broken
straw causes vacuum to draw on the gas supply line) confirming the structural
integrity of the tracking detector straws. Visual flow indicators are also used
for observing gas flow rates at the tracking detector module exit. After passing
through the second isolation solenoid, the eight module lines are recombined
in a manifold to a sequence of exhaust bubblers. The first bubbler in the sequence is empty to
provide protection from drawing mineral oil into the vacuum chamber. In the
event of a catastrophic failure, the empty bubbler provides the PLC with
sufficient time to close solenoids before the mineral oil could be drawn into
the tracking detector and damage other components inside the vacuum chamber. The second bubbler is filled
with mineral oil at a sufficient height ($\sim10$ cm) to maintain a
positive pressure difference with atmosphere of $\sim14.4$~psia
at the ventilation exit. The used tracking detector gas is vented to atmosphere outside
of the experimental hall.

As part of both start-up and safety protocols, the multiple PLC valve connections permit tracking detector
module isolation in the event of a straw failure and gas selection of either dry nitrogen (from the $g-2$ main LN$_{2}$ supply) or the operational Ar:C$_2$H$_6$ 50:50 mixture to flow within the lines. Dry nitrogen is
used rather than the Ar:C$_2$H$_6$ 50:50 mixture during routine maintenance to systems,
or when the vacuum system is brought up to atmospheric pressure. The dry
nitrogen system permits purging the flammable ethane gas in
the event that a tracking detector module needs to be replaced, and is more economical if
there is a significant interruption to data collection. Importantly, 
the ability to transition rapidly from Ar:C$_2$H$_6$ 50:50 to nitrogen reduces the
potential for damage to the components inside of the vacuum chamber from the
flammable gas. A full gas volume change in all eight modules of a tracking detector station,
including transitioning from nitrogen to the operational Ar:C$_2$H$_6$ 50:50 mixture,
 takes approximately ten minutes.
The valve isolation of the gas occurs at multiple levels: (i)
the building supply from the storage shelter can be closed; (ii) each tracking detector
module can be filled with either nitrogen or Ar:C$_2$H$_6$, as needed for
repairs, or diagnostic tasks; and (iii) individual modules can be
isolated to prevent gas flowing into the vacuum chamber using the PLC
solenoids.

The overall leak rate of a tracking detector station, including permeation through the straws and from the gas distribution system, is $3 \times 10^{-5}$ Torr L/s, below the storage ring vacuum load requirement of $<5 \times 10^{-5}$ Torr L/s. Under normal operation of Ar:C$_2$H$_6$ 50:50, the contribution to the total leak rate from either permeation through the straws or from the gas distribution system is comparable.

\subsection{Installation}

The first straw tracking detector modules were installed into Muon $g-2$ Experiment in June 2016. Before installation, each module underwent a noise-threshold scan (described in Section~\ref{sec:performance}), ensuring that the connections to the ASDQs remained and no noise was observed above a $4.5$~fC threshold. Each module was then placed in the test stand (see Section~\ref{sec:testStand}), raised to the operating voltage and tested again with a $^{55}$Fe source to ensure the channels were operating as desired. 

\begin{figure}[!t]
\centering
\includegraphics[width=0.6\textwidth]{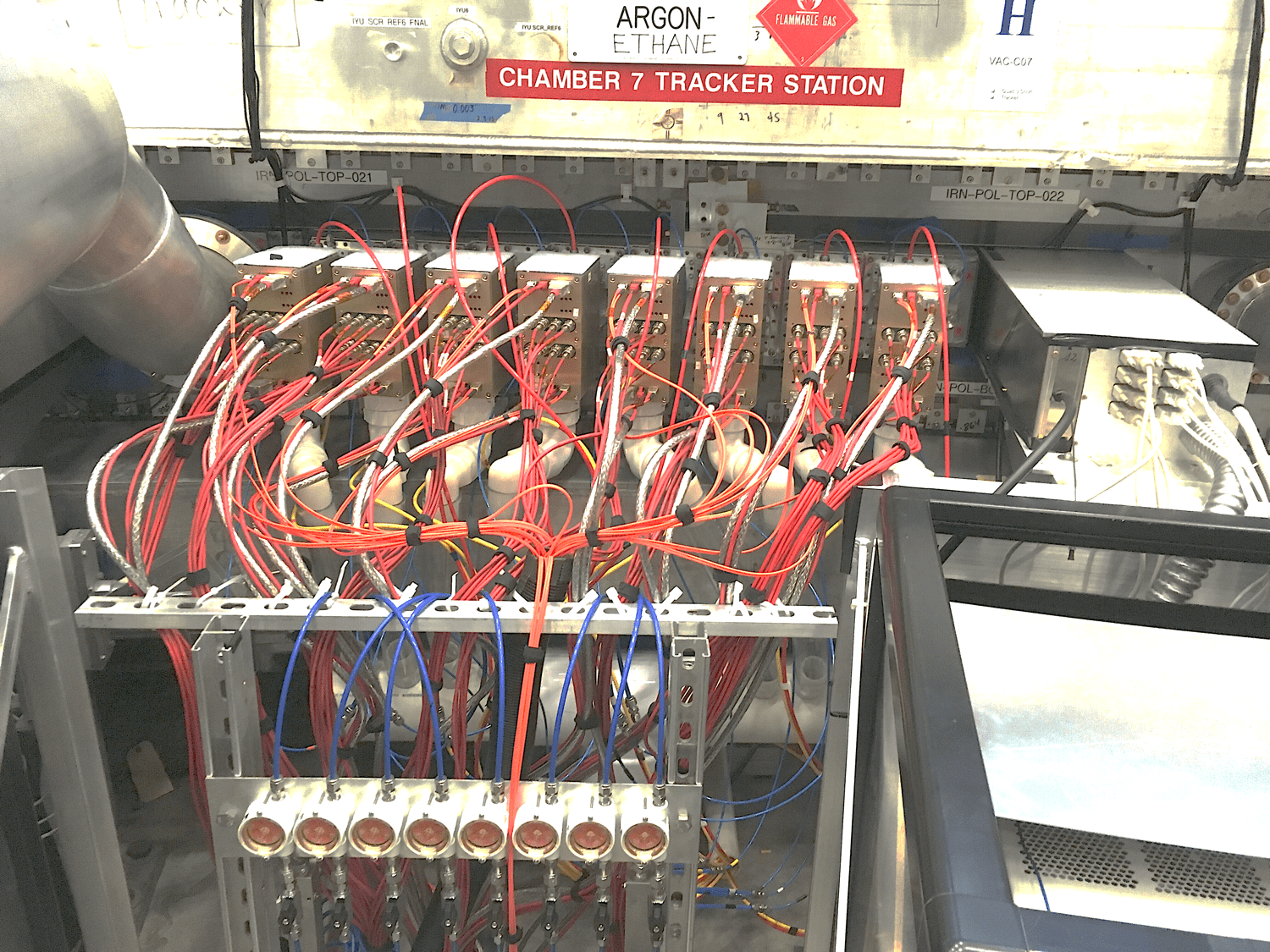}
\caption{One of the two tracking detector stations installed in the muon storage ring vacuum chamber. Eight modules can be seen, connected to red HV cables, orange optical fibers and silver power cables. Red gas tubes enter at the top of each module, air cooling is attached with white tube below. At the front of the photo, the flow meters for the water-cooling system are visible.}\label{Trackers-installed} 
\end{figure}

Additionally, modules were again vacuum tested to check for any leaks that could have occurred in transit to Fermilab or developed during storage---none were found. The surfaces of the manifolds were cleaned using ethyl alcohol to remove impurities that would jeopardize the quality of the vacuum. During this process, the straws were kept behind the custom Perspex shields to ensure they did not come into contact with the alcohol. After passing the vacuum test, each module to be installed was purged with nitrogen for a minimum of 4 hours and sealed during the short transport from the test area to the experimental hall. The modules were bolted into the modified vacuum chambers through matching holes in the vacuum flanges and tightened using a torque-wrench to ensure consistent seals. The vacuum seal is provided by a nitrile-rubber o-ring, which sits between the module flange and the vacuum chamber.
One of the installed tracking detector stations connected to all its services is shown in Figure~\ref{Trackers-installed}.

It is possible to swap the modules in and out of the different positions in the vacuum chamber. However, in general, swapping is kept to a minimum so as not to risk damaging the delicate straws. In order to install a module, the vacuum inside the storage ring must be at atmosphere. During any period when the storage ring is at atmosphere, the storage ring is purged with nitrogen for a minimum of 2 hours. This is done to minimize impurities on the walls of the vacuum chamber and to avoid water contamination on the chamber walls being attracted to the straws, which would otherwise affect the leak rate and quality of the vacuum. Deionized water is used to cool the electronics in the manifolds and is supplied to each station through plastic water inlets. The system is designed as a closed loop cooling system that keeps the supply flow to the tracking detector at an ambient temperature. The flow of this cooling water to each of the tracking detector stations is monitored with flowmeters to ensure that the flow is consistent, and the electronics do not overheat (the temperatures of the ASDQs are read out during operation).

The HV and slow control boards housed on the inside of the storage ring are air-cooled, with one pump per station that sucks air through all the modules, splitting to each individual module under the storage ring (the filters are replaced every three months). Initially, the modules were installed into the T2 position (as shown in Figure~\ref{fig:ring_layout}) and filled from the position closest to the calorimeter first, with the later modules installed into the T1 position. All 16 modules, comprising 2 tracking detector stations, were in place by November 2017 which was before the experiment's first data-taking run.

The HV is provided by two CAEN Mod. SY 127 power supply systems per station, filled with CAEN A333 HV boards. The two systems together can supply 80 channels, which allows sufficient spares for the 64 channels required per station. The HV is monitored throughout data-taking and recorded every 5 minutes. The trip currents are set to $2~\mu$A. In general, the ability of the tracking detector to hold HV has been exceptional---only two HV trips were recorded during the first year of data taking.
%A single trip removes 16 straws until the HV is recovered.
The operating voltages are set to $1650~$V as described in Section~\ref{sec:testStand}. The supplied voltage typically varies by $\pm2~$V and each channel is calibrated separately.

After installation and periodically thereafter, the global alignment of the tracking detector is measured with dedicated laser surveys.
A high-precision inter-module alignment is extracted using fitted tracks, which establishes 10-$\mu$m-level
corrections to the assumed detector position. To this end, the \texttt{Millipede-II} alignment framework~\cite{millipede2} is employed for each 
data-taking run of the experiment and whenever a significant positional change 
to a component of the tracking detector occurs (\textit{e.g.} when replacing a tracking detector module).
The use of this software for the $g-2$ tracking detector has been fully validated by simulation. 
Performing both radial and vertical alignment of the tracking detector modules results in 
an increased number of reconstructed tracks, as well as improvements in the track fits. 
An in-depth description of the internal alignment of the $g-2$ trackers is given in~\cite{GlebThesis}.

\section{Commissioning and Performance}
\label{sec:performance}
\subsection{Straw Performance}

\begin{figure}[!t]
\centering
\begin{subfigure}{0.49\textwidth}
\includegraphics[width=\textwidth]{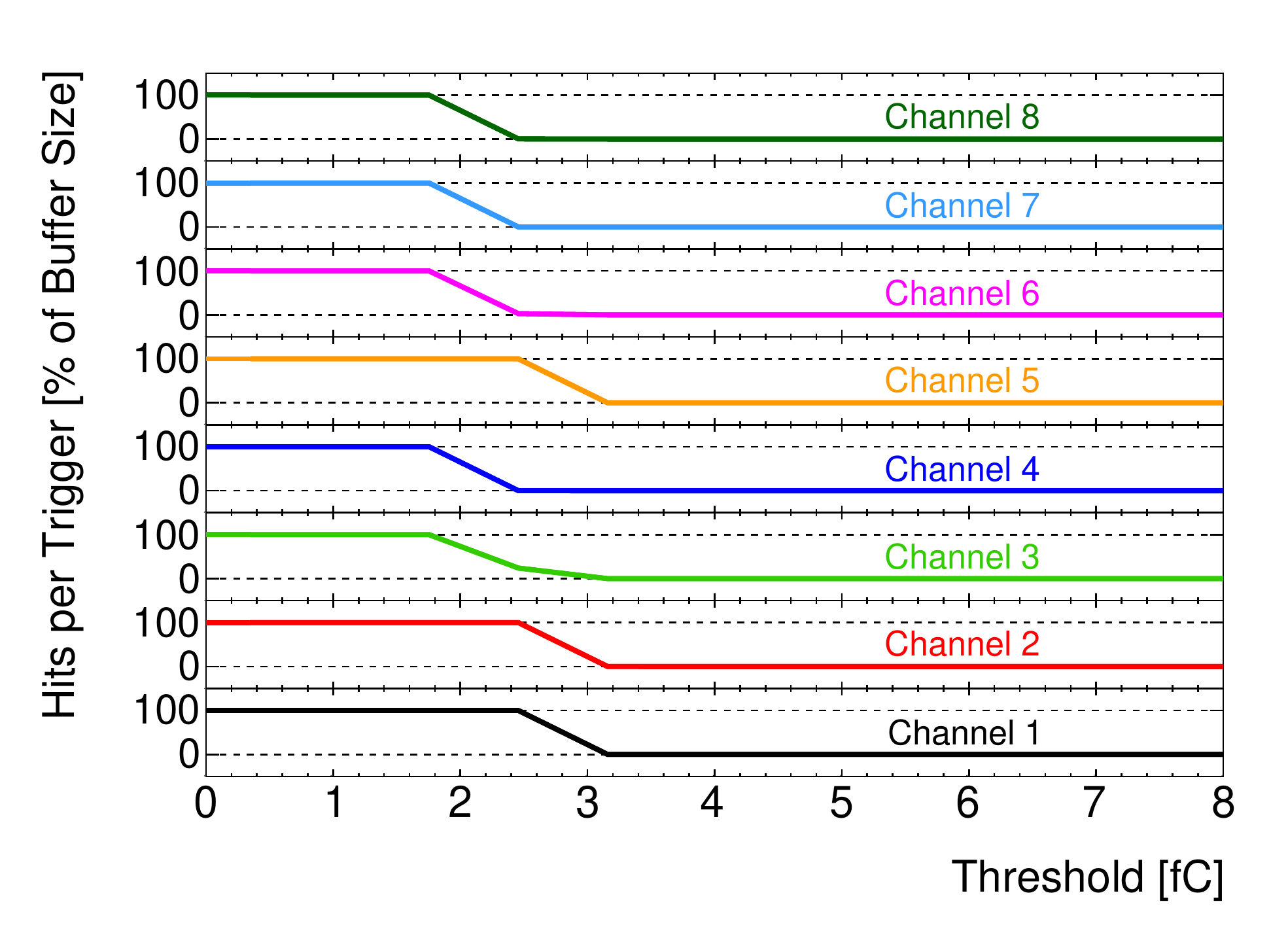}
\caption{}
\label{fig:Noise_HitChannels_a}
\end{subfigure}
\begin{subfigure}{0.49\textwidth}
\includegraphics[width=\textwidth]{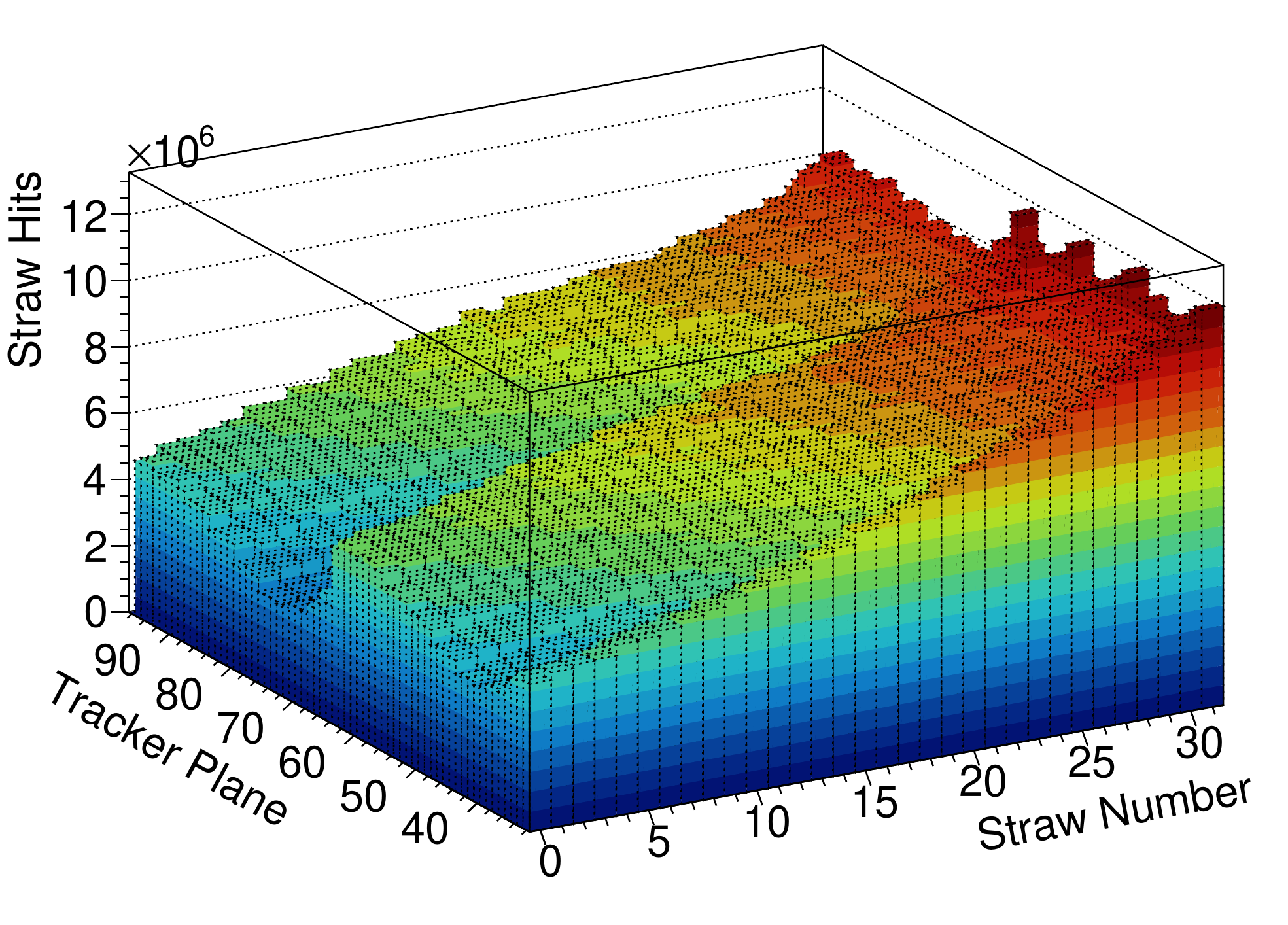}
\caption{}
\label{fig:Noise_HitChannels_b}
\end{subfigure}
\caption{ (a) The number of hits per trigger as a function of the ASDQ discriminator threshold.
At low thresholds, where noise dominates, the hit rate is large enough to fill the storage buffers.
Above 3.5~fC the noise rate is negligible.
(b) The number of hits per channel for all straws over a three-day data-taking period.
No channels have significantly more than any other, demonstrating sufficiently low noise levels.}
\label{fig:Noise_HitChannels} 
\end{figure}

With a straw tracking detector, the lowest possible detection threshold is desirable
as this increases the likelihood of detecting the first ionization cluster arrival which in turn improves the
individual hit resolution. The limitation for the detection threshold is the
electronics noise floor. In order to assess the noise level in each channel,
the hit rate was measured as a function of discriminator threshold. A
typical result is shown in Figure~\ref{fig:Noise_HitChannels_a}.
Setting the threshold at a level of 3.5~fC (21,850 $e^{-}$) results in a noise level of $<< 1$ 
hit per straw per muon fill.

As previously mentioned, six straws contained no active gas due to high leak rates.
The other 2042 out of 2048 channels of the detector had uniform performance
in terms of efficiency and noise. The number of hits in each channel over a
three-day data-taking period is shown in Figure~\ref{fig:Noise_HitChannels_b}.
The differences in the channels are due to the geometry of the detector and
proximity of each straw to the stored muon beam. No channels have many more
or many fewer hits than their neighbors, indicating that the noise level
achieved is satisfactory.

Crosstalk occurs when unwanted electrical couplings in the system generate
fake hits in the straws. This can cause issues in track reconstruction and
should therefore be minimized. Crosstalk occurs dominantly in the straw 
components (e.g., straw-to-straw coupling), but also in the electronics (e.g. 
ASDQ boards). Crosstalk is identified in $\sim6$\% of all hits ($\sim4$\% 
from the straws, $\sim2$\% from the electronics). These are removed in analysis
by identifying tracks that have neighboring hits and removing those hits. Any 
remaining instances of crosstalk following this are removed by a $\chi^2$ cut. 

\begin{figure}[!t]
\centering
\includegraphics[width=0.6\textwidth]{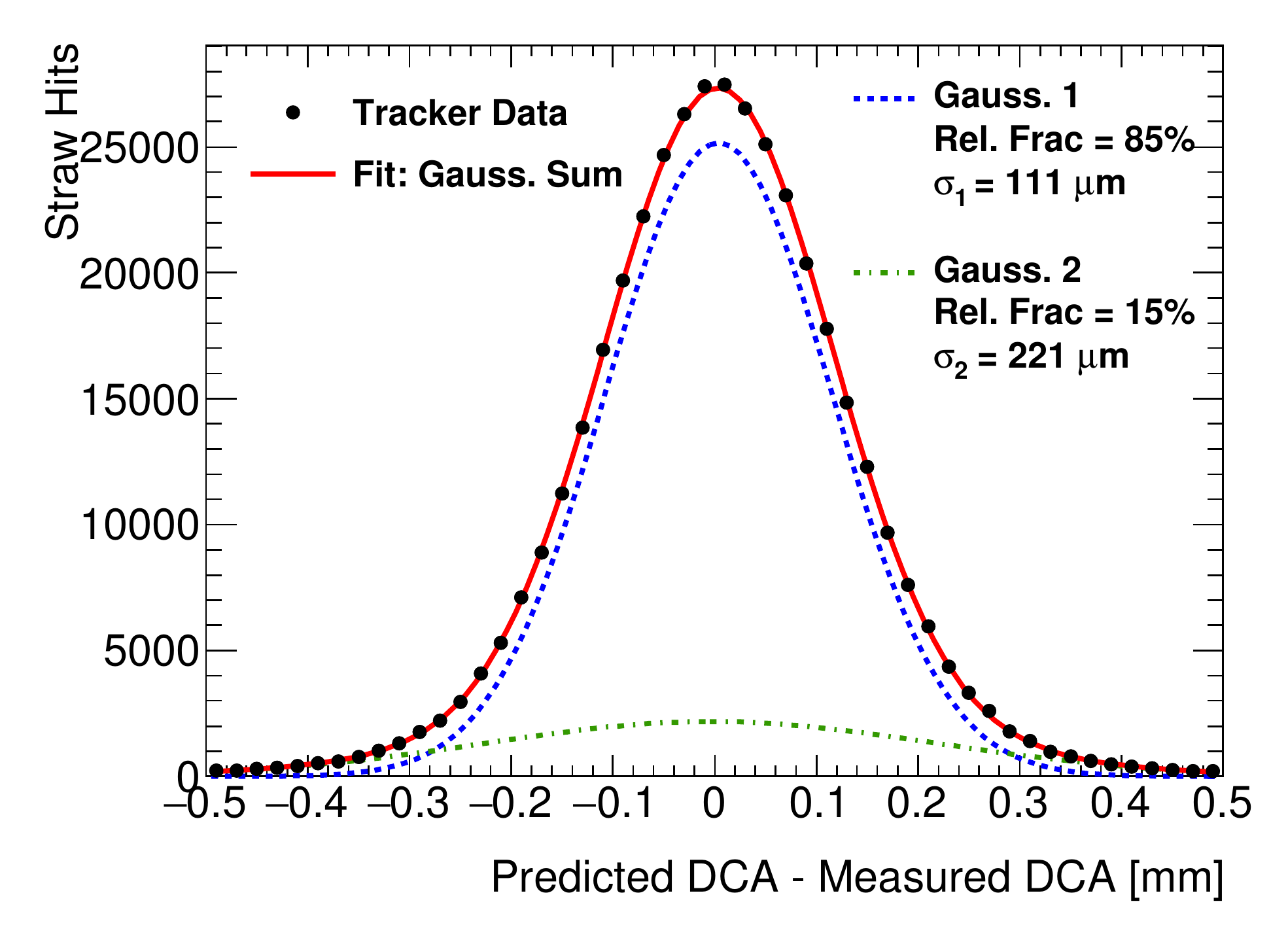}
\caption{ Residuals between the straw measurement position and
interpolated track position for straws that are removed from the fitted track.
For inclusion in the distribution, the predicted track position must be larger than 0.5\,mm
to avoid a left-right ambiguity near the wire. The data is fit with the sum of two Gaussians (red solid line).
The individual Gaussians are also plotted showing that the sum is dominated by a core with smaller rms (dotted blue line)
and a smaller distribution with longer tails (dashed green line). Removal of the track prediction resolution leads to
an estimated straw hit resolution of $110\,\mu$\si{\meter}.
\label{fig:Resolution}
}
\end{figure}

The best estimate of hit resolution of each straw comes from a comparison of
the measurement position with the prediction from interpolated tracks
reconstructed using the algorithm described in Section
\ref{sec:TrackRecon}. The distribution of these residuals for straws that are
excluded from the fit is shown in Figure~\ref{fig:Resolution}.
The data are consistent with a mixture of two Gaussian distributions: a main core
with $\sigma = 110\,\mu$\si{\meter} and a smaller contribution from tails with $\sigma = 220\,\mu$\si{\meter}.
Including both contributions, the rms of the distribution is $133\,\mu$\si{\meter}.
This measured value is broadened by the resolution of the fitted track prediction,
which is estimated using Monte Carlo simulations to be approximately $75\,\mu$\si{\meter}.
Removing this extra resolution leads to an estimated average straw hit resolution of $110\,\mu$\si{\meter}.
This value can vary by up to $\pm20$~$\mu$m depending on the straw layer and analysis choices.

\begin{figure}[!t]
\centering
\includegraphics[width=0.6\textwidth]{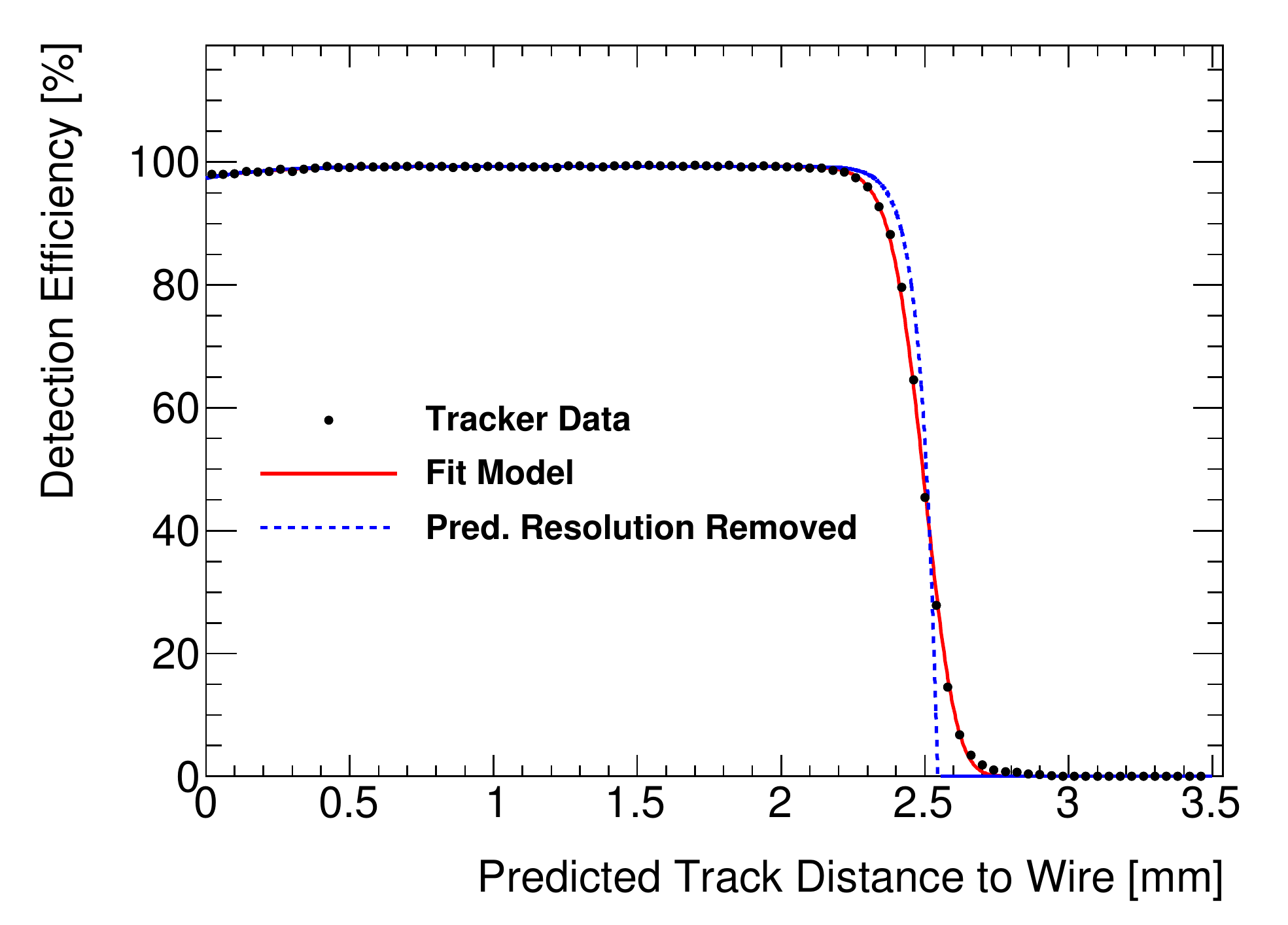}
\caption{ Straw detection efficiency measured as a function of distance of
the interpolated track prediction from the wire. The red solid line is a fit
model of exponential decreases from a constant efficiency at the wire and straw edges,
and includes the intrinsic resolution from the track position prediction.
The blue dotted line is the same function after deconvolution of this resolution.}
\label{fig:StrawEfficiency}
\end{figure}

Similarly to hit resolution, the position of the interpolated tracks can be
used to estimate the straw detection efficiency as a function of track distance
from the wire. The measured straw efficiency for one layer of straws is shown in
Figure~\ref{fig:StrawEfficiency}. The efficiency decreases at the edge of the
straw, where a sharp edge is degraded by the track prediction resolution. If a track
passes within $2580\,\mu$\si{\meter} of the wire, then the detection
efficiency is greater than 97\%. This drop is dominated by missed tracks at
the edge of the straw, where the path length through the gas is too short for
ionization. The efficiency in the main body of the straw is 99.3\%,
with a decrease of less than 2\% at the wire. This is attributed to
multiple ionizations that do not arrive close enough in time to jointly pass
the discrimination threshold. 

The tracker operates in a high-rate environment. No performance degradation is observed up
to the hit rates of standard operation. No signs of aging have been observed to date, and the efficiency and resolution are the same as during
commissioning.

\subsection{Track Reconstruction and Extrapolation}
\label{sec:TrackRecon}

The track reconstruction flow undergoes the following stages:
(i) track finding where straw hits are grouped into track candidates,
(ii) estimating the distance of closest approach for each hit based on the measured time,
(iii) track fitting of candidates,
and (iv) extrapolating the optimal track back to the muon decay point.

\begin{figure}[!t]
\centering
\includegraphics[width=0.8\textwidth]{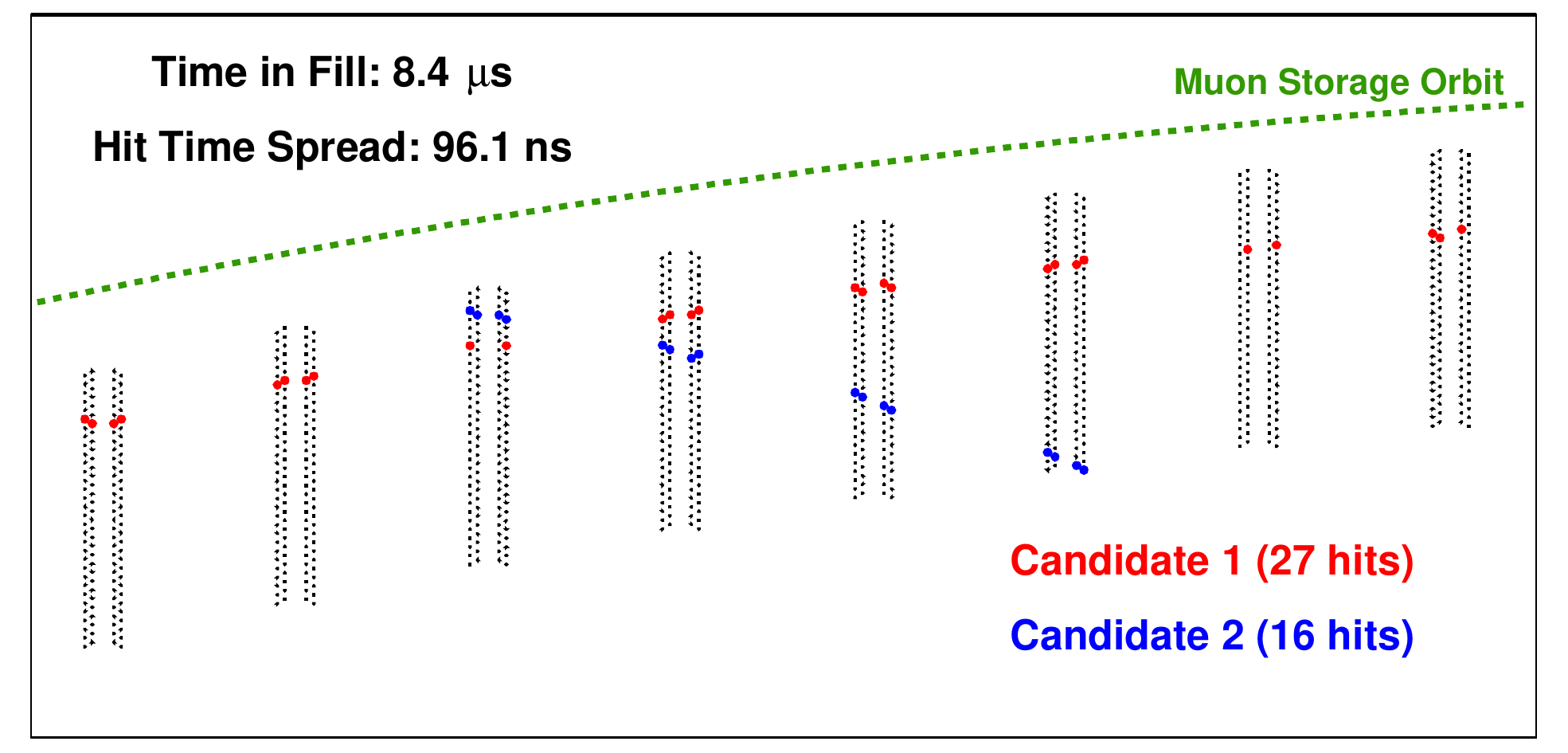}
\caption{ An event display of a time island with multiple track candidates in the tracking detector.
Each wire is marked with a black dot and the hits are marked with larger circles.
The color of each hit designates the candidate to which it has been assigned. The longer,
straighter track (red circles) is likely a muon lost from the storage region, and the shorter, more curved
track (blue circles) is a decay positron.}
\label{fig:TimeIsland}
\end{figure}

The track finding stage uses pattern recognition routines
in order to group individual hits into separate sets corresponding to individual
tracks. It starts by first grouping hits into \emph{time islands} where the difference
between hits must be less than 80\,ns, which is more than the maximum drift time for a straw hit.
Within the time islands, spatial and temporal information are used to group hits into
track candidates. After a first pass, a second iteration further splits the candidates
if there are large gaps in time or space between the hits.
A typical example of multiple track candidates from the same time island is shown in Figure~\ref{fig:TimeIsland}.
If hits are shared among multiple track candidates, then these hits are dropped from all candidates.
All the identified track candidates are passed to the next stage of reconstruction where the drift distance from each hit is estimated.

\begin{figure}[!t]
\begin{center}
\begin{subfigure}[b]{0.7\textwidth}
\includegraphics[width=\textwidth]{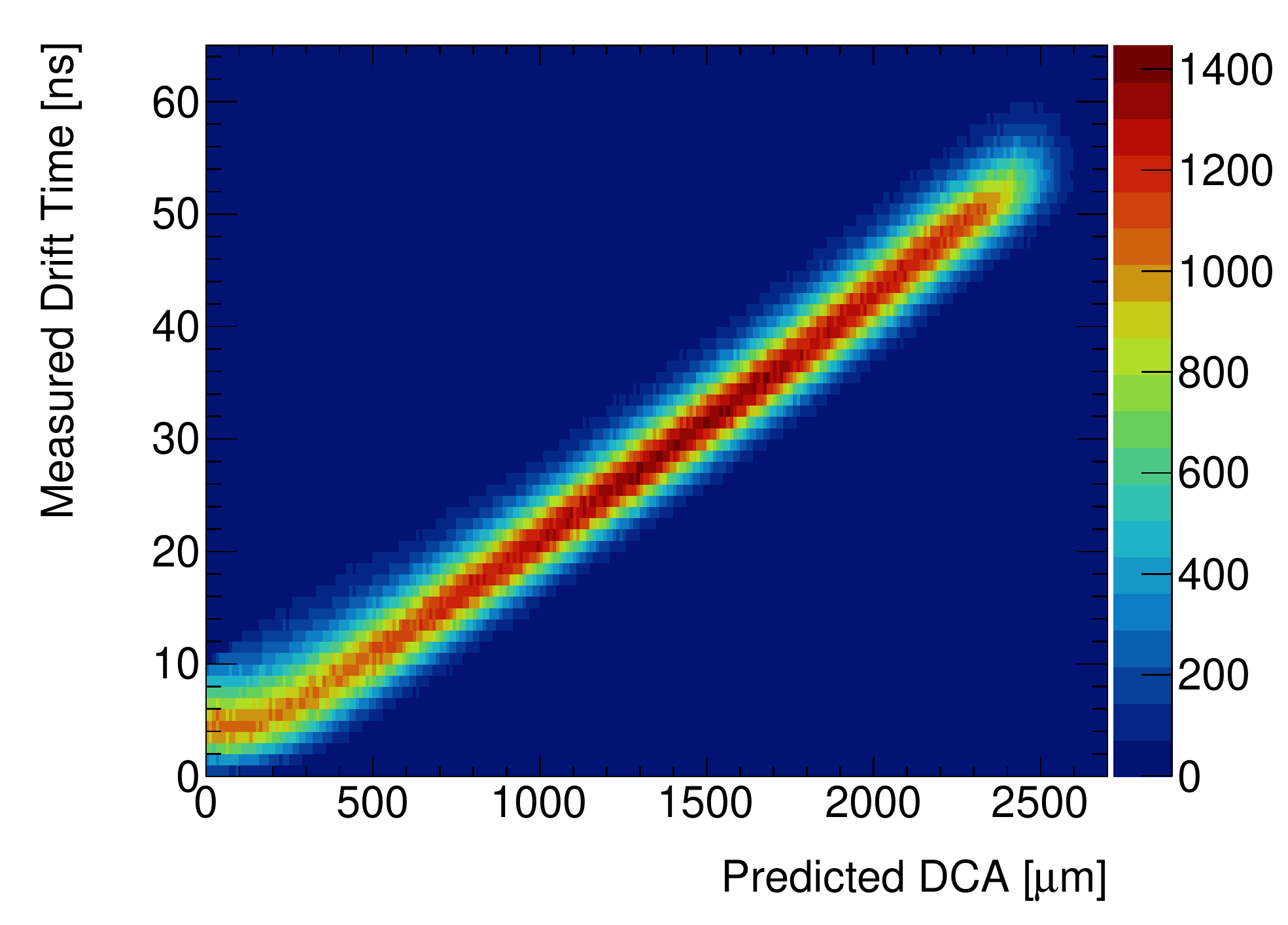}
\end{subfigure}
\end{center}
\caption{ A distribution of the measured drift time for different distances of the track from the wire.
This distance is estimated using interpolated tracks that do not use the straw hit.
An entire layer of 32 straws is shown in this plot, whereas the time-to-distance parameterization used
in the reconstruction is performed on a straw-by-straw basis.}
\label{fig:DriftTimeVsDCA}
\end{figure}

The track arrival time is measured for each candidate by exploiting the correlation between drift time
values in adjacent tracking detector layers. An approximation to the angle of the track is used along with
the sum of hit times in adjacent layers to estimate an entry time for each pair of hits. The mean of these
gives the arrival time estimate, which has a typical resolution of $\sim 1$~ns.
 Using this time, the drift time for each hit can be calculated to give an
estimate of the distance of closest approach of the track to the wire.\footnote{As the straws are short (< 10~cm), propagation of the signal along the length of the wire is not a concern. This is because the signal propagation time is small compared to the hit time resolution, and calibration is performed using tracks from data which, to first order, averages out any bias between the top and bottom of the straws.}
The drift-time-to-distance calibration is performed once per experimental data run using interpolated tracks where the hit in question has been
removed from the fit. Each straw is parameterized independently using data similar to that in Figure~\ref{fig:DriftTimeVsDCA}.


The track fitting stage~\cite{NickThesis} is based off the \texttt{GEANE}
method~\cite{Innocente:1991md} (Geometry and Error
Propagation) that was used successfully in E821.
A set of
track hits defining objects such as transport matrices, error
matrices, and predicted parameter vectors are generated within the full $g-2$
\texttt{Geant4}~\cite{Allison:2006ve,GEANT4:2002zbu,Allison:2016lfl} simulation from initial guesses. This has the advantage of direct
access to both the geometry and material of the tracking detector as well as the
magnetic field
that the decay positrons experience as they curl inward. This is important as
the tracking detector is located in a region of high field non-uniformity as shown in
Figure~\ref{fig:fringeField}.
These track objects are then passed to a global
chi-squared minimization algorithm that produces an optimal state vector
at the track entry point. This minimization is performed in the same
reference frame as the straw stereo angle, which reduces parameter
correlations and improves the fitting. The final optimized state vector is then
passed to the extrapolation stage.

The track extrapolation stage~\cite{SaskiaThesis} utilizes a fourth-order Runge-Kutta Nystr\"{o}m
algorithm~\cite{extrap1,extrap2} to extrapolate tracks either back through the varying
magnetic field into the storage region until the approximate position of the
muon decay point is reached, or forwards to the calorimeter face. This is
done within the full $g-2$ \texttt{Geant4}~\cite{Allison:2006ve,GEANT4:2002zbu,Allison:2016lfl} simulation utilizing the same geometry
and field information as the track fitting. Track
parameters from the fitting and the corresponding covariance matrix are propagated
in discrete steps. At each step, the position of the extrapolated track is
compared to the geometry in order to ensure that tracks that pass through material are reconstructed taking into consideration the degradation in resolution at the storage region. Because there is no fixed interaction point in the storage region, tracks are extrapolated backwards to the point
of tangency where the radial momentum is equal to zero, which is a reliable proxy for the decay point.
Extensive simulation studies were done to verify that this approximation for the muon decay point was sufficient. By applying a momentum-independent $1.1$~mm radial correction to the point of
tangency, the muon decay point can be determined~\cite{SaskiaThesis}.

\subsection{Measurement of the Stored Muon Beam}
\begin{figure}[!t]
\centering
\begin{subfigure}{0.49\textwidth}
\includegraphics[width=\textwidth]{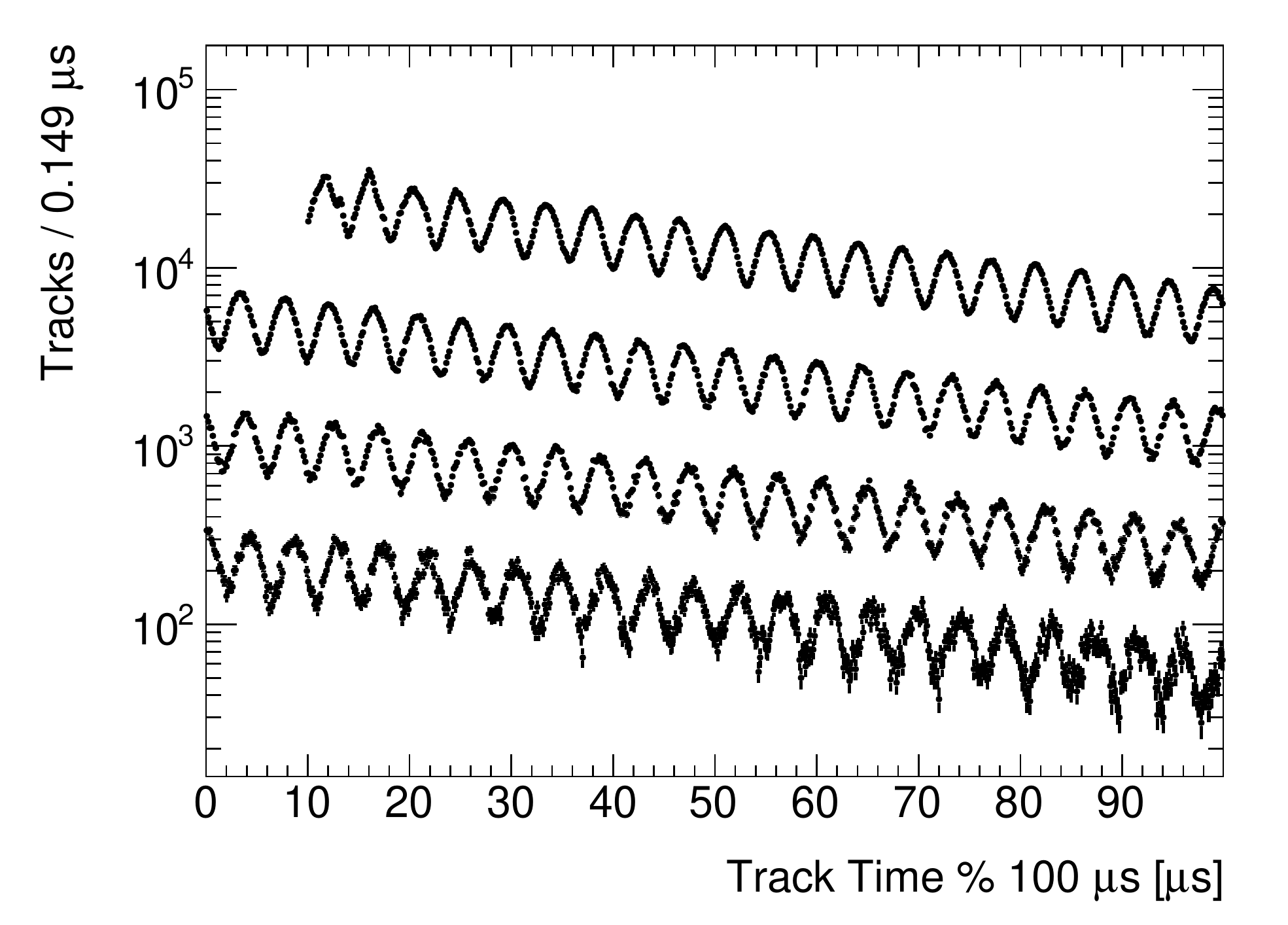}
\caption{}
\label{fig:BeamMeasurements_a}
\end{subfigure}
\begin{subfigure}{0.49\textwidth}
\includegraphics[width=\textwidth]{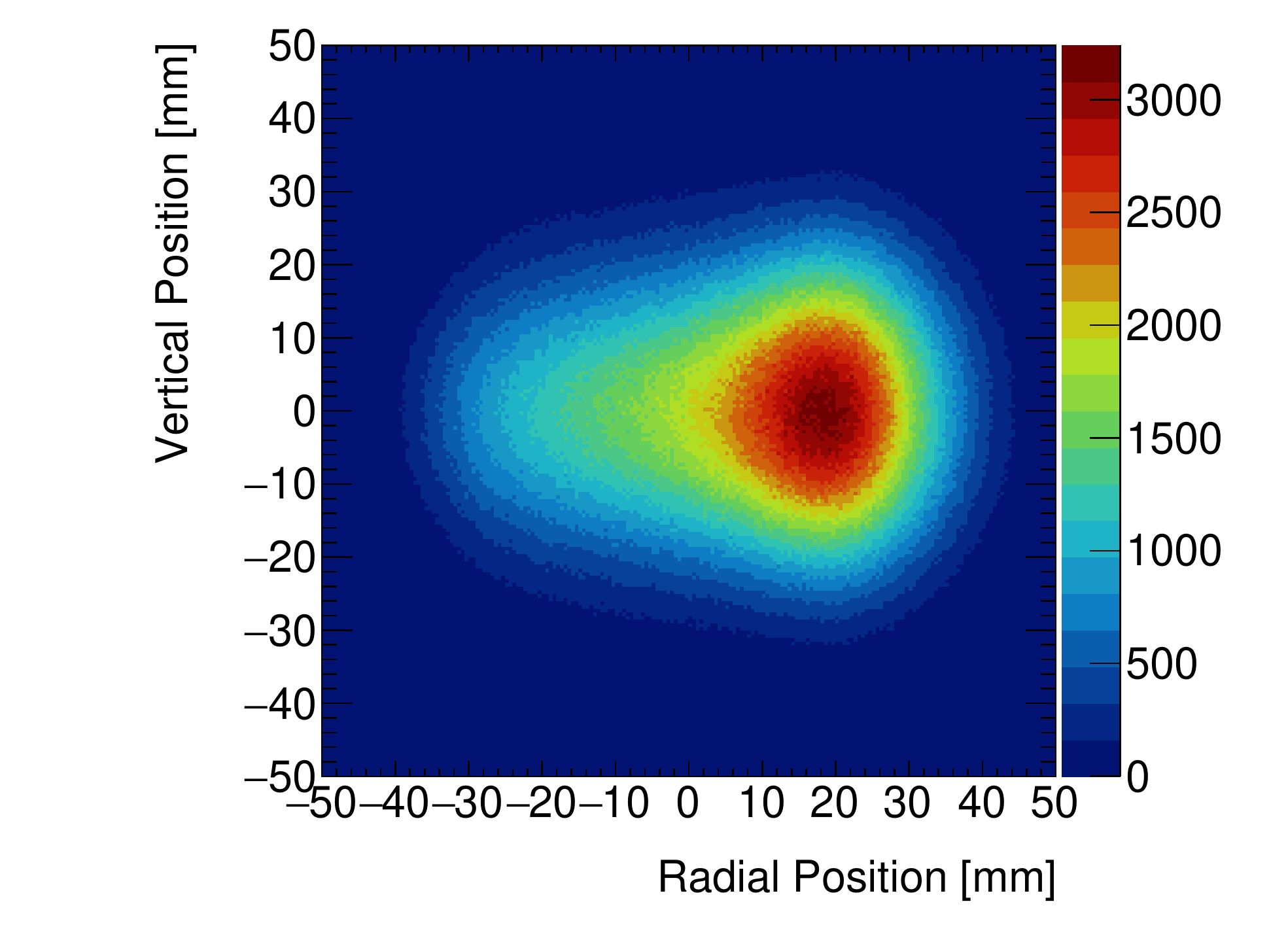}
\caption{}
\label{fig:BeamMeasurements_b}
\end{subfigure}
\begin{subfigure}{0.49\textwidth}
\includegraphics[width=\textwidth]{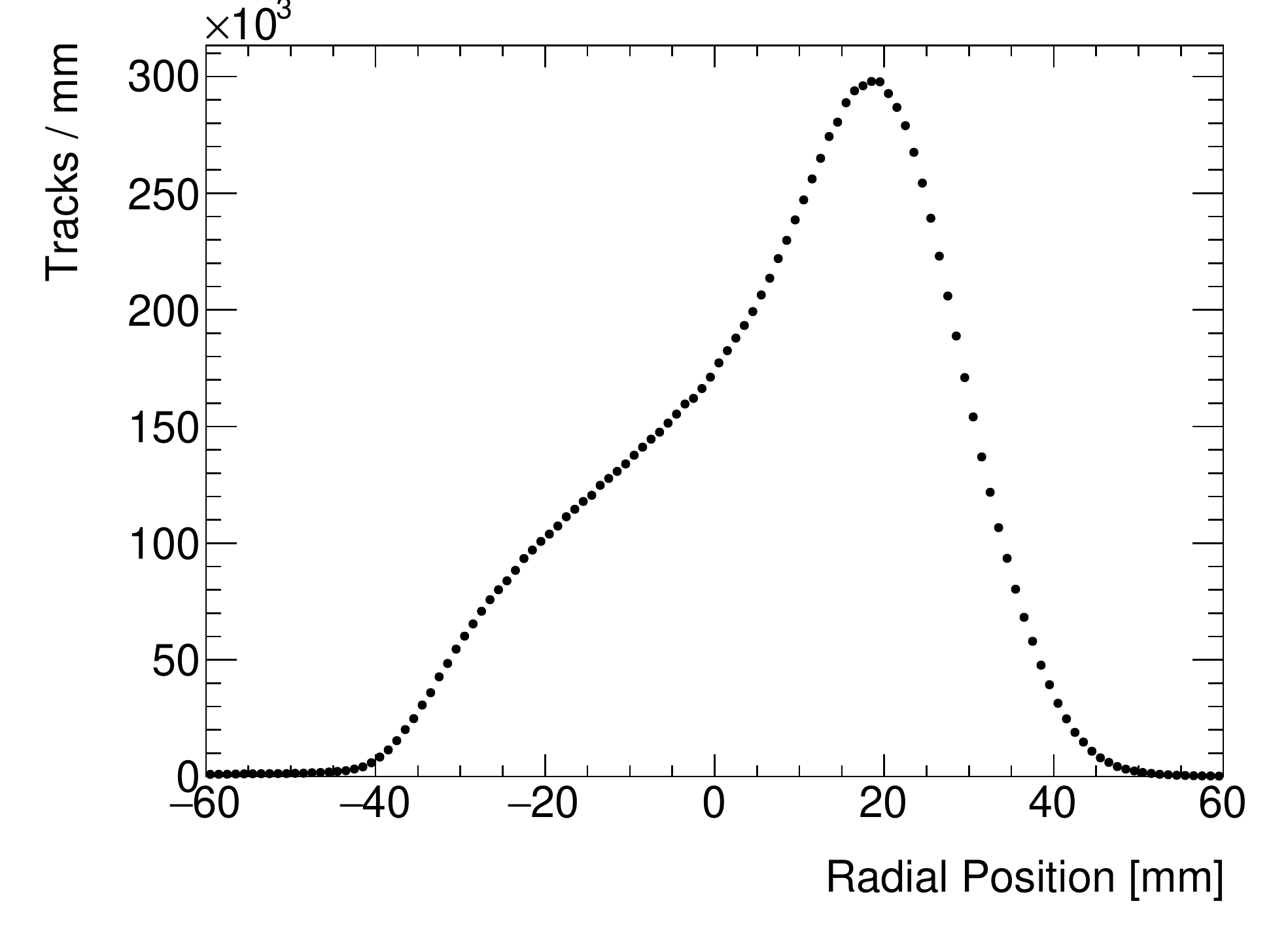}
\caption{}
\label{fig:BeamMeasurements_c}
\end{subfigure}
\begin{subfigure}{0.49\textwidth}
\includegraphics[width=\textwidth]{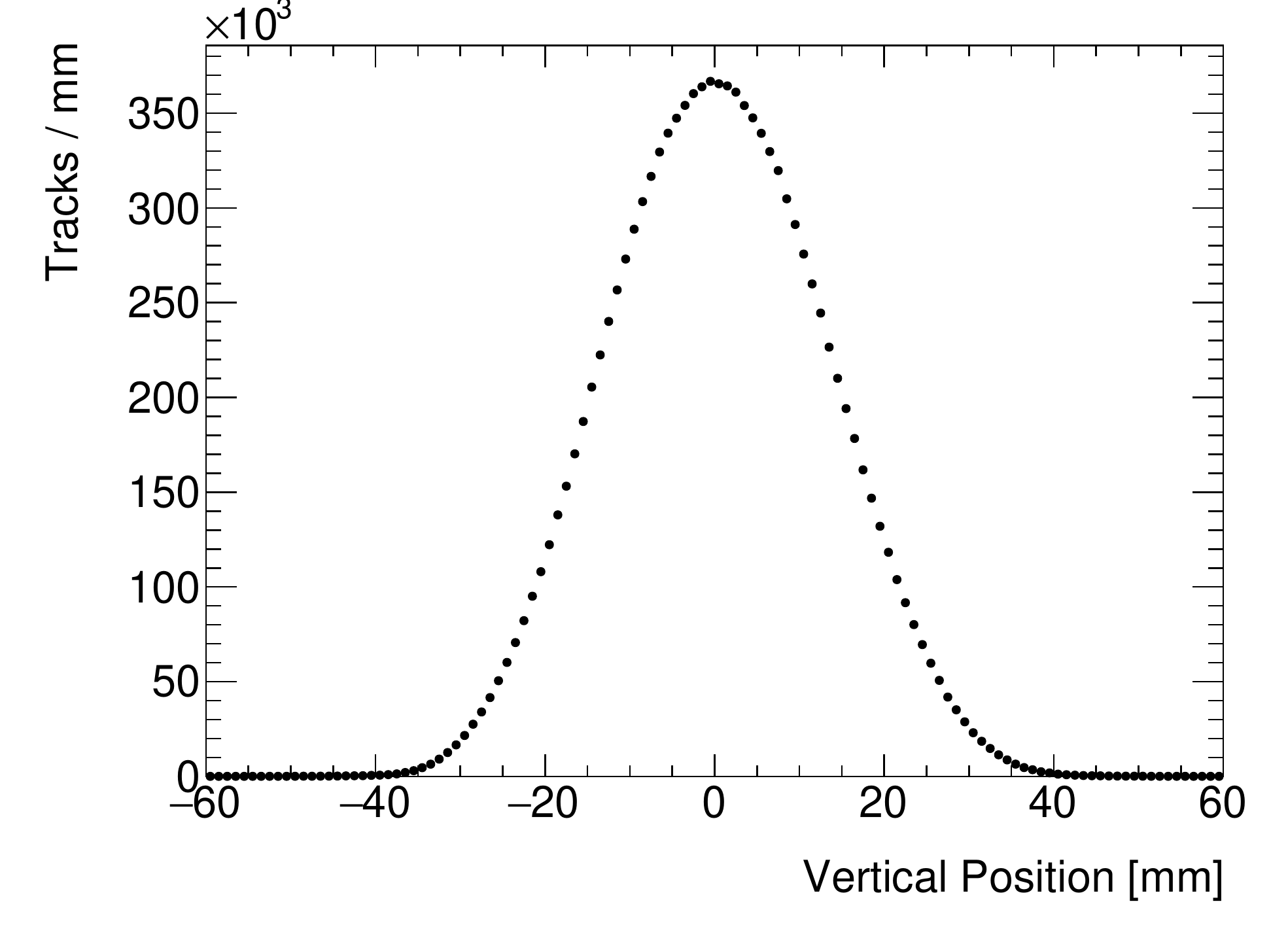}
\caption{}
\label{fig:BeamMeasurements_d}
\end{subfigure}
\caption{Tracking detector data from a three-day period during Run-1.
(a) The number of tracks (in 0.149 $\mu$s bins) as a function of the track arrival time (modulo 100~$\mu s$) with respect to the muon beam injection, for tracks with momentum greater than 1800~MeV. The oscillation in number of tracks is due to the precession of the muon spin.
(b) A cross-section of the stored muon beam from extrapolated tracks after the start of the measurement time at 30~$\mu$\si{\second}. The radial position is measured relative to the design radius of 7112~mm.
(c) and (d) The radial and vertical projections, respectively, of the beam cross-section.}
\label{fig:BeamMeasurements}
\end{figure}

The data from the tracking detector are passed through the reconstruction stages described above to measure distributions of the stored muon beam.
Examples of the most important distributions are shown in Figure~\ref{fig:BeamMeasurements}.
The arrival time of tracks with momentum above 1.8~GeV (Figure~\ref{fig:BeamMeasurements_a}) shows exponential decay with a boosted muon lifetime of $\sim64$~$\mu$s and the characteristic $g-2$ oscillation due to precession of the muon spin~\cite{Run1wa}.
The main role of the tracking detector is to measure the muon beam's spatial profile, which is shown in Figure~\ref{fig:BeamMeasurements_b} along with projections in the radial and vertical directions (Figures~\ref{fig:BeamMeasurements_c} and \ref{fig:BeamMeasurements_d} respectively). These distributions are used to extract the correctly weighted magnetic field experienced by the muons~\cite{Run1field}.
The vertical distribution is also used to extract the \textit{Pitch} correction to the measured value of $\omega_a$~\cite{Run1BD}.
The stored beam has a larger average radius than the design goal, which would be at 0~mm in Figure~\ref{fig:BeamMeasurements_c}, due to a lower strength magnetic field provided by the injection kicker. This leads to an average stored muon momentum fractionally higher than intended~\cite{Run1BD}.

\begin{figure}[!t]
\centering
\begin{subfigure}{0.49\textwidth}
\includegraphics[width=\textwidth]{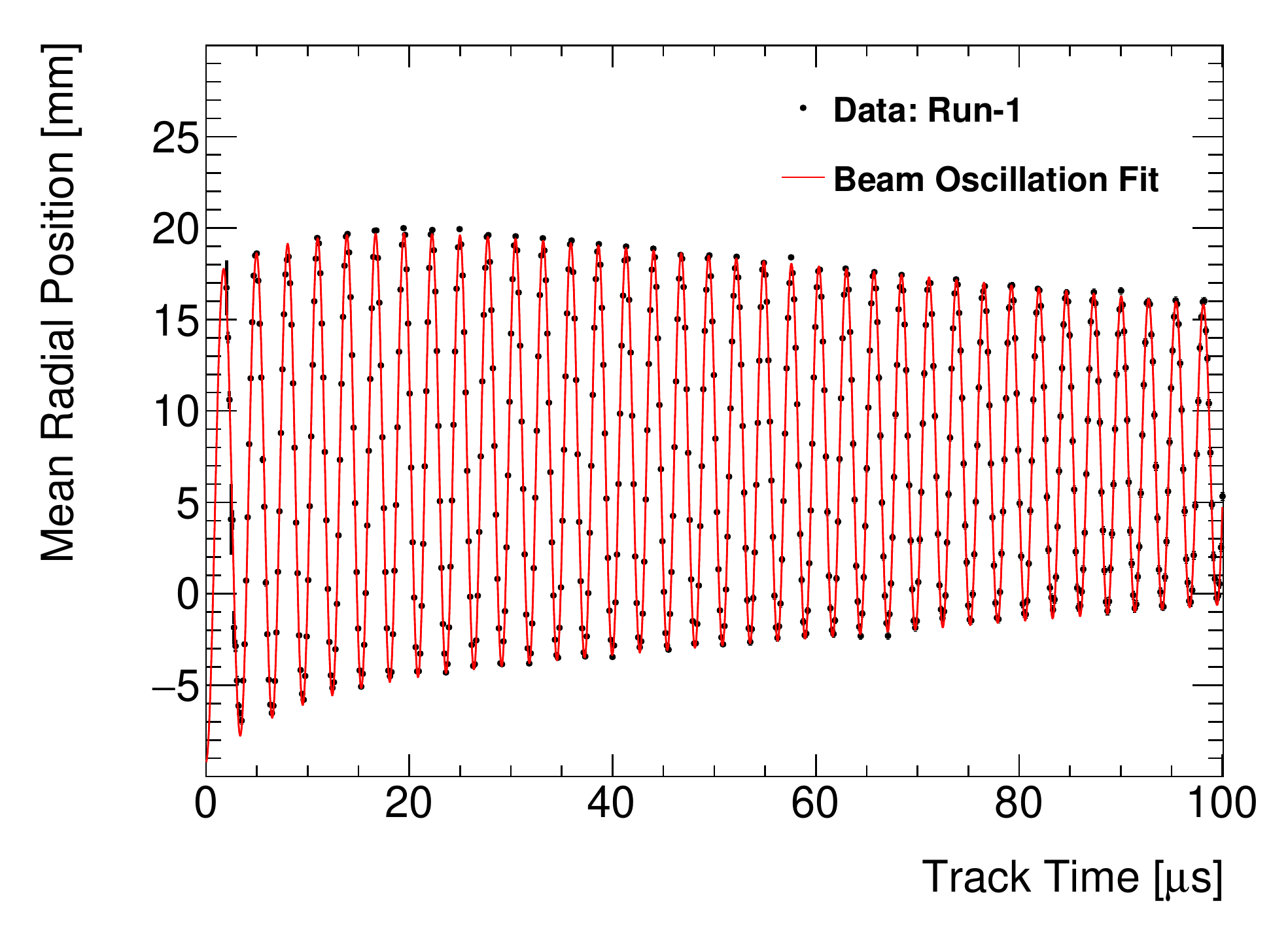}
\caption{}
\label{fig:CBO_a}
\end{subfigure}
\begin{subfigure}{0.49\textwidth}
\includegraphics[width=\textwidth]{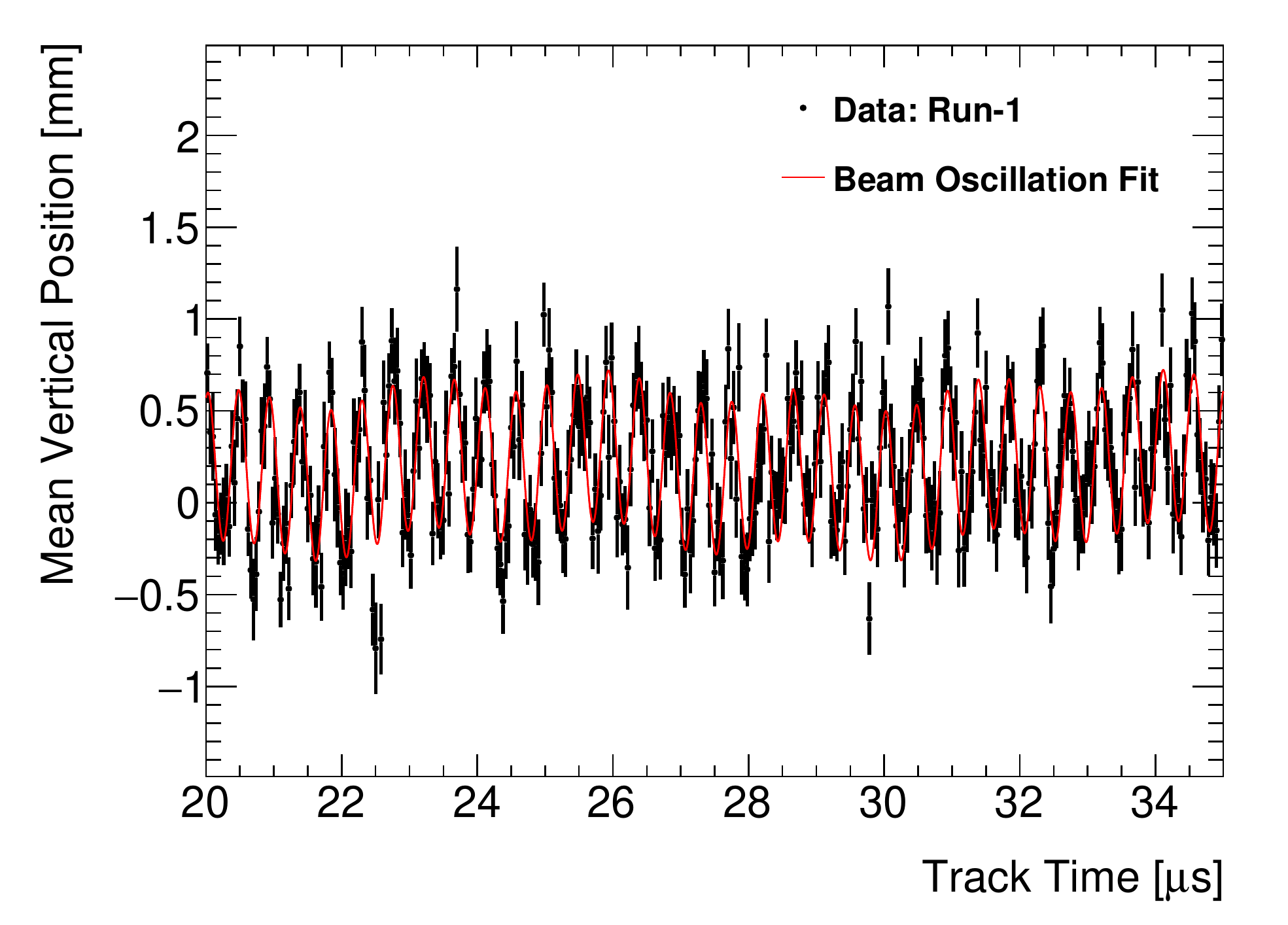}
\caption{}
\label{fig:CBO_b}
\end{subfigure}
\caption{Tracking detector data from a three-day period during Run-1.
(a) The average radial position as a function of time for tracks arriving in the first 100~$\mu s$ of muon beam storage.
The beam oscillations are due to coherent betatron oscillations, and the amplitude decreases over time due to decoherence of the slightly different betatron frequencies for different muons.
(b) The average vertical position as a function of time for tracks arriving between 25-30~$\mu s$ of muon beam storage, showing faster and smaller amplitude oscillations.}
\label{fig:CBO}
\end{figure}

The beam position is not constant in time, but rather oscillates at the betatron oscillation frequency~\cite{Run1BD}. The tracking detector measures these oscillations in both the radial and vertical directions as shown in Figure~\ref{fig:CBO}.
During the measurement period, it is intended that this oscillation frequency is constant. However, during Run-1 of the Muon $g-2$ Experiment, analysis of the reconstructed beam data from the tracking detector revealed that this
frequency was changing over the course of each muon fill. Subsequent investigation revealed the cause to be
faulty resistors in the electrostatic quadrupole system. The same issue also caused the beam to move slowly downwards
over the course of each fill. This couples to detector acceptance and causes a systematic shift in the extraction of the anomalous precession frequency~\cite{Run1BD}.
Tracking detector measurements of the changing oscillation frequency and the motion of the beam,
in addition to the average spatial profile, were therefore essential inputs to analysis of Run-1 data and
will provide vital cross-checks of the data quality in the other runs.

\section{Conclusions}
\label{sec:conclusions}
The Muon $g-2$ Experiment’s straw tracking detector is essential to the experiment achieving its high precision uncertainty goals. The experiment employs two tracking detector stations which are located inside of the muon beam vacuum chamber upstream of a calorimeter and inside the magnetic field region of the muon storage ring. Each station consists of eight modules with four layers of 32 straws at a stereo angle of $\pm7.5^\circ$. 

The tracking detector reconstructs positron trajectories from muon decays and extrapolates those trajectories to estimate the position of the parent muon at the time
of decay. In doing so, the detector provides non-destructive, time-dependent measurements of the muon beam-profile. The tracking detector detects as many decay positrons as possible and reconstructs their tracks with enough fidelity to determine the muon beam profile at the millimeter level after extrapolation. It functions in a vacuum of $<10^{-6}$ Torr and has a vacuum load on the system below $5 \times 10^{-5}$ Torr L/s. The detector performs in a non-uniform storage ring magnetic field without affecting the reconstruction algorithm and without perturbing magnetic field itself. It has multiple detection planes to ensure an adequate acceptance (covering the beam aperture of $\pm 45$ mm and sensitive to 100~mm in height at the storage region) with material below $0.5\%$ radiation length so as not to disrupt the muon beam or degrade the performance of the calorimeters. 

All design choices for the tracking detector (e.g., straw material, straw gas, operating voltage, electronics) were assessed prior to and after construction. All stages of tracking detector module construction were performed in an ISO 5 (Class 100) clean room, with rigorous testing and quality assurance (e.g., gas leak rate) procedures at every stage. Transport to Fermilab was achieved without damage to any components. On arrival at Fermilab, all modules were subject to a dedicated test stand and radioactive source tests for data quality assurance. For the straw wires, these tests resulted in an operating voltage of 1650~V for the Ar:C$_2$H$_6$ 50:50 gas mixture occupying the straws.

The gas delivery system was designed and tested to mitigate and reduce flammable gas and oxygen deficiency hazard (ODH) risks inside the Muon $g-2$ experimental hall. Both the ethane and argon supply lines penetrate the experimental hall for mixing, routing, and delivery. The system, controlled by programmable control logic, is designed to provide a continuous flow to the tracking detector during gas canister replacement without interruption to experimental operation, and to isolate single points of failure such as a burst straw. In case of significant interruption to data collection, the need to purge the flammable ethane gas, or when the vacuum system is brought up to atmospheric pressure, the system is designed to flow dry nitrogen instead of Ar:C$_2$H$_6$ 50:50. This, in particular, is done to minimize impurities on the walls of the vacuum chamber and to avoid water contamination which would otherwise affect the quality of the vacuum. On installation, all modules again undergo significant testing and quality assurance procedures and are globally and internally aligned. 

The Muon $g-2$ straw tracking detector has operated successfully since installation in 2017 and performed excellently. Electronic noise levels are found to be adequate across all channels, with crosstalk assessed and minimized. The average straw hit resolution is $(110 \pm 20) \,\mu$\si{\meter}. The detection efficiency ranges from 97\% at the edge of the straw to 99.3\% in the main body of the straw, with a decrease of less than 2\% at the wire. Since the beginning of operation, no degradations in performance or signs of aging have been observed. 

The tracking detector's measurements of the spatial and temporal motion and the average spatial profile of the stored muon beam have been invaluable to the Muon $g-2$ Experiment’s first result~\cite{Run1PRL,Run1BD,Run1wa,Run1field}. In particular, they have contributed significantly to the overall understanding of the necessary systematic corrections and uncertainties. The determination of two of the largest systematic corrections described in~\cite{Run1PRL,Run1BD}: the pitch correction, $C_p = (180 \pm 13)$~ppb, and the phase-acceptance correction, $C_{pa} = (-158 \pm 75)$~ppb, together comprise of 25\% of the total Run 1 systematic uncertainty of 157 ppb and would not have been possible without the tracking detector. Furthermore, a detailed understanding of the beam motion during the first few rotations of the storage ring when the majority of the muons decay is critical to enhancing the statistical precision of the measurement. Without the information from the tracking detectors of this early motion, the overall uncertainty of the measurement would be larger since it would be necessary to either ignore the early data, thus increasing the statistical uncertainty, or rely on simulation to extract the details of the beam motion with an attendant increase in the systematic uncertainty. As such, the tracker detector will continue to be crucial for all future results from the Muon $g-2$ Experiment at Fermilab, particularly when the measurement transitions to becoming dominated by the systematic uncertainties.

\acknowledgments

We would like to pay our gratitude and respect to our friend and colleague, Dr.~Barry~T.~King of the University of Liverpool, who sadly passed away in January 2019. He and his unwavering enthusiasm were integral to the successes of this work. We would also like to thank the dozens of BU and NIU undergraduate students who contributed to this work and the staff of the Fermilab Test Beam Facility.

This document was prepared by the Muon $g-2$ Experiment using the resources of the Fermi National Accelerator Laboratory (Fermilab), a U.S. Department of Energy, Office of Science, HEP User Facility. Fermilab is managed by Fermi Research Alliance, LLC (FRA), acting under Contract No. DE-AC02-07CH11359. Additional support for this work was provided by the
Department of Energy offices of HEP and NP (USA),
the National Science Foundation (USA), the Science and
Technology Facilities Council (UK) and the European Union's Horizon 2020 research and
innovation programme under the Marie Sk\l{}odowska-Curie grant agreement
No. 690835 and No.734303 (MUSE).

%\paragraph{Note added.} This is also a good position for notes added
%after the paper has been written.


%%%%%%%%%%%%%%%%%%%%%%%%%%%%%%%%%%%%%%%%

%%%%%%%%%%%%%%%%%%%%%%%%%%%%%%%%%%%%%%%%

\end{document}